\documentclass[onecolumn, floatfix, preprintnumbers, eqsecnum, letterpaper, superscriptaddress, nofootinbib]{revtex4}
\usepackage{graphicx}
\usepackage{microtype}
\usepackage{amsmath}
\usepackage{amssymb}
\usepackage{subfigure}
\usepackage{hyperref}
\usepackage{url}
\usepackage{xcolor}
\usepackage{color}
\usepackage{mathrsfs}
\usepackage{calrsfs}
\usepackage{amsfonts}
\usepackage{eufrak}
\usepackage{tabularx}
\usepackage{eucal}
\usepackage{latexsym}
\usepackage{ragged2e}
\usepackage{epsfig}
\usepackage{textcomp}

\usepackage{caption}
\usepackage{subfigure}

\usepackage{marvosym}
\usepackage{ifsym}

\usepackage{lipsum}

\usepackage{tablefootnote}
\usepackage{booktabs}
 \usepackage{multirow}
 \usepackage[normalem]{ulem}

\DeclareCaptionJustification{justified}{\leftskip=0pt \rightskip=0pt \parfillskip=0pt plus 1fil}
\definecolor{vividviolet}{rgb}{0.62, 0.0, 1.0}
\definecolor{amaranth}{rgb}{0.9, 0.17, 0.31}
\definecolor{palatinateblue}{rgb}{0.15, 0.23, 0.89}
\definecolor{brightpink}{rgb}{1.0, 0.0, 0.5}
\definecolor{cornflowerblue}{rgb}{0.39, 0.58, 0.93}
\definecolor{deepcarminepink}{rgb}{0.94, 0.19, 0.22}
\definecolor{radicalred}{rgb}{1.0, 0.21, 0.37}

\hypersetup{ linktoc=all,
	colorlinks, linkcolor={palatinateblue},
	citecolor={brightpink}, urlcolor={amaranth}
}

\graphicspath{{Images/}}

\graphicspath{{Images/}}

\def\@fnsymbol#1{\ensuremath{\ifcase#1\or \ddagger \or  $\textleaf$  \or \dagger
		\else\@ctrerr\fi}}%

\makeatother

\def\sideremark#1{\ifvmode\leavevmode\fi\vadjust{\vbox to0pt{\vss
			\hbox to 0pt{\hskip\hsize\hskip1em
				\vbox{\hsize1.3cm\tiny\raggedright\pretolerance10000
					\noindent #1\hfill}\hss}\vbox to8pt{\vfil}\vss}}}%
\def\beq{\begin{equation}}
	\def\eeq{\end{equation}}

\begin{document}
\title{Investigating shadow images and rings of the charged Horndeski black hole  illuminated by various thin accretions}

\author{Xiao-Jun Gao}\email{xjgao2020@nuaa.edu.cn}
\address{College of Physics, Nanjing University of Aeronautics and Astronautics, Nanjing 211106, China}

\author{Tao-Tao Sui}\email{taotaosui@nuaa.edu.cn}
\address{College of Physics, Nanjing University of Aeronautics and Astronautics, Nanjing 211106, China}

\author{}
\address{Key Laboratory of Aerospace Information Materials and Physics (NUAA), MIIT, Nanjing 211106, China}

\author{Xiao-Xiong Zeng}\email{xxzengphysics@163.com}
\address{State Key Laboratory of Mountain Bridge and Tunnel Engineering, Chongqing Jiaotong University, Chongqing 400074, China}
\address{Department of Mechanics, Chongqing Jiaotong University, Chongqing 400074, China}

\author{Yu-Sen An}\email{anyusen@nuaa.edu.cn}
\address{College of Physics, Nanjing University of Aeronautics and Astronautics, Nanjing 211106, China}

\author{Ya-Peng Hu}\email{huyp@nuaa.edu.cn (corresponding author)}
\address{College of Physics, Nanjing University of Aeronautics and Astronautics, Nanjing 211106, China}
\address{Key Laboratory of Aerospace Information Materials and Physics (NUAA), MIIT, Nanjing 211106, China}

\begin{abstract}
In this paper, we investigate the shadows and rings of the charged Horndeski black hole illuminated by accretion flow that is both geometrically and optically thin. We consider two types of accretion models: spherical and thin-disk accretion flow. We find that in both types of models, the size of the charged Horndeski black hole shadow decreases with the increase of the charge, and it decreases more slowly for the Reissner-Nordstr\"om (RN) black hole.
In the spherical accretion flow model, we find that the increase of the charge of Horndeski black hole brightens the light ring around it, and it brightens more significantly in comparison with RN black hole. Due to the Doppler effect, the charged Horndeski black holes with accretion flow of radial motion have darker shadows than those with the static accretion flow, but the size of the shadow is not affected by accretion flow motion. In the thin disk-shaped accretion flow model, we find that the brightness of the light ring around the charged Horndeski black hole is dominated by the direct emission from the accretion flow, and the contribution from lensed rings is relatively small, and that from the photon rings is negligible. We also find that the ring brightness decreases as the charge of Horndeski black hole increases, and the decrease is more significant than that in the RN black hole case. Moreover, the radiation position of the accretion flow can affect the shadow size and the ring brightness of the charged Horndeski black hole.

\end{abstract}

\maketitle



\section{Introduction}
Black holes are thought of as the most mysterious one of celestial bodies predicted by General Relativity (GR). Recently, the Laser-Interferometer Gravitational Wave-Observatory (LIGO) has successfully detected the gravitational waves from the merger of two black holes \cite{LIGOScientific:2016aoc} and the Event Horizon Telescope (EHT) collaboration directly observed the ultra-high angular resolution images of supermassive black holes in the $M87^*$ \cite{EHT:2019dse,EHT:2019ths,EHT:2019ggy} and Sagittarius $A^*$ \cite{EHT:2022wkp,EHT:2022wok}. These observations strongly confirm the existence of black holes in our universe. Particularly, the black hole images present that there is a dark central region surrounded by a bright ring, which are called shadow and photon ring of the black hole, respectively. The formation of black hole shadow is due essentially to the deflection of light in the strong gravity field \cite{Synge:1966okc,Bardeen:1972fi,Bozza:2010xqn,Gao:2021lmo}, its image may carry some valuable information of the geometry around the black hole. Consequently, the investigation of the black hole shadow may provide a new window to constrain different gravity models \cite{Mizuno:2018lxz,Stepanian:2021vvk,Perlick:2021aok,KumarWalia:2022aop,Gao:2023ltr}.

Black holes can obtain the angular momentum via the gravitational collapse of a massive star. Thus we should think the astrophysical black holes are rotating black holes in realistic astrophysical settings. Bardeen illustrated that the shadow shape of the Kerr black hole is the deformation due to the drag effect \cite{Bardeen:1973tla}. Wang $et~ al.$ investigated the shadow of a Konoplya-Zhidenko rotating non-Kerr black hole with an extra deformation parameter \cite{Wang:2017hjl}. Haroon $et$ $al.$ studied the effects of perfect fluid dark matter and a cosmological constant on the shadow of a rotating black hole \cite{Haroon:2018ryd}. Wei $et$ $al.$ explored the nature of Gauss-Bonnet gravity by four-dimensional rotating black hole shadow \cite{Wei:2020ght}. The shadow of the rotating black hole has also been comprehensively investigated with many interesting results \cite{Cunha:2015yba,Zhu:2019ura,Meng:2022kjs,Kuang:2022ojj,Badia:2021kpk,Wang:2021ara,Long:2019nox,Li:2020drn}. The energy and angular momentum for a rotating black hole can be extracted by the Penrose superradiance process \cite{Penrose:2002} or the Blandford and Znajek mechanism \cite{Blandford:1977ds}, which would lead to a rotating black hole degenerating into a non-rotating one. Thus we cannot exclude the possibility that static black holes exist in the universe. The shadow of a static spherically symmetric black hole also is extremely important for studying the properties of the physics of a black hole, and for recent the literature see \cite{Junior:2021dyw,Meng:2023wgi,Ling:2021vgk,Cunha:2018acu,Wang:2019tjc,Hu:2020usx,Guo:2022nto}. It is well known that the accretion matter exists around a black hole, so their contribution is always very important and interesting, when we in detail study shadows and optical images of such black holes in the Universe.

Considering the astrophysical black hole is surrounded by a luminous accretion flow, a few pioneering works have investigated the nature of the shadow and photon ring of the black hole. Luminet investigated the optical appearance of a spherical black hole surrounded by thin accretion disk and pointed out that the shadow and photon ring observation characteristics are related to the position and profile of accretion flow \cite{Luminet:1979nyg}. Bambi studied apparent images of the Schwarzschild black hole and the static wormhole to distinguish black holes from wormholes \cite{Bambi:2013nla}. Gralla $et$ $al.$ primarily investigated the simple case of emission
from an optically and geometrically thin disk near a Schwarzschild black hole and proposed that the bright ring near the black hole shadow consists of the direct emission, lensed ring, and photon ring \cite{Gralla:2019xty}, which are determined by the times of the intersects between the light ray and the thin accretion disk. Narayan $et$ $al.$ explored a simple spherical model of optically thin accretion on a Schwarzschild black hole, and shown that the size of the observed shadow is hardly influenced by accretion flow \cite{Narayan:2019imo}.  Cunha $et$ $al.$ studied the lensing and the shadow of the Schwarzschild black hole with a thin and heavy accretion disk \cite{Cunha:2019hzj}. Zeng $et$ $al.$ investigated influence of the Gauss-Bonnet coupling parameter or quintessence dark energy on the optical appearance of the black hole with static/infalling spherical accretion flows \cite{Zeng:2020vsj,Zeng:2020dco}. Guo $et$ $al.$ showed the feature of the observed shadows and rings of the Hayward black hole depend on the accretion flow property and the black hole magnetic charge
\cite{Guo:2021bhr}. The photon ring and observational appearances of black holes have been extensively studied in \cite{Chakhchi:2022fls,Zeng:2022pvb,Hu:2022lek,Saleem:2023pyx,Wang:2023vcv,Zeng:2021mok,Okyay:2021nnh,Gan:2021xdl,Gan:2021pwu,Guerrero:2021ues,Zeng:2022fdm},
which disclosed the images of black holes surrounded by various accretions beyond GR.

Although GR successfully describes the gravitational interaction at the galactic and cosmological scales, it cannot give a satisfactory description for the accelerated expansion of our Universe \cite{Super:1998fmf,Super:1998vns}. Therefore, it is generally thought that GR might need to be modified to give a satisfactory explanation of this phenomena \cite{Clifton:2011jh}. The scalar-tensor theories are regarded as a simplest nontrivial modification of GR, which contain a scalar field $\varphi$ nonminimal couplings with a metric tensor $g_{\mu\nu}$. Horndeski has constructed the most famous four-dimensional scalar-tensor theories in 1974 \cite{Horndeski:1974wa}, which is called Horndeski gravity, by inspiration of the work of Lovelock \cite{Lovelock:1971yv}. Horndeski gravity has been extensively studied in P-V criticality \cite{Hu:2018qsy} and thermodynamics \cite{Walia:2021emv,Miao:2016aol} of these black holes, holographic applications \cite{Feng:2015oea,Kuang:2016edj,Baggioli:2017ojd,Feng:2018sqm} and other interesting features (please see \cite{Kobayashi:2019hrl} for a review). In addition, Horndeski theory might be able to evade the no-hair theorems \cite{Hui:2012qt, Sotiriou:2013qea,Sotiriou:2014pfa,Antoniou:2017acq}.

In this paper, we mainly focus on the shadows and photon rings of a charged Horndeski black hole with distinctive thin accretion flows, whose metric is obtained from the Horndeski theory with the scalar field coupled to the Einstein tensor in the presence of an electric field. Its thermodynamical properties were explored in \cite{Feng:2015wvb}, the weak and strong deflection gravitational lensing by a charged Horndeski black hole were analyzed in \cite{Wang:2019cuf}.
A black hole can have a positive net electric charge due to the twisting of magnetic field lines \cite{Wald:1974np} and the balance between the Coulomb and gravitational forces for charged particles near the surface of the compact object \cite{Bally:1978}. Hence, it is very important to study the properties of charged black holes. By the investigation of this paper, we find that the influence of the electric charge of the Horndeski black hole on event horizon, photon sphere, shadow size and brightness of light ring are more significantly in comparison with Reissner-Nordstr\"om (RN) black hole cases as the charge increases. If future observations show that black hole shadows are produced by charged black holes, this result may have important signification for verifying GR and Horndeski theory.

Our paper is organized as follows: In Sec. \ref{section2}, we discuss the effective potential and shadow radius of the charged Horndeski black hole, and study the trajectories of photons by utilizing the ray-tracing method. In Sec. \ref{section3} and \ref{section4}, we consider that the charged Horndeski black hole illuminated by spherical and thin disk accretion flows, respectively,  and explore the shadow contour, photon rings as well as the corresponding observed luminosity for a distant observer. The Sec. \ref{section5} is the conclusions and discussions. Throughout this paper we use the geometric units with $G=c=1$.

\section{Null geodesic and shadow radius of the charged Horndeski black hole}
\label{section2}
The most general action of Horndeski gravity has been constructed in \cite{Horndeski:1974wa}. In our paper, we only investigate a special case in Horndeski gravity with the scalar field $\varphi$ couple to the Einstein tensor $G_{\mu\nu}$ controlled by parameter $\eta$ in the presence of an electromagnetic field $F^{\mu\nu}$, and the corresponding action is written as \cite{Feng:2015wvb,Wang:2019cuf,Cisterna:2014nua}
\begin{align}
S=\dfrac{1}{16\pi}\int d^4x\sqrt{-g}\left[R+\dfrac{\eta}{2}G_{\mu\nu}\nabla^\mu\varphi\nabla^\nu\varphi-\dfrac{1}{4}F_{\mu\nu}F^{\mu\nu}\right],\label{Hornsec-new}
\end{align}
The variation of the action (\ref{Hornsec-new}) with respect to the metric tensor, the scalar field, and the electric field yields
\begin{align}
&G_{\mu\nu}=\dfrac{1}{2}\left(\eta T_{\mu\nu}+E_{\mu\nu}\right),\label{HornfirstfieldEq}\\
&\nabla_\mu\left[\eta G^{\mu\nu}\nabla_\nu\varphi\right]=0,\label{HornsecondfieldEq}\\
&\nabla_\mu F^{\mu\nu}=0,\label{HornthirdfieldEq}
\end{align}
respectively, where $T_{\mu\nu}$ and $E_{\mu\nu}$ are defined as
\begin{align}
T_{\mu\nu}=&\dfrac{1}{2}\nabla_{\mu}\varphi\nabla_{\nu}\varphi R-2\nabla_{\rho}\varphi\nabla_{(\mu}\varphi R^{\rho}_{\nu)}-\nabla^{\rho}\varphi\nabla^{\lambda}\varphi R_{\mu\rho\nu\lambda}\notag\\
&-(\nabla_{\mu}\nabla^{\rho}\varphi)(\nabla_{\nu}\nabla_{\rho}\varphi)+(\nabla_{\mu}\nabla_{\nu}\varphi)\Box\varphi
+\dfrac{1}{2}G_{\mu\nu}(\nabla\varphi)^2\notag\\
&-g_{\mu\nu}\left[-\dfrac{1}{2}(\nabla^{\rho}\nabla^{\lambda}\varphi)(\nabla_{\rho}\nabla_{\lambda}\varphi)+\dfrac{1}{2}(\Box\varphi)^2
-\nabla_{\rho}\varphi\nabla_{\lambda}\varphi R^{\rho\lambda}\right],\label{HornTmunu}\\
E_{\mu\nu}=&F_{\mu\sigma}F_{\nu}~^{\sigma}-\dfrac{1}{4}g_{\mu\nu}F_{\sigma\tau}F^{\sigma\tau}.\label{electricTmunu}
\end{align}

In this paper, we focus on an asymptotically flat solution describing the charged Horndeski black hole to (\ref{HornfirstfieldEq})-(\ref{HornthirdfieldEq}), while the metric, scalar field and Maxwell field are obtained \cite{Feng:2015wvb,Cisterna:2014nua}
\begin{align}
&ds^2=-A(r)dt^2+B(r)dr^2+r^2(d\theta^2+\sin^2\theta d\phi^2),\label{4Dspacetime}\\
&\varphi=\varphi(r),~~~~~~A=\Psi(r)dt,\label{varphiPsi}
\end{align}
where
\begin{align}
A(r)=&1-\dfrac{2M}{r}+\dfrac{4Q^2}{r^2}-\dfrac{4Q^4}{3r^4},~~B(r)=\dfrac{(8r^2-16Q^2)^2}{64r^4}A^{-1}(r),\label{Hornmetricvalue}\\
&\varphi'(r)=\sqrt{-\dfrac{8Q^2B(r)}{\eta r^2}},~~~~~~~~~\Psi(r)=\Psi_0-\dfrac{4Q}{r}+\dfrac{8Q^3}{3r^3},
\end{align}
where $M$ denotes the black hole mass, $Q$ is the total electric charge \footnote{Note that the $Q$ generally represents the electric charge parameter, but it is the total electric charge in \cite{Feng:2015wvb}. Their relation has been given in  the thermodynamical investigation of the back hole \cite{Feng:2015wvb}. }, and $\Psi_0$ is an integration constant. The (\ref{Hornmetricvalue}) are consistent with those of the Schwarzschild black hole as $Q=0$, while they can't reduce to the RN black hole metrics due to $\eta$ can not be zero\footnote{Using the equation of motion for the scalar field, one can obtain the equation (3.1) in reference \cite{Feng:2015wvb}. From this equation, it clearly shows that $\eta$ can't be zero. Therefore, we cannot switch off the scalar field within this solution family, which implies that this solution is usually not continuously connected with the RN solution \cite{Hu:2018qsy}.}.
Utilizing the (\ref{Hornmetricvalue}), we analytical derive the  event horizon radius
\begin{align}
r_+=&\frac{M}{2}+\frac{1}{2} \sqrt{M^2+\frac{2}{3} \left(\sqrt[3]{64 Q^6-18 M^2 Q^4}-4 Q^2\right)}\notag\\
&+\frac{1}{2} \sqrt{2 M^2-\frac{16 Q^2}{3}-\frac{1}{3} 2 \sqrt[3]{64 Q^6-18 M^2 Q^4}+\frac{2 \left(M^3-4 M Q^2\right)}{\sqrt{\frac{2}{3} \left(\sqrt[3]{64 Q^6-18 M^2 Q^4}-4 Q^2\right)+M^2}}}.\label{Hornoutr}
\end{align}
The $Q$ need to satisfy the condition to ensure the existence of the event horizon \cite{Feng:2015wvb,Wang:2019cuf}
\begin{align}
0<Q<\dfrac{3M}{4\sqrt{2}}.\label{chargecondition}
\end{align}
The charged Horndeski black hole (\ref{Hornmetricvalue}) is distinct from the RN black hole \cite{Reissner,Nordstr,Eiroa:2002mk}
\begin{align}
A_{RN}(r)=B^{-1}_{RN}(r)=1-\dfrac{2M}{r}+\dfrac{Q^2}{r^2};~~ 0<Q<M,\label{RNmetric}
\end{align}
by $A(r)B(r)\neq 1$ as well. Therefore, it is important that we analyze the different characteristics of two charged metric solutions via studying the shadow and optical appearance of the black hole.

\begin{figure}[!h]
\centering
\includegraphics[scale=0.50]{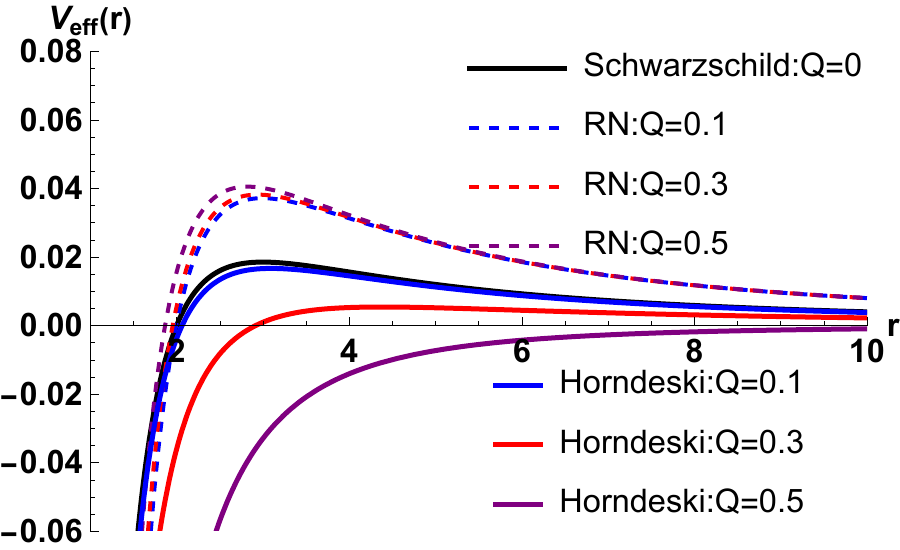}
\caption{Effective potential $V_{eff}(r)$ as a function of black hole radius $r$ for different electric charges with $b=M=1$.}
\label{potenfig}
\end{figure}

The Euler-Lagrange equation is given as follows:
\begin{align}
\dfrac{d}{d\lambda}\left(\dfrac{\partial \mathcal{L}}{\partial\dot{x}^\mu}\right)=\dfrac{\partial \mathcal{L}}{\partial x^\mu},\label{ELequation}
\end{align}
where $\lambda$ is the affine parameter of the light trajectory, and $\dot{x}^\mu$ denotes the four-velocity of the photon. The Lagrangian $\mathcal{L}$ of the photon can be written as
\begin{align}
\mathcal{L}=\frac{1}{2}g_{\mu \nu }\dot{x}^{\mu }\dot{x}^{\nu }=\frac{1}{2}\left[-A(r)\dot{t}^2+B(r)\dot{r}^2+r^2(\dot{\theta}^2+\sin^2\theta\dot{\phi}^2)\right].\label{Lagrangian}
\end{align}
Taking the spherical symmetry into account, we only investigate, without loss of generality, the light traveling on the equatorial plane of black hole, i.e. $\theta=\pi/2$ and $\dot{\theta}=0$. The spacetime metric in (\ref{4Dspacetime}) does not depend explicitly on time $t$ and azimuthal angle $\phi$. Hence, there are two conserved quantities in this spacetime correspond to the total energy and angular momentum of the photon, which read as
\begin{align}
E =- \frac{\partial \mathcal{L}}{\partial \dot{t}} = A(r)\dot{t} , \qquad L = \frac{\partial \mathcal{L}}{\partial \dot{\phi}} = r^2 \dot{\phi}.\label{twoconserved}
\end{align}
For the trajectory of the light, one impose the null condition $ds^2=0$. According to (\ref{4Dspacetime}) and (\ref{twoconserved}), one can easily derive the following equations of motion of photon:
\begin{align}
\dfrac{dt}{d\lambda}&=\dfrac{1}{bA(r)},\label{time}\\
\dfrac{d\phi}{d\lambda}&=\dfrac{1}{r^2},\label{phi}\\
\dfrac{dr}{d\lambda}&=\sqrt{\dfrac{1}{b^2A(r)B(r)}-\dfrac{1}{r^2B(r)}},\label{radial}
\end{align}
where the affine parameter $\lambda$ has been replaced with $\lambda L$, the impact parameter $b\equiv L/E$. The effective potential $V_{eff}(r)$ for the motion of a photon can be defined as
\begin{align}
V_{eff}(r)\equiv-\dfrac{1}{2}\dot{r}^2+C=-\dfrac{1}{2}\left[\dfrac{1}{b^2A(r)B(r)}-\dfrac{1}{B(r)r^2}\right]+\dfrac{1}{2b^2},\label{potential}
\end{align}
which has been demonstrated in the Appendix~\ref{A}. 
Submitting (\ref{Hornmetricvalue}) (or (\ref{RNmetric})) into (\ref{potential}), the charged Horndeski black hole (or RN black hole) effective potential as a function of radius is shown in Fig.\ref{potenfig} for different $Q$.
It is found that a larger electric charge leads to a smaller peak effective potential at the larger radius in the charged Horndeski black hole, and to the opposate situation in the RN black hole. In addition, the $V_{eff}(r)$ is affected more significantly in the charged Horndeski black hole in comparison with the RN black hole case with the increase of the $Q$ in Fig.\ref{potenfig}.

\begin{figure}[!h]
\centering
\includegraphics[scale=0.38]{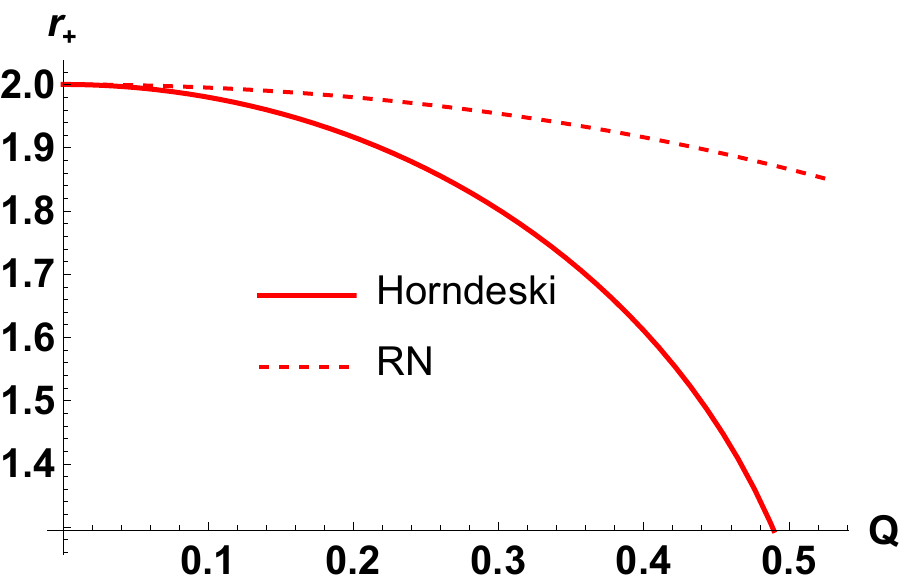}
\includegraphics[scale=0.38]{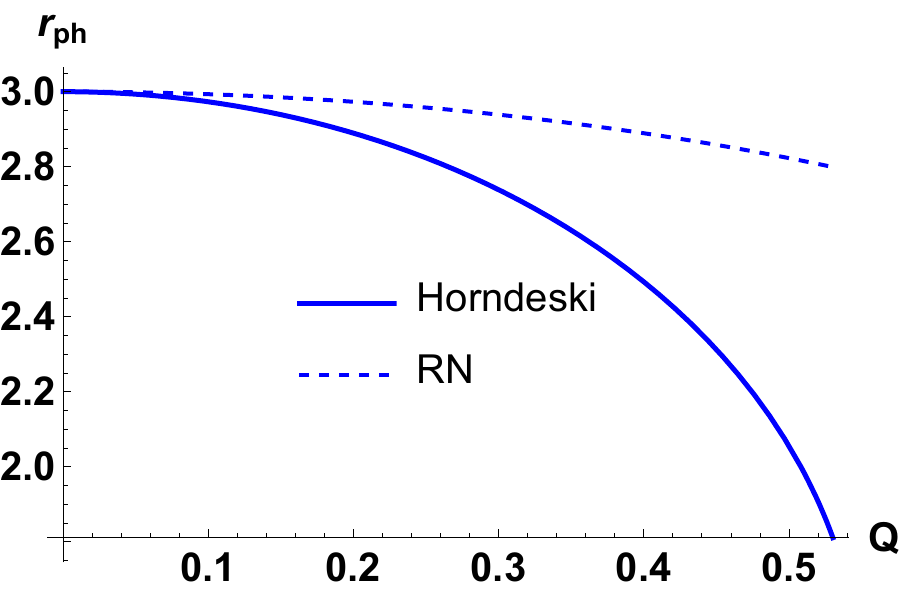}
\includegraphics[scale=0.38]{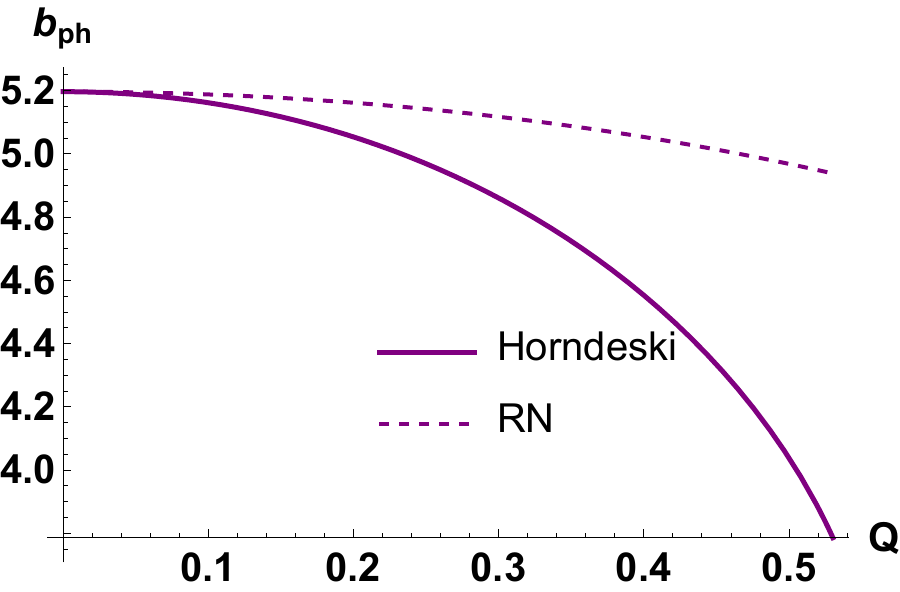}
\caption{The event horizon radius $r_+$, the radius of the photon sphere $r_{ph}$ and the critical impact parameter $b_{ph}$ as the function of the electric charge $Q$ with $M=1$.}
\label{outrphb}
\end{figure}

The motion of the photon should satisfy $\dot{r}=0$ and $\ddot{r}=0$ at the photon sphere, which result in from (\ref{potential})
\begin{align}
V_{eff}(r_{ph})=\dfrac{1}{2b_{ph}^2},\qquad V'_{eff}(r_{ph})=0,\label{effecteq}
\end{align}
where $b_{ph}\equiv b(r_{ph})$ is the critical impact parameter, and $r_{ph}$ is the radius of the photon sphere. From (\ref{effecteq}), one easily obtain
\begin{align}
r^2_{ph}=&A(r_{ph})b^2_{ph}, \label{phequation1}\\ b^2_{ph}A^2(r_{ph})\left[2B(r_{ph})+r_{ph}B'(r_{ph})\right]=&r^3_{ph}\left[B(r_{ph})A'(r_{ph})+B'(r_{ph})A(r_{ph})\right].\label{phequation}
\end{align}
Substituting (\ref{Hornmetricvalue}) into (\ref{phequation1}) and (\ref{phequation}), the sphere photon and the critical impact parameter are derived respectively as follows
\begin{align}
r_{ph}=&\frac{3 M}{4}+\frac{1}{2} \sqrt{\frac{9 M^2}{4}+\frac{1}{3} \left(\sqrt[3]{2} \delta +\frac{8\ 2^{2/3} Q^{4}}{\delta }-16 Q^2\right)}\notag\\
&+\frac{1}{2}\sqrt{\frac{9 M^2}{2}-\frac{32 Q^2}{3}-\frac{\sqrt[3]{2} \delta }{3}-\frac{8\ 2^{2/3} Q^4}{3 \delta }+\frac{27 M^3-96 M Q^2}{4 \sqrt{\frac{9 M^2}{4}+\frac{1}{3} \left(\sqrt[3]{2} \delta +\frac{8\ 2^{2/3} Q^4}{\delta }-16 Q^2\right)}}},\label{Hornrph}\\
b_{ph}=&\sqrt{\dfrac{r^2_{ph}}{1-\dfrac{2M}{r_{ph}}+\dfrac{4Q^2}{r^2_{ph}}-\dfrac{4Q^4}{3r^4_{ph}}}},\label{Hornbph}
\end{align}
where
\begin{align}
\delta=\sqrt[3]{Q^4 \left(-243 M^2-3 \sqrt{6561 M^4-44928 M^2 Q^2+76800 Q^4}+832 Q^2\right)}.
\end{align}
Using (\ref{Hornoutr}), (\ref{Hornrph}) and (\ref{Hornbph}), the charged Horndeski black hole outside event horizon radius $r_+$, the radius of the photon sphere $r_{ph}$ and the critical impact parameter $b_{ph}$ as the function of the electric charge $Q$ are plotted respectively in Fig.\ref{outrphb}. It clearly show that the $r_+$, $r_{ph}$ and $b_{ph}$ are gradually decrease with the $Q$ increasingly compared with the Schwarzschild back hole (Q=0) in the $Q$ region (\ref{chargecondition}), and the influences of the $Q$ of the Horndeski black hole are more significant than those in the RN black hole cases.

Based on (\ref{phi}) and (\ref{radial}), the trajectory of light can be given by
\begin{align}
\dfrac{dr}{d\phi}=r^2\sqrt{\dfrac{1}{b^2A(r)B(r)}-\dfrac{1}{r^2B(r)}}.\label{drdphiEq}
\end{align}
For convenience later, we introduce a parameter $u\equiv 1/r$, and then the (\ref{drdphiEq}) is rewritten as
\begin{align}
\dfrac{du}{d\phi}=\sqrt{\dfrac{1}{b^2A(u)B(u)}-\dfrac{u^2}{B(u)}}.\label{drdphiEq-new}
\end{align}
\begin{figure}[!h]
\centering
\includegraphics[scale=0.32]{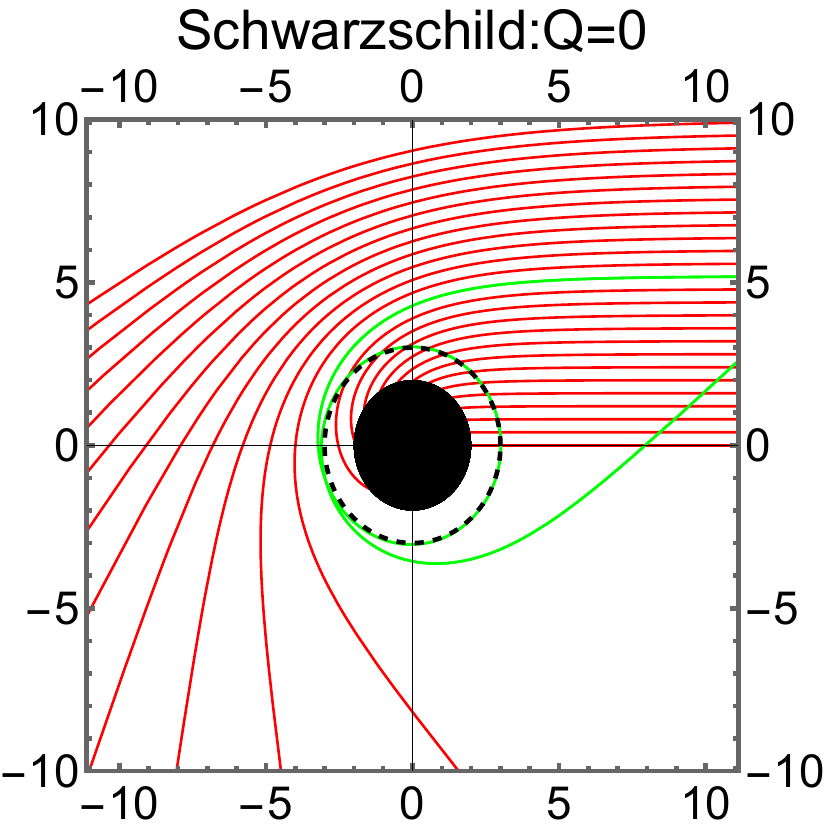}
\includegraphics[scale=0.32]{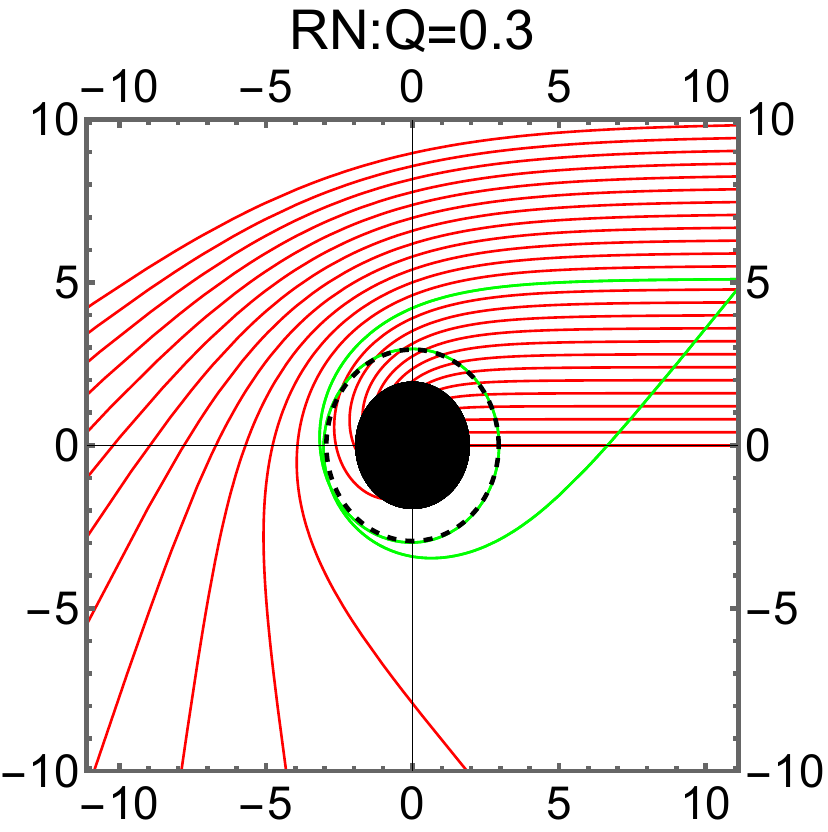}
\includegraphics[scale=0.32]{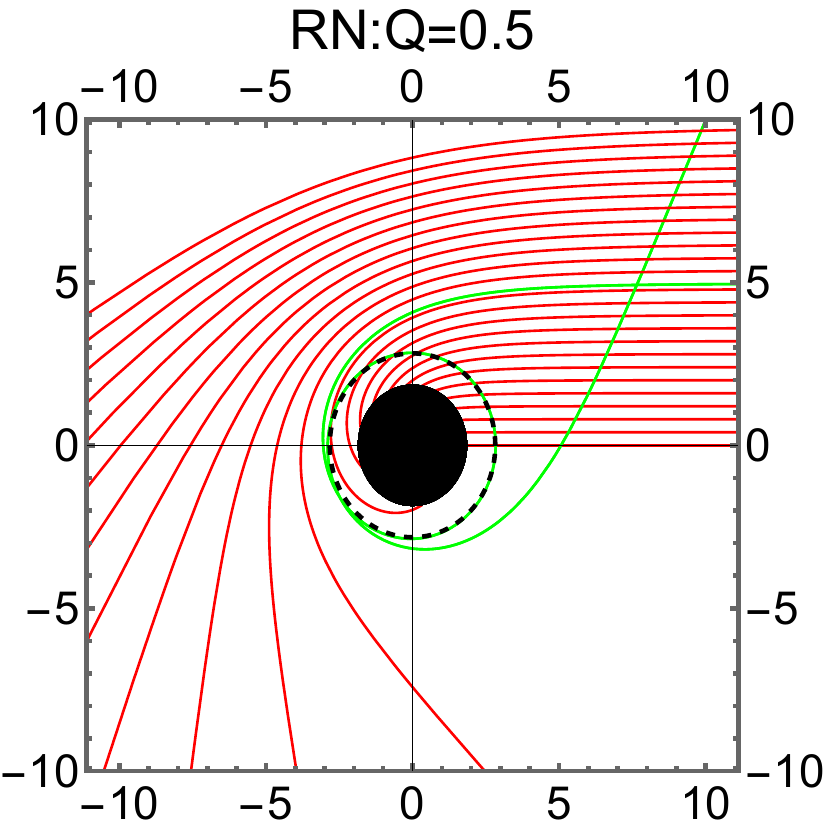}\\
\includegraphics[scale=0.32]{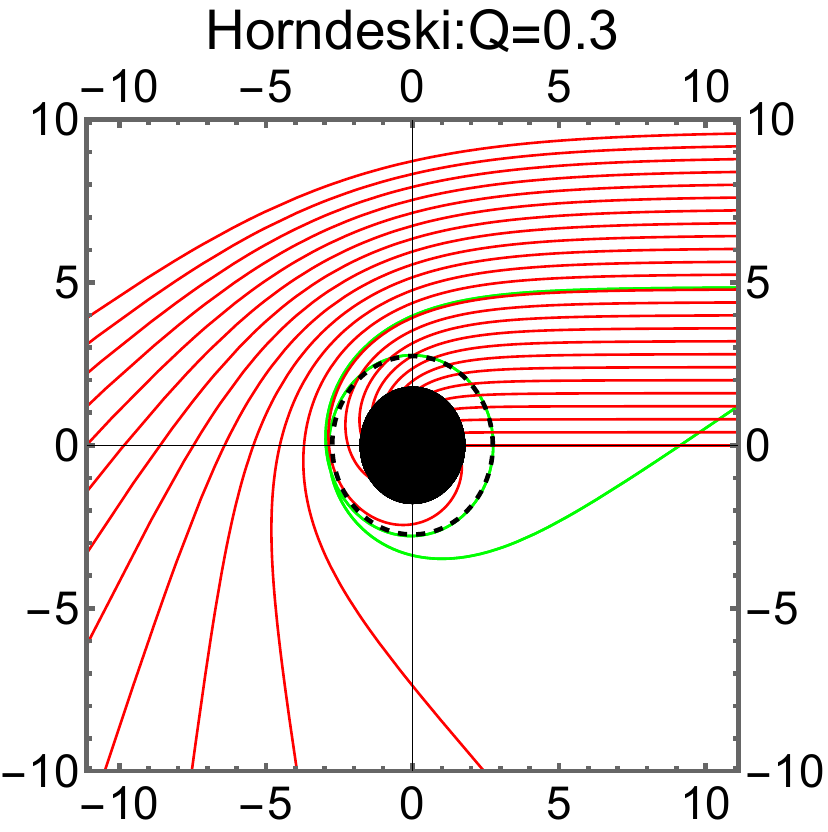}
\includegraphics[scale=0.32]{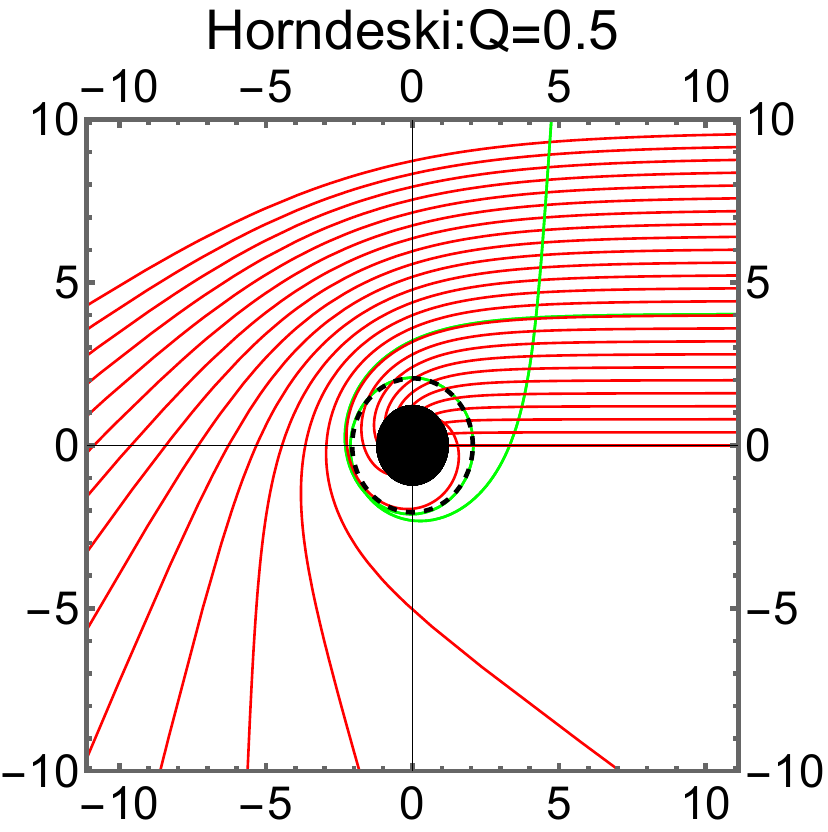}
\caption{Trajectories of light rays around black hole of different $Q$ in the polar coordinates $(r, \phi)$ with $M=1$.}
\label{orbitfig}
\end{figure}
Utilizing the ray-tracing code, in Fig.\ref{orbitfig}, we have plotted the trajectories of the photons around the charged Horndeski black hole for different values of $Q$. In Fig.\ref{orbitfig}, the black dashed line represents the photon sphere ($r=r_{ph}$), and the projection of the event horizon surface is regarded as a black disk. On the green line ($b=b_{ph}$), the light rays will make a circular motion around the black hole infinitely many times without perturbations; the area above the green line ($b>b_{ph}$), the light rays will encounter the potential barrier (with the right hand of peak in Fig.\ref{potenfig}) and then are scattered to infinite after passing through the turn point; the area under the green line ($b<b_{ph}$), the light rays will move in the inward direction (with the left hand of peak in Fig.\ref{potenfig}) and are absorbed by the black hole eventually. This kind of light rays cannot be observed by the distant observer, so a shadow forms in the observational sky. Compared with the Schwarzschild back hole (Q=0), the Fig.\ref{orbitfig} shows that the photon sphere and the black disk are smaller with the increase of the charge of the Horndeski black hole, expect the deflection angle increases, and their changes are more slowly for the RN black hole case.

\section{Shadows and rings of the charged Horndeski black hole  illuminated by different spherical accretion flows}
\label{section3}
There is plenty of free-moving matters in the Milky Way that might be accreted by its central black hole. The light rays radiated by these substances are the light source that illuminates the black hole. In this section, we will continue to investigate the properties of the shadows and photon rings of the charged Horndeski (or RN) black hole with static and infalling spherical accretion flows, by regarding them as optically and geometrically thin.

\subsection{The static spherical accretion flow model}
We firstly consider a static spherically symmetric accretion flow distributed outside the event horizon of a charged Horndeski (or RN) black hole. For an observer at infinity, the observed specific intensity (measured in ${\rm erg}~ {\rm s}^{-1}~ {\rm cm}^{-2}~ {\rm str}^{-1}~ {\rm Hz}^{-1}$) of photon with a frequency $v^s_o$ is determined
by \cite{Bambi:2013nla,Jaroszynski:1997bw}
\begin{align}
I(v^s_o)=\int g^{s 3}j(v^s_{em})dl_{prop},\label{obsintenEq}
\end{align}
where $v^s_{em}$ is the intrinsic photon frequency; $g^{s}$ is the red-shift factor, which can be defined by $g^{s}\equiv v^s_o/v^s_{em}$; $dl_{prop}$ represents the infinitesimal proper length; $j(v^s_{em})$ is the emissivity per unit volume in the rest frame of the emitter. Integrating (\ref{obsintenEq}) over all the observed frequencies, the total observed intensity is given by
\begin{align}
I^s_{obs}=\int I(v^s_o)dv^s_o=\int\int g^{s 4}j(v^s_{em})dl_{prop}dv^s_{em}.\label{totalintenstate}
\end{align}

From the four-dimensional spherical symmetric black hole spacetime (\ref{4Dspacetime}), the red-shift factor can be written as $g^{s}=A(r)^{1/2}$. In addition, assuming a simple case that the radiation of light is monochromatic with fixed a frequency $v_f$, thus the specific emissivity takes the form
\begin{align}
j(v^s_{em})\propto \dfrac{\delta(v^s_{em}-v_f)}{r^2},\label{specemissEq}
\end{align}
and the proper length measured in the rest frame of the emitter can be expressed as
\begin{align}
dl_{prop}=\sqrt{B(r)+r^2\left(\dfrac{d\phi}{dr}\right)^2}dr,\label{properlengthEq}
\end{align}
where $d\phi/dr$ is given from (\ref{drdphiEq}). Putting (\ref{specemissEq}) and (\ref{properlengthEq}) into (\ref{totalintenstate}), the total photon intensity measured
by the distant observer can be written as
\begin{align}
I^s_{obs}=\int\dfrac{A(r)^{2}}{r^2}\sqrt{B(r)+\dfrac{A(r)B(r)b^2}{r^2-b^2A(r)}}dr.\label{stotalinten}
\end{align}

\begin{figure}[!h]
\centering
\includegraphics[scale=0.50]{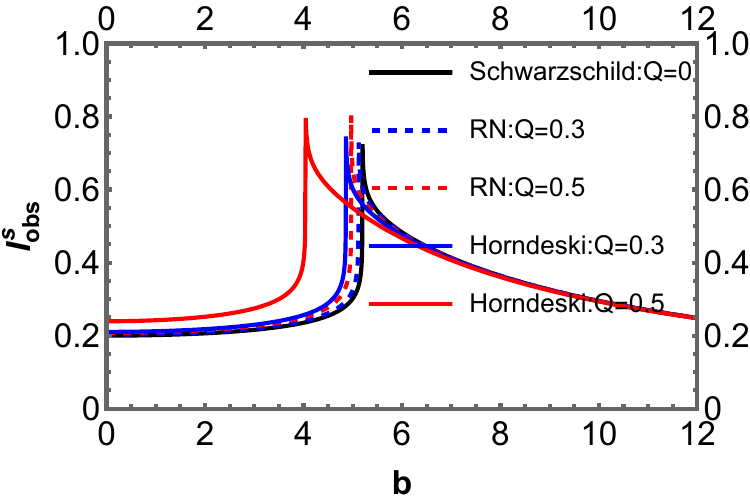}
\caption{The total observed intensities $I^s_{obs}$ as a function of the impact parameter $b$ with a static spherical accretion flow for $M=1$.}
\label{stateIb}
\end{figure}
\begin{figure}[!h]
\centering
\includegraphics[scale=0.46]{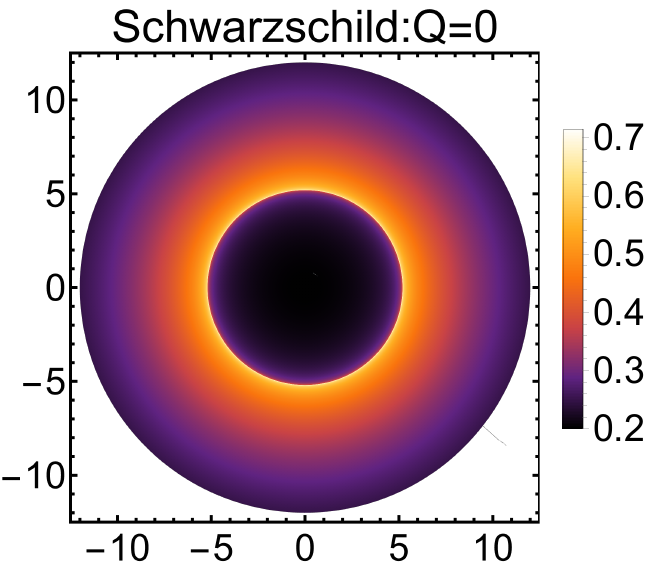}
\includegraphics[scale=0.46]{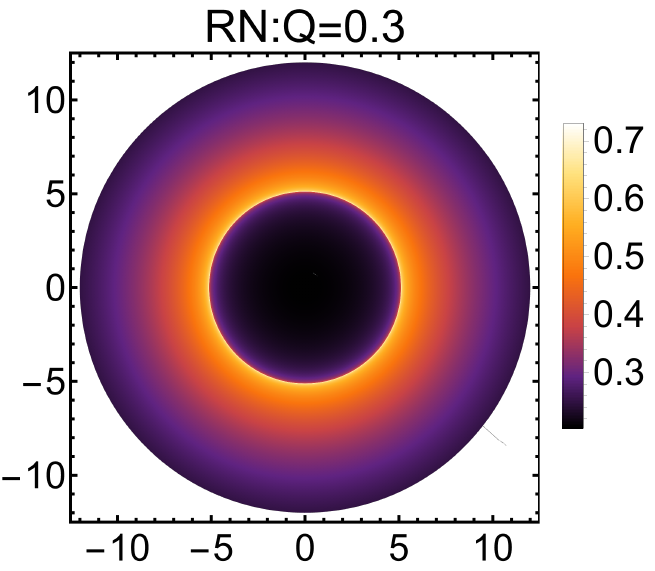}
\includegraphics[scale=0.46]{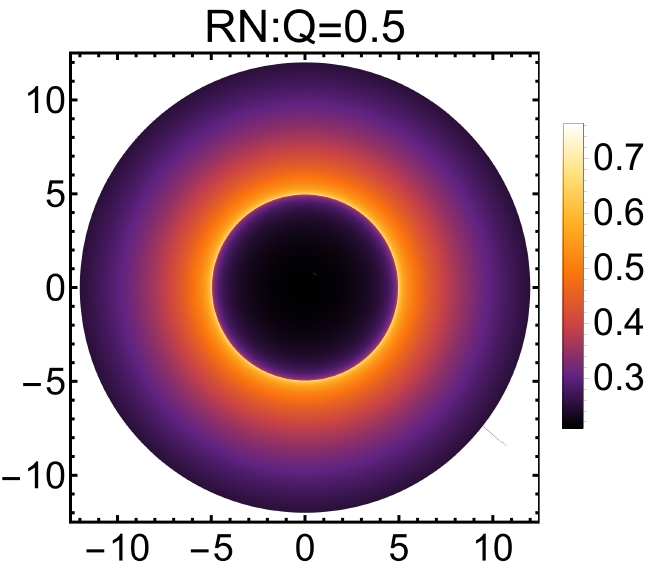}\\
\includegraphics[scale=0.46]{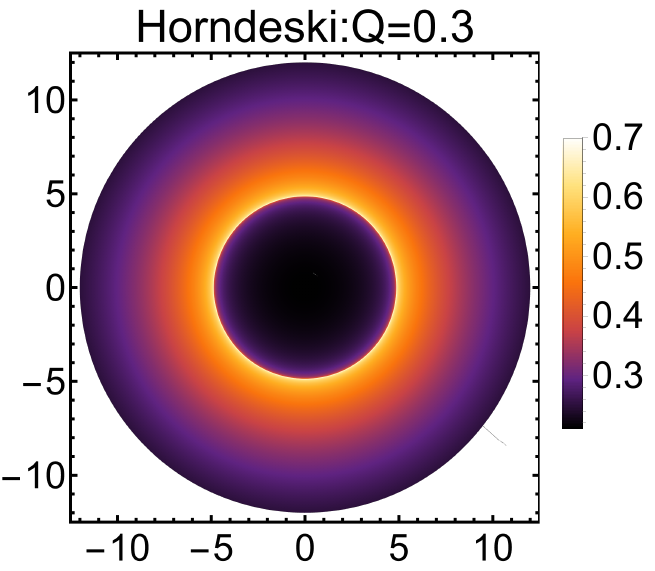}
\includegraphics[scale=0.46]{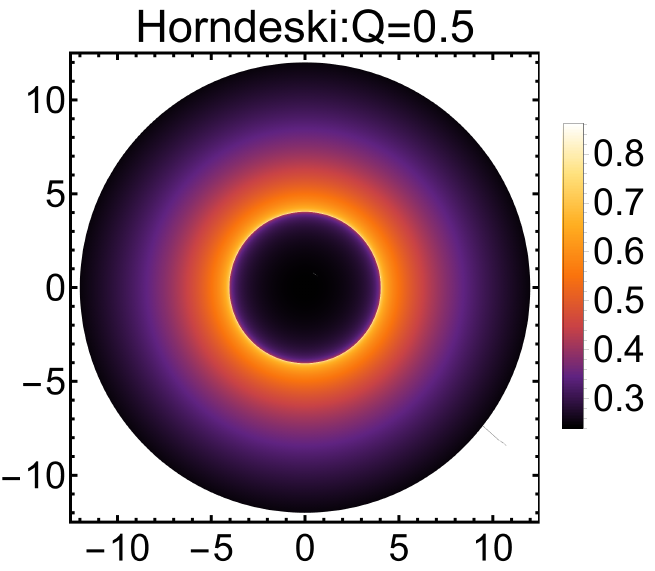}
\caption{Two-dimensional images of shadows and photon rings of the black holes with a static spherical accretion flow for $M=1$.}
\label{stateintenfig}
\end{figure}
The total observed intensities $I^s_{obs}$ as a function of the impact parameter $b$ are shown for several representative values of electric charge in Fig.\ref{stateIb}. One can observe that the intensity curve sharply peaks at $b_{ph}$, which corresponds to the photon rings. Meanwhile, the observed intensity is stronger as a larger $Q$ in the charged Horndeski black hole, but the corresponding $b_{ph}$ gets smaller. We observe that the observed intensity is more obvious in the charged Horndeski black hole than those of the RN black hole with the increase of $Q$ in Fig.\ref{stateIb}. The Fig.\ref{stateintenfig} demonstrates that the two-dimensional shadows cast of the charged Horndeski (or RN) black holes in the celestial coordinates. One can see clearly that the bright ring outside the shadow of black hole is the photon sphere, here the luminosity is strongest. Furthermore, the specific intensity among dark shadow does not completely vanish but has a small finite value. The reason is that the tiny fraction of radiating gas behind the black hole can escape to infinity due to spherically symmetric accretion flow.

\subsection{The infalling spherical accretion flow model}
In realistic astrophysical settings, the substances with the intrinsic initial velocity may move toward the centre of celestial body.
For example, the accretion flows fall the centre of the supermassive black hole in $M87^{*}$\cite{Falcke:1999pj}. For simplicity, we consider a charged Horndeski (or RN) black hole that is surrounded by radial infalling spherical accretion flow in this section. In the scenario, the (\ref{totalintenstate}) still is valid to calculate the total observed intensity, but the red-shift factor is related to the velocity of the accretion flow, thus it rewritten as \cite{Bambi:2013nla}
\begin{align}
g^i=\dfrac{k_\mu u^{\mu i}_o}{k_\nu u^{\nu i}_{em}},\label{infalingfac}
\end{align}
where $k_\mu$ is the four-velocity of the photon; $u^{\mu i}_o$ is the four velocity of an observer; $u^{\nu i}_{em}=(u^{i t}_{em},u^{i r}_{em},0,0)$ is the four-velocity of the accretion flow. Utilizing (\ref{time})-(\ref{radial}), one easily obtain
\begin{align}
k_t=\dfrac{1}{b},~~k_r=\pm\sqrt{\dfrac{B(r)}{b^2A(r)}-\dfrac{B(r)}{r^2}},\label{ktkr}
\end{align}
and then their ratio
\begin{align}
\dfrac{k_r}{k_t}=\pm\sqrt{\dfrac{1}{A(r)}\left(B(r)-\dfrac{b^2A(r)B(r)}{r^2}\right)},\label{ktkrratio}
\end{align}
in which the symbol $\pm$ corresponds to the photon is approaching ($+$) or away ($-$) from the black hole. For the simplicity of analysis, we assume that the distant
observer is stationary with $u^{\mu i}_o=(1,0,0,0)$, and the four-velocity of the infalling spherical accretion flow $u^{\nu i}_{em}$ are given by
\begin{align}
u^{i t}_{em}=\dfrac{1}{A(r)},~~u^{i r}_{em}=-\sqrt{\dfrac{1}{A(r)B(r)}-\dfrac{1}{B(r)}},~~u^{i \theta}_{em}=u^{i \phi}_{em}=0,\label{utrthetaphi}
\end{align}
respectively. For the scenario, the red-shift factor in (\ref{infalingfac}) is rewritten as
\begin{align}
g^i=\left(u^{i t}_{em}+\dfrac{k_r}{k_t}u^{i r}_{em}\right)^{-1}
\end{align}
and the proper distance is expressed by
\begin{align}
dl_{prop}=k_\sigma u^{\sigma i}_{em}d\lambda=\dfrac{k_t}{g^i|k_r|}dr.\label{infallingprop}
\end{align}
Finally, the total observed intensity of the infalling spherical accretion flow in (\ref{totalintenstate}) is obtained
\begin{align}
I^i_{obs}=\int\dfrac{g^{i 3}}{r^2}\dfrac{1}{\sqrt{\dfrac{1}{A(r)}\left(B(r)-\dfrac{b^2A(r)B(r)}{r^2}\right)}}dr.\label{infallinginten}
\end{align}

\begin{figure}[htbp]
\centering
\includegraphics[scale=0.50]{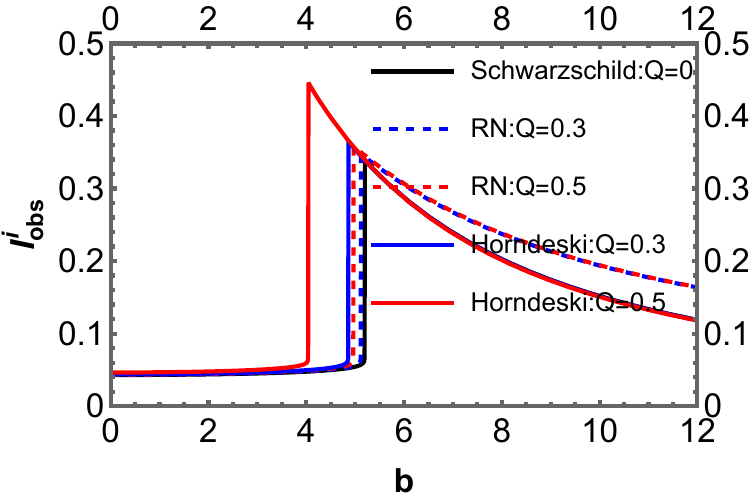}
\caption{The observed specific intensities $I^i_{obs}$ as a function of the impact parameter $b$ with an infalling spherical accretion flow for $M=1$.}
\label{infallingIb}
\end{figure}

Fig.\ref{infallingIb} plots that the total observed intensity $I^i_{obs}$ of the charged Horndeski (or RN) black hole surrounded by an infalling accretion flow as a function of the impact parameter $b$. 
\begin{figure}[!h]
\centering
\includegraphics[scale=0.48]{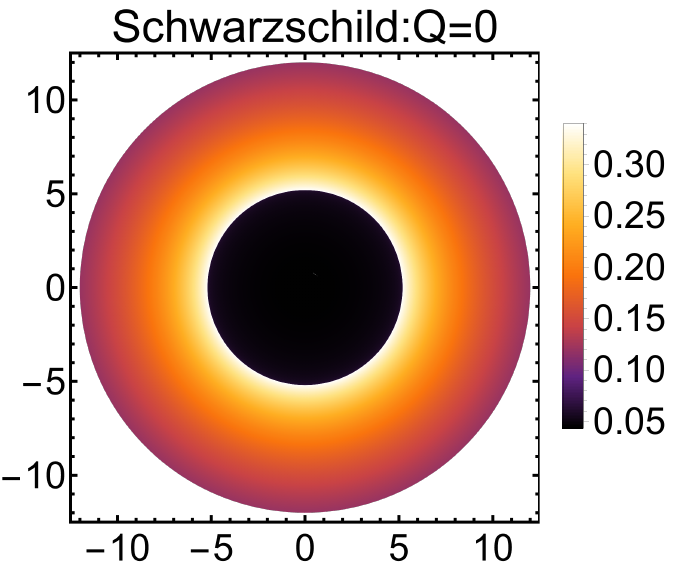}
\includegraphics[scale=0.48]{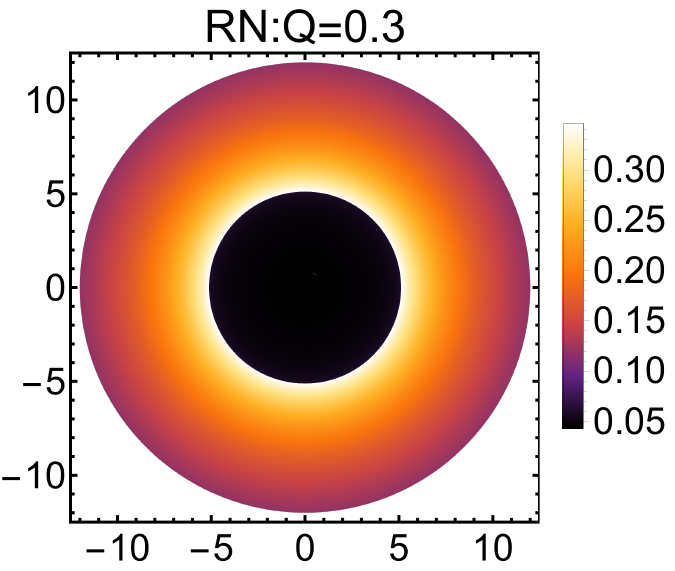}
\includegraphics[scale=0.48]{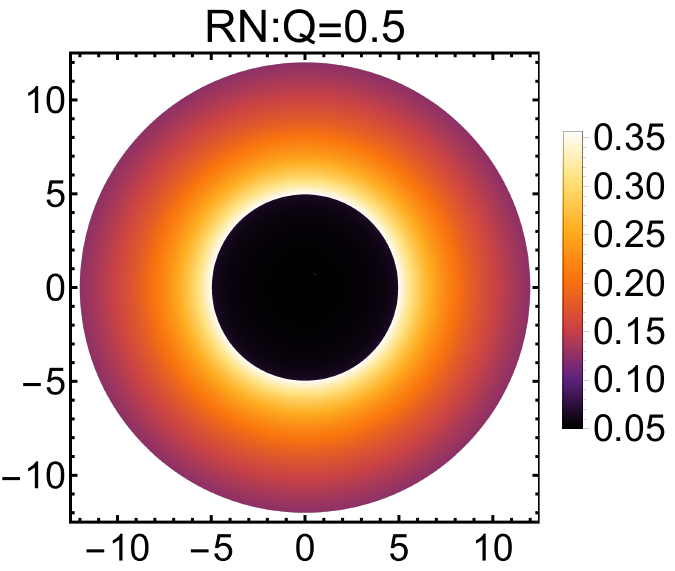}\\
\includegraphics[scale=0.48]{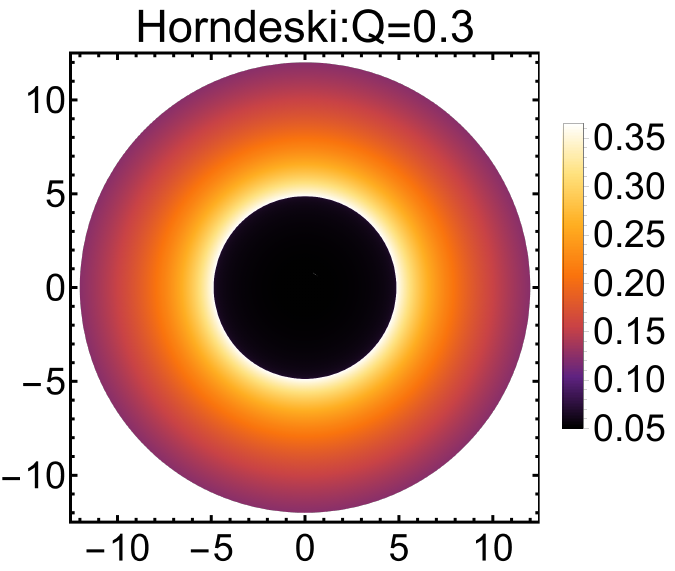}
\includegraphics[scale=0.48]{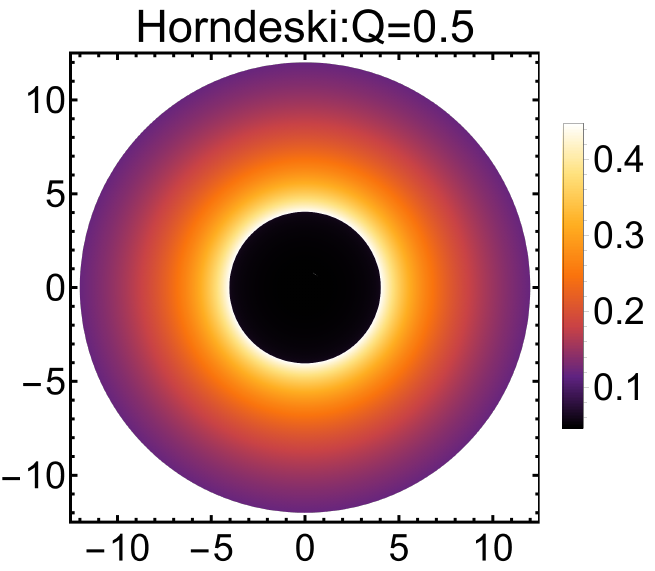}
\caption{Two-dimensional images of shadows and photon rings of the black holes with an infalling spherical accretion flow for $M=1$.}
\label{infintenfig}
\end{figure}
The features of the function curves are similar to the Fig.\ref{stateIb}, except the intensity has an extremely sharp rise before the peak. Fig.\ref{infintenfig} shows that the two-dimensional shadows cast on the celestial coordinates in this scenario. In the same way, the observed intensity and shadow size are increase faster with the increase of the $Q$ in the charged Horndeski black hole, while it is slower in the RN black hole.
One can intuitively see that the black hole shadows with state accretion flow of the Fig.\ref{stateintenfig} are brighter than those with the infalling accretion flow of the Fig.\ref{infintenfig} due to the Doppler effect, but the size of the black hole shadows does not change with the same electric charge. The results manifest the size of the black hole shadow depends on the spacetime geometry, while the luminosity of the black hole shadow relies on the accretion flow model.

\section{Shadows and rings of the charged Horndeski black hole illuminated by thin disk accretion flow}
\label{section4}
The compact celestial body (i.e. white dwarf, neutron star and black hole) can accrete various matters around it that forms an accretion disk in our Universe. There are some accretion disk models near the compact object have been investigated in \cite{Shakura:1972te,Shapiro:1976fr,Paczynski:1979rz,Narayan:1994is}. The accretion substances can be regarded as an light source that illuminates the compact object. Therefore, the investigation of the optical appearance around compact object may help us to constraint the accretion disk model and explore the distribution of matters around it.
In this section, considering a simple model that the substances around the black hole are regarded as an optically and geometrically thin disk-shaped accretion disk \cite{Shapiro:1976fr}, and one mainly focus on the emission from the accretion disk is isotropic in the rest frame of static world-lines, the disk is located on the equatorial plane and the observer is at the north pole.

\subsection{Direct emission, lensed ring and photon ring}
The literature \cite{Gralla:2019xty} manifested that the black hole shadow with the thin-disk accretion is surrounded by photon rings and lensed rings. Based on the definition of the total number of light orbits, i.e. $n\equiv\phi/2\pi$, the light rings near the black hole can be divided into three types~\cite{Gralla:2019xty}:

\textbf{(i)}~$n<3/4$ is the direct emission, implying the light trajectories only intersect with the thin disk once.

\textbf{(ii)}~$3/4<n<5/4$ is lensed ring, meaning the light trajectories of intersect with the the thin disk at twice.

\textbf{(iii)}~$n>5/4$ is photon ring , denoting the light trajectories intersect the equatorial plane at least three times.
\begin{table*}[h]
\setlength{\abovecaptionskip}{0.2cm}
\setlength{\belowcaptionskip}{0.3cm}
\caption{The range of impact parameter corresponding to direct emission, lensed ring and photon ring of the charged Horndeski (or RN) black hole, where the black hole mass as $M=1$, and the electric charge taking as $Q=0,0.1,0.3,0.5$.}
\renewcommand\arraystretch{1.3}
\setlength{\tabcolsep}{2.5mm}{
\begin{tabular}{|c|c|c|c|c|}
\hline
Q                & Black hole & Direct emission & Lensed ring & Photon ring \\ \hline
0 & Schwarzschild &$b < 5.01514$; $b > 6.16757$ &$5.01514 < b < 5.18781$; $5.22794 < b < 6.16757$ &$5.18781 < b < 5.22794$  \\ \hline
\multirow{3}{*}{0.1} & Horndeski &$b <4.97992$; $b > 6.13461$ &$4.97992 < b <5.15283$; $5.19330< b < 6.13461$ &$5.15283< b < 5.19330$   \\ \cline{2-5}
                      & RN  &$b<5.00600$;$b>6.16009$ &$5.00600 < b < 5.17908$; $5.21940 < b <6.16009$ &$5.17908<b<5.21940$ \\ \hline
\multirow{3}{*}{0.3} & Horndeski &$b<4.67449$;$b>5.85601$ &$4.67449< b < 4.85088$; $4.89503 < b <5.85601$ &$4.85088< b <4.89503$  \\ \cline{2-5}
                      & RN  &$b<4.93142$;$b>6.09956$ &$4.93142 < b < 5.10795$; $5.14990 < b <6.09956$ &$5.10795<b<5.14990$ \\ \hline
\multirow{3}{*}{0.5} & Horndeski &$b<3.80100$;$b>5.18139$ &$3.80100 < b <4.02122$; $4.09456 < b < 5.18139$ &$4.02122< b <4.09456 $  \\ \cline{2-5}
                      & RN  &$b<4.77294$;$b>5.97448$ &$4.77294 < b < 4.95793$; $5.00390 < b <5.97448$ &$4.95793<b<5.00390$ \\ \hline
\end{tabular}}
\label{table2}
\end{table*}

\begin{figure}[htbp]
\centering
\includegraphics[scale=0.55]{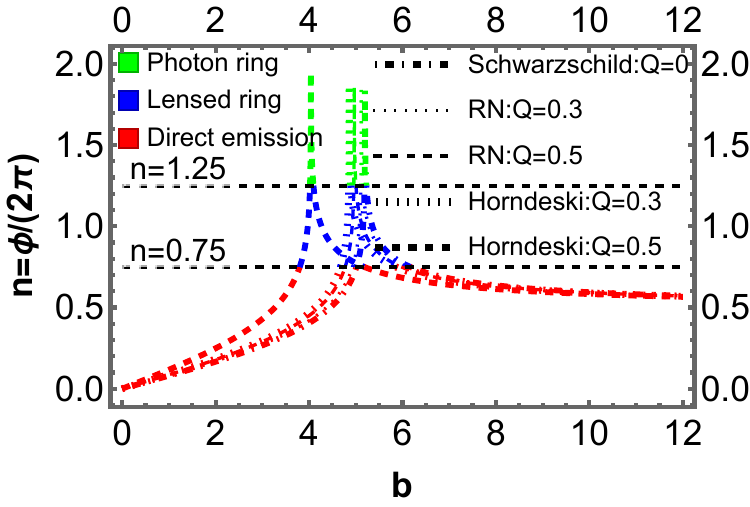}
\caption{The orbit numbers $n$ as the functions of impact parameter $b$ for different $Q$ with $M=1$.}
\label{Thindisknb}
\end{figure}
\begin{figure}[!htbp]
\centering
\includegraphics[scale=0.32]{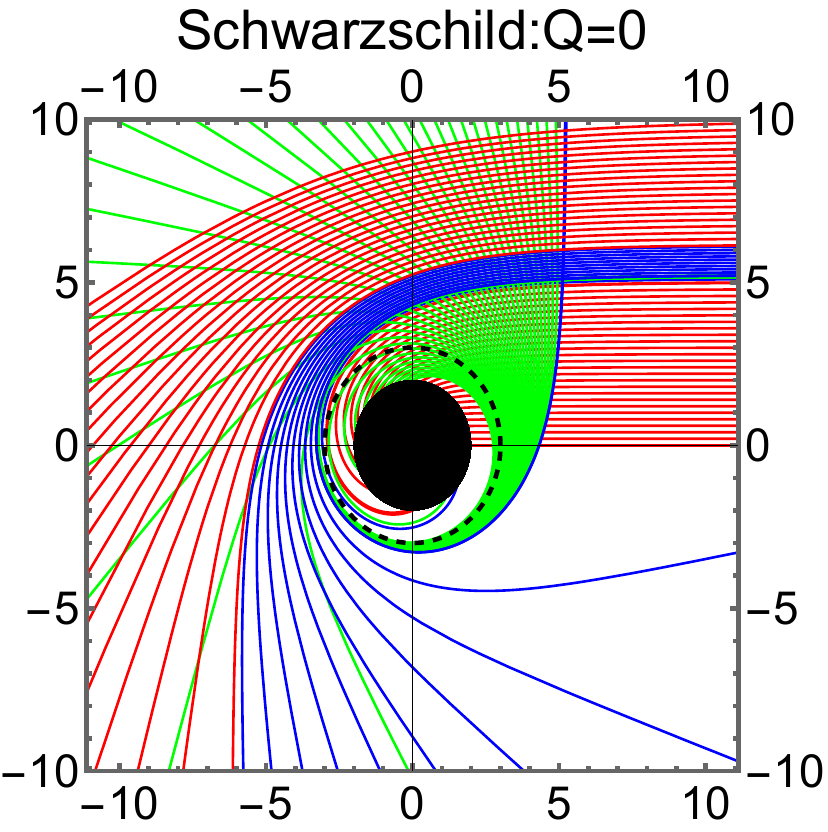}
\includegraphics[scale=0.32]{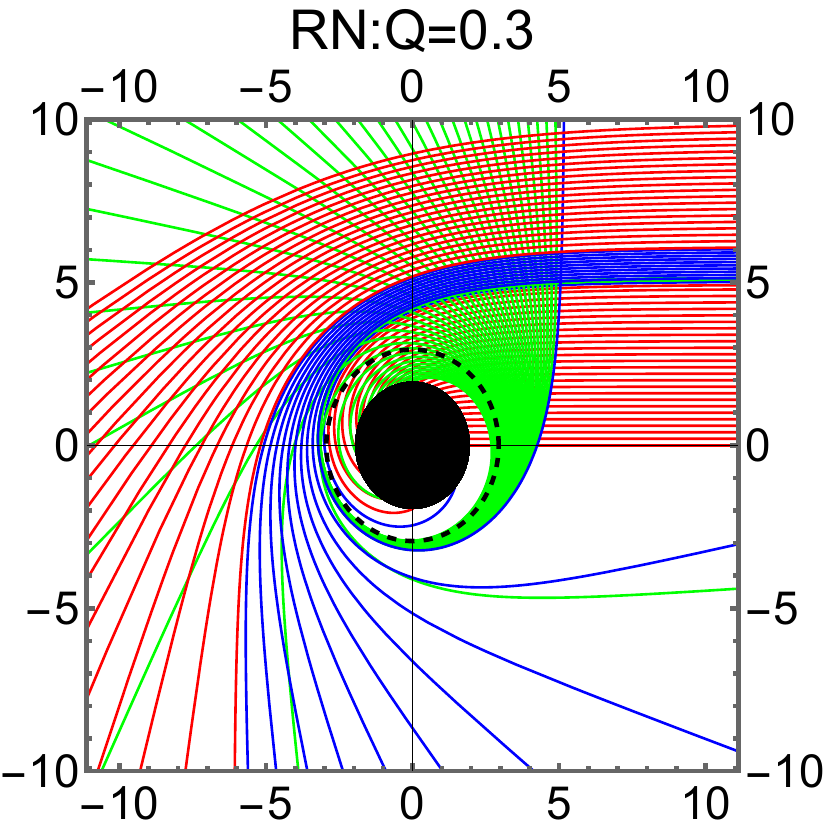}
\includegraphics[scale=0.32]{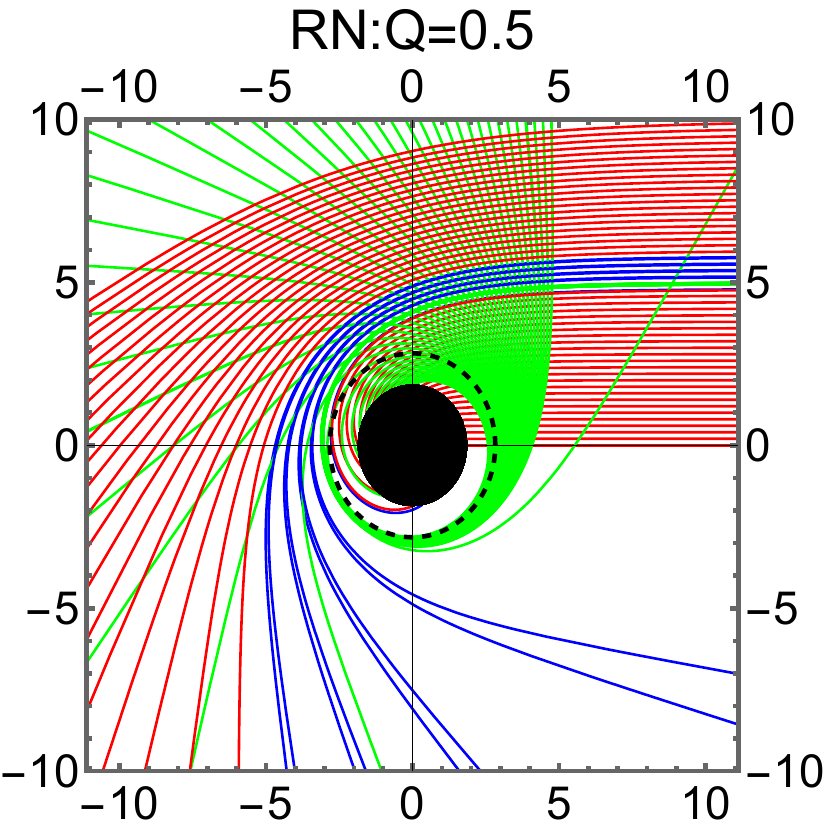}\\
\includegraphics[scale=0.32]{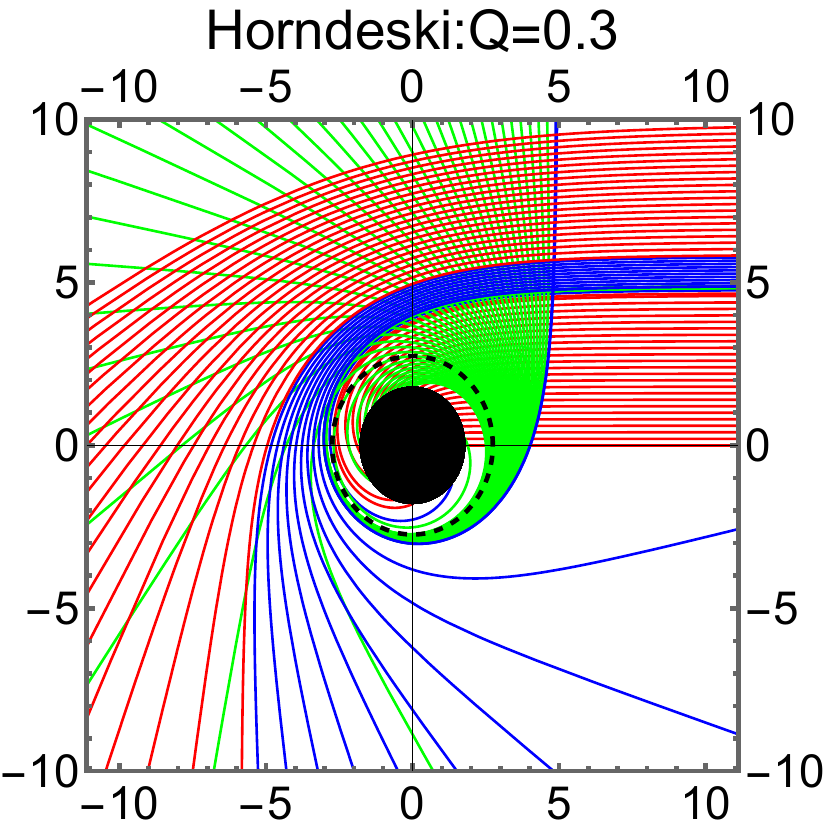}
\includegraphics[scale=0.32]{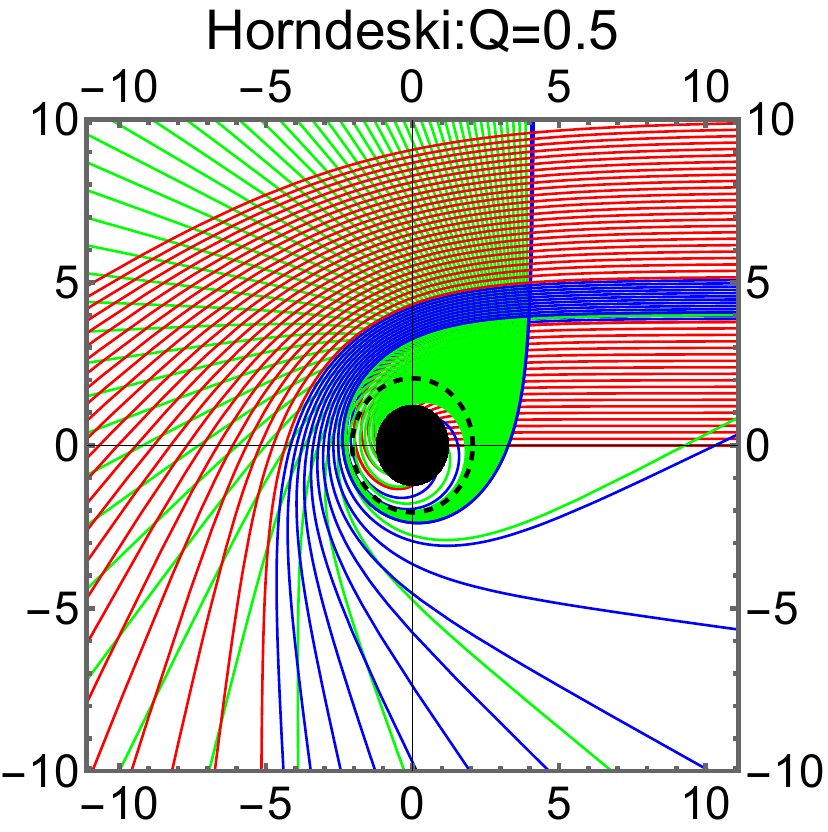}
\caption{The selection of associated photon trajectories for the black hole in the polar coordinates ($r,\phi$). The black dashed line is the photon sphere, the black disk is a black hole. These red, blue and green line represent the light trajectories intersect with the thin disk once, twice and least three times, respectively.}
\label{Thinphib}
\end{figure}
For various values of the $Q$, the ranges of the $b$ of the direct emission, lensed ring, and photon ring are respectively shown in Table~\ref{table2}. In addition, in order to depict these three regions in a more intuitive way, the $Q=0, 0.3, 0.5$ are chosen as examples. Fig.\ref{Thindisknb} presented the total number of orbits as a function of the impact parameter, and Fig.\ref{Thinphib} shows that the corresponding photon trajectories around the black hole as well. The black dashed circle is the photon sphere and the black disk stands for the black hole. Combining with Table~\ref{table2} and Fig.\ref{Thinphib}, we can find that the ranges of $b$ of the lensed rings and photon rings in the charged Horndeski black hole spacetime are narrower compared with the Schwarzschild black hole, meaning the thickness of the lensed rings and photon rings are getting thinner with existence of the electric charge. In addition, the $b$ regions of the lensed rings and photon rings in the charged Horndeski black hole shrink faster than those of the RN black hole as a larger $Q$.

\subsection{Total observed intensity and transfer functions}
The specific intensity and frequency of the emitted light from the thin accretion disk are denoted as $I_{e}$ and $v_e$. Based on the Liouville's theorem, the $I_e/(v_e)^3$ is an invariant in the direction of light propagation, thus one can give
\begin{align}
\dfrac{I_e}{v_e^3}=\dfrac{I_o}{v_o^3},\label{Liouville}
\end{align}
where the $I_o$ is the observed specific intensity measured by the infinity observer, and $v_o=\sqrt{A(r)}v_e$ is corresponding to the red-shifted frequency. The observed intensity for a specific frequency can be expressed as
\begin{align}
I_o(r)=A(r)^{3/2}I_e(r).\label{Thininten}
\end{align}
Therefore, the total observed intensity can be obtained by an integral over all the observed photon frequencies
\begin{align}
I_{obs}(r)=\int I_o(r)dv_o=\int A(r)^2 I_e(r)dv_e=A(r)^2 I_{em}(r),\label{Thintotalobsint}
\end{align}
where the $I_{em}(r)\equiv \int I_e(r)dv_e$ denotes the total emitted specific intensity on the thin accretion disk.

One merely consider that the intensity of light is emitted by the thin disk accretion flows, and the absorption and reflection of light are ignored due to optically and geometrically thin. The light can pick up an additional brightness for each intersection between the light and the accretion disk on the equatorial plane.
As previous discussions, for $3/4<n<5/4$, the light ray will bend around the black hole and falls on the back of the accretion disk (See blue lines in Fig.\ref{Thinphib}). The light will pick up the additional brightness from the second crossing between the light and the accretion flow on thin disk. For $n>5/4$, the light again arrives at the front side of the thin disk (See green lines in Fig.\ref{Thinphib}), which will add more brightness from the third passage through the thin disk. Consequently, the sum of the intensity at each intersection is the total observed intensity, which should be written as
\begin{align}
I_{obs}(b)=\sum_{n}A(r)^2 I_{em}(r)\mid_{r=r_n(b)},\label{nbtotalinten}
\end{align}
\begin{figure}[htbp]
\centering
\includegraphics[scale=0.32]{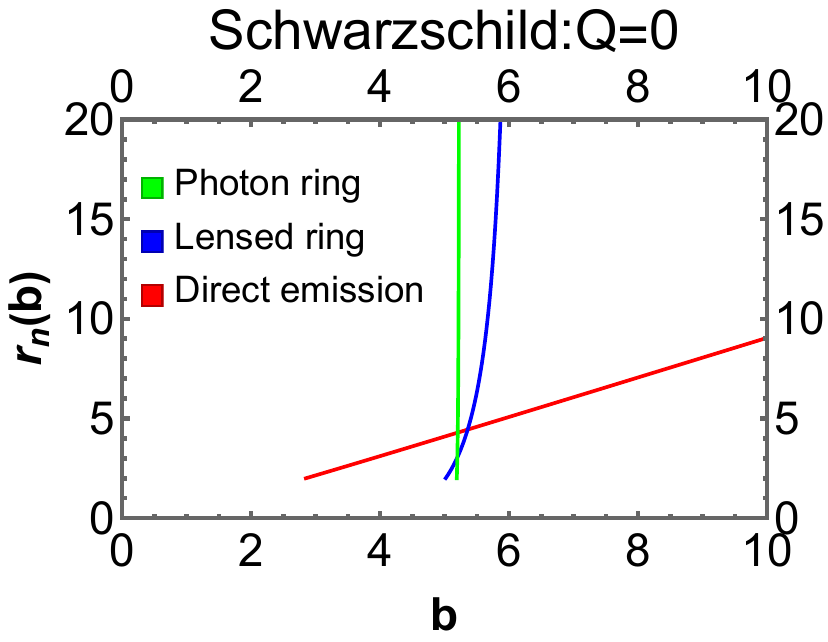}
\includegraphics[scale=0.32]{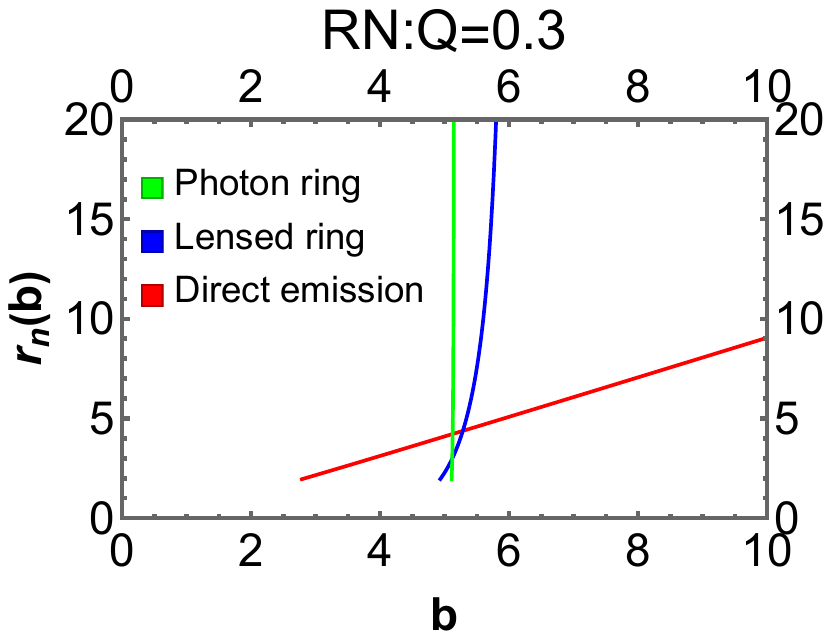}
\includegraphics[scale=0.32]{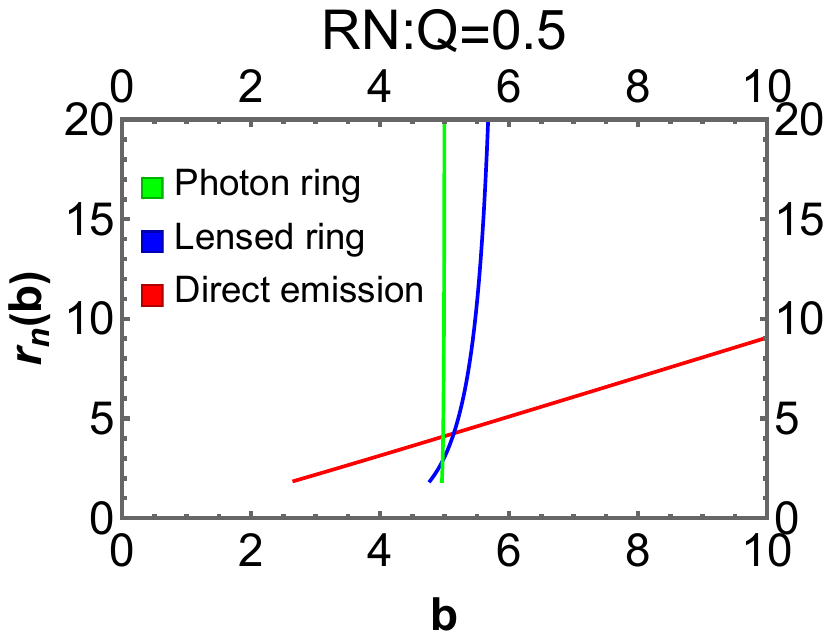}\\
\includegraphics[scale=0.32]{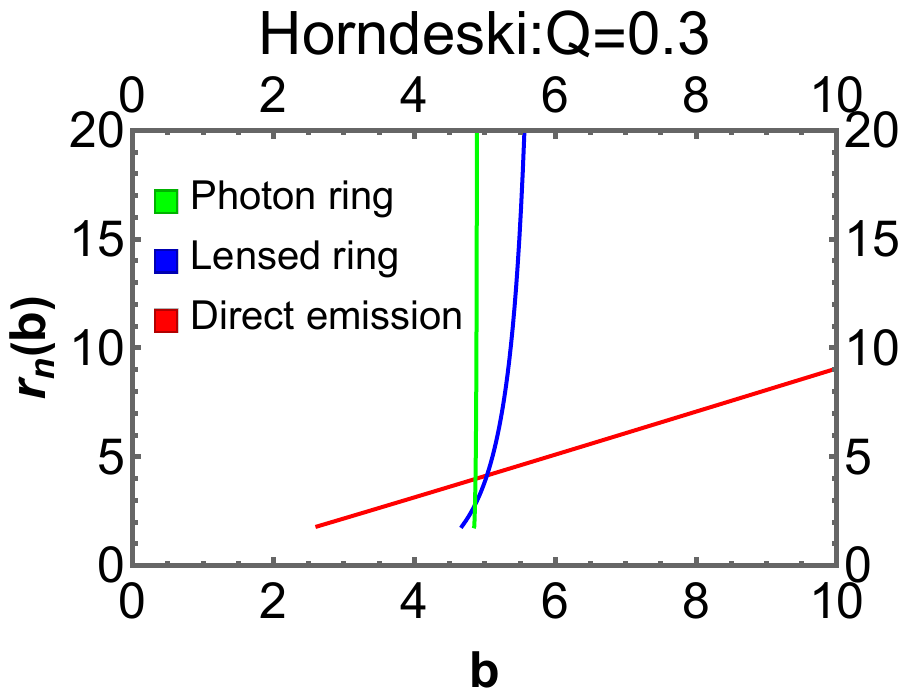}
\includegraphics[scale=0.32]{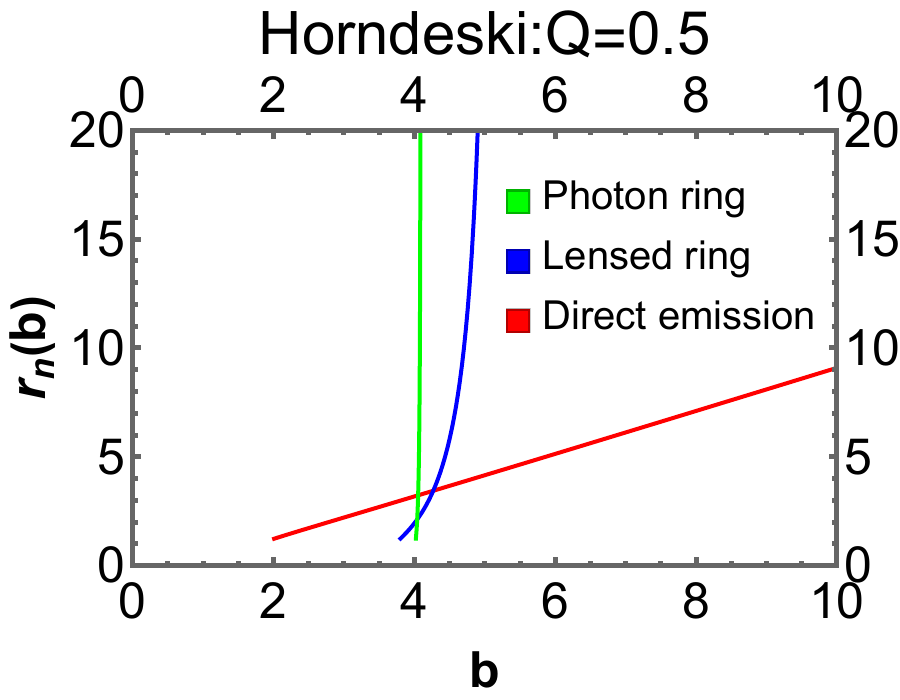}
\caption{The first three transfer functions of the black hole as a function of the impact parameter for different $Q$ with $M=1$.}
\label{Thinrb}
\end{figure}
where the $r_n(b)$ is the so-called transfer function, representing the radial position of the $n_{th}$ intersection between the light and the thin disk. The slope of the transfer function $dr/db$ describes the demagnification factor at each $b$ \cite{Gralla:2019xty}. The transfer functions as a function of the impact parameter are shown with the different values of the electric charge in Fig.\ref{Thinrb}.

In Fig.\ref{Thinrb}, the red lines denote the first ($n=1$) transfer function, corresponding to the direct image. The slope of the red curve is approximately equal to 1, thus the direct image profile is the gravity red-shifted source profile. Then, the blue lines are associated with the second ($n=2$) transfer function, representing the lensed ring image. In the case, the image of the backside of the disk will be demagnified due to the slope of the blue curve much larger than 1. Finally, the green lines are the third ($n=3$) transfer function, relating to the photon ring image. Because the slope of the green curve tends to be infinite for the case, the image of the front side of the disk will be extremely demagnified. Therefore, its contribution can be hardly ignored to the total brightness of the image. In addition, compared with the transfer functions of the Schwarzschild black hole, we clearly see that the electric charge results in toward left migration of the transfer functions in Fig.\ref{Thinrb}, while the effects of the $Q$ on the transfer function in the charged Horndeski black hole are larger than those of the RN black hole in the region (\ref{chargecondition}).

\subsection{Observed appearance of the charged Horndeski black hole}
\begin{figure}[htbp]
\centering
\subfigure[Schwarzschild:Q=0]{
\includegraphics[scale=0.24]{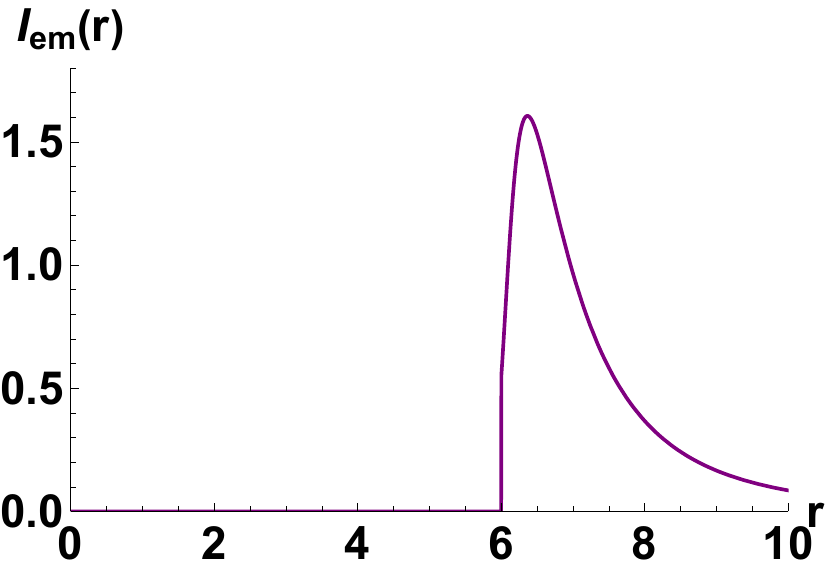}}
\subfigure[RN:Q=0.3]{
\includegraphics[scale=0.24]{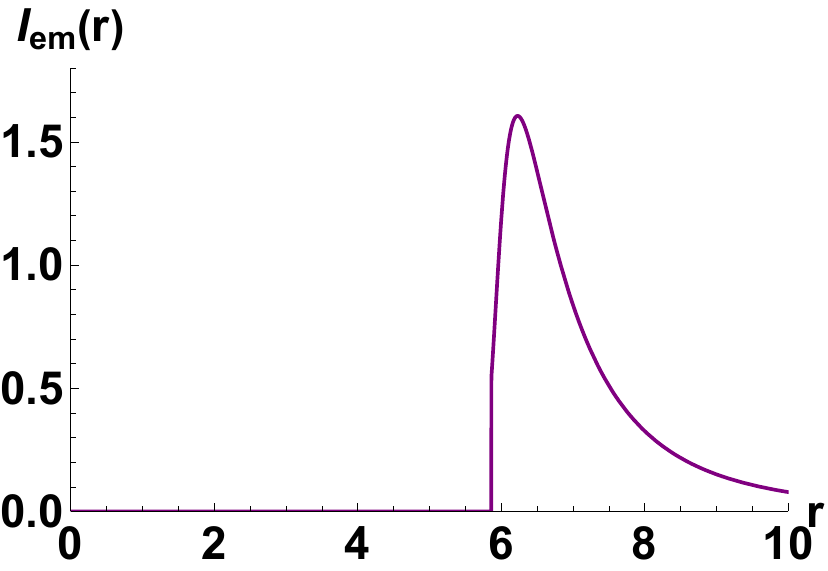}}
\subfigure[RN:Q=0.5]{
\includegraphics[scale=0.24]{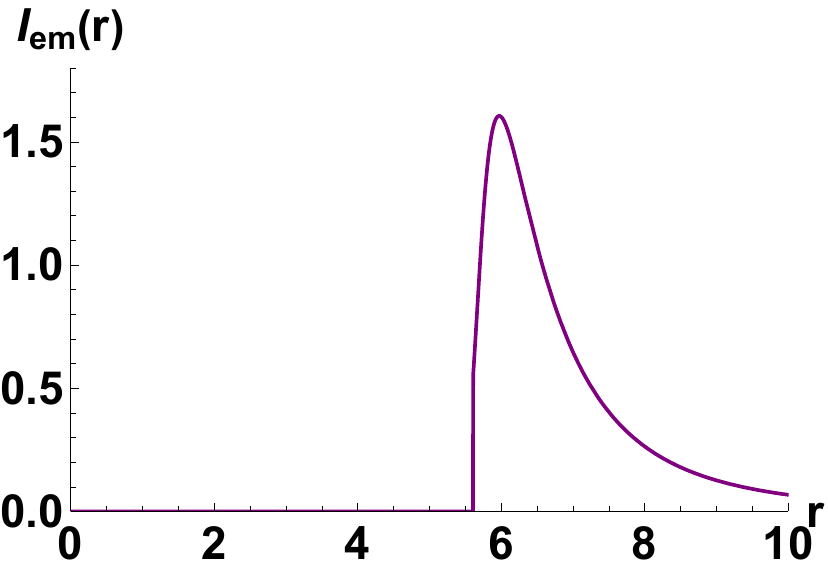}}
\subfigure[Horndeski:Q=0.3]{
\includegraphics[scale=0.22]{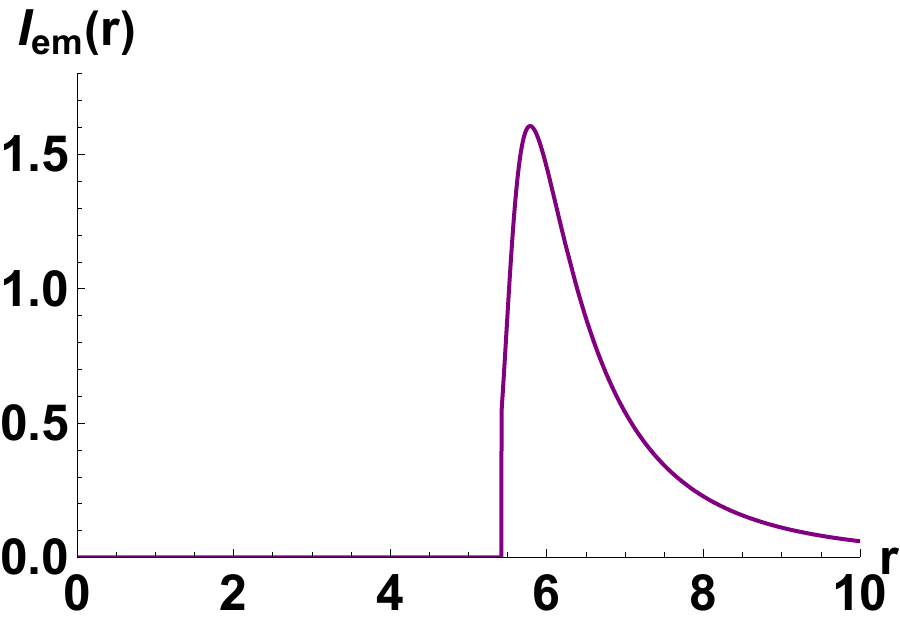}}
\subfigure[Horndeski:Q=0.5]{
\includegraphics[scale=0.22]{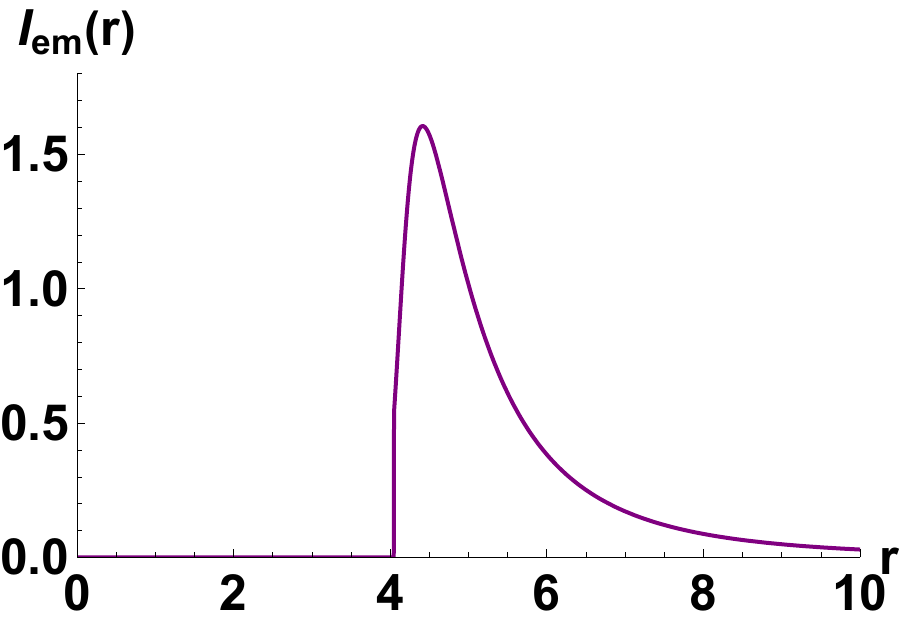}}
\caption{\textbf{Case I}: the emitted intensities $I_{em}(r)$ as a function of the radius $r$ with $Q=0, 0.3, 0.5$ for $M=1$.}
\label{emitted-intenCase1}
\end{figure}

\begin{figure}[!h]
\centering
\subfigure[Schwarzschild:Q=0]{
\includegraphics[scale=0.24]{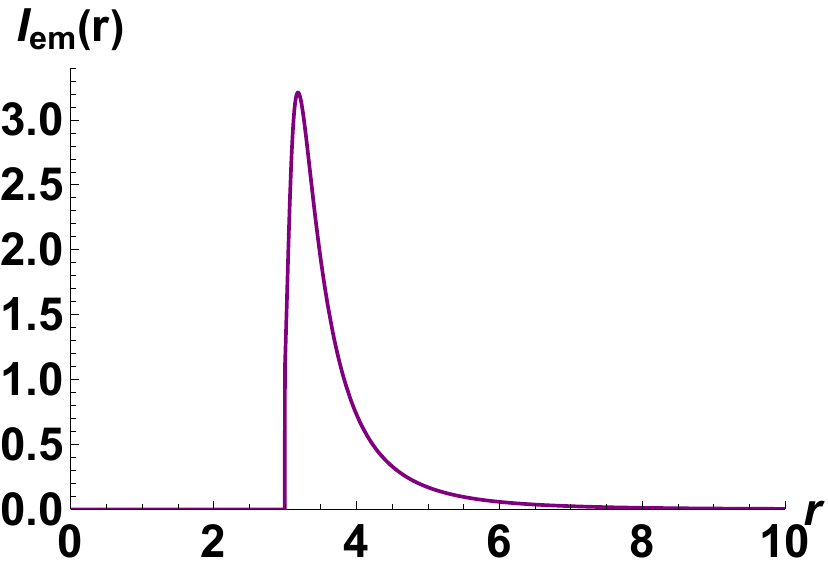}}
\subfigure[RN:Q=0.3]{
\includegraphics[scale=0.24]{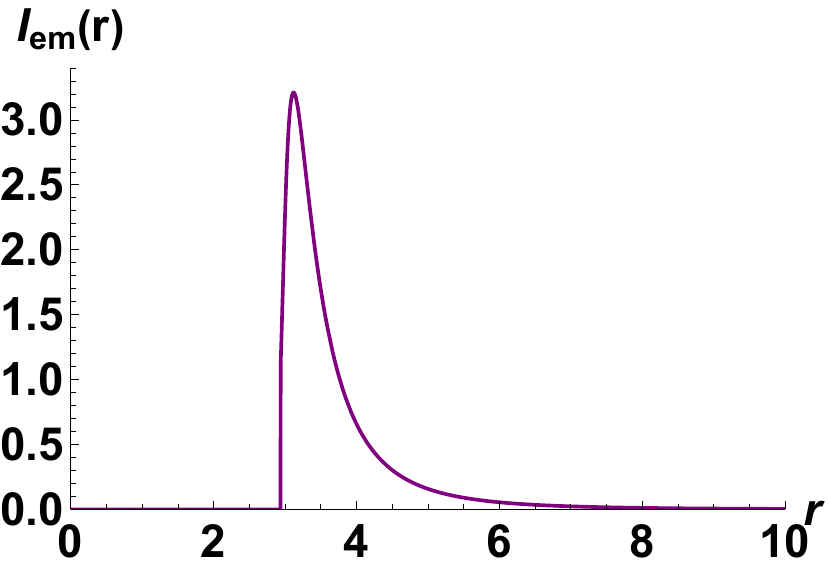}}
\subfigure[RN:Q=0.5]{
\includegraphics[scale=0.24]{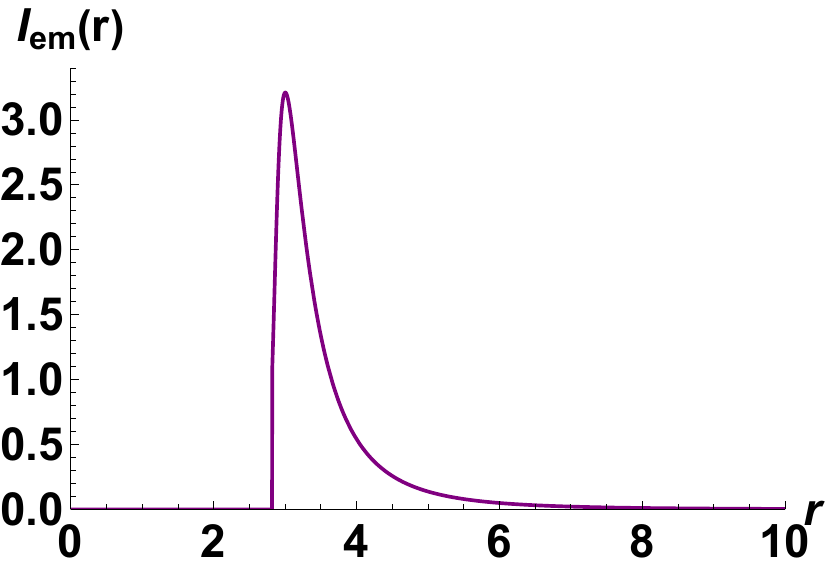}}
\subfigure[Horndeski:Q=0.3]{
\includegraphics[scale=0.22]{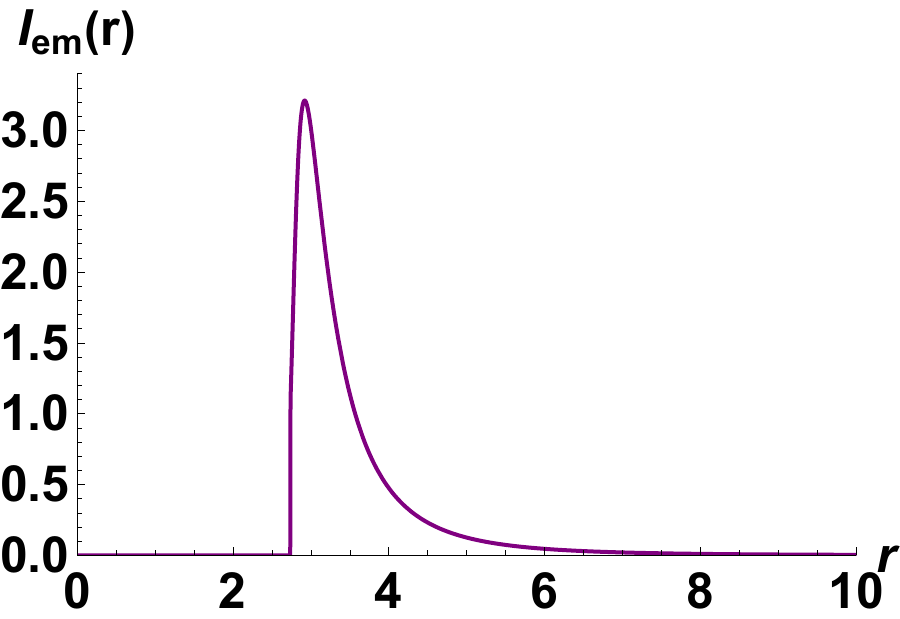}}
\subfigure[Horndeski:Q=0.5]{
\includegraphics[scale=0.22]{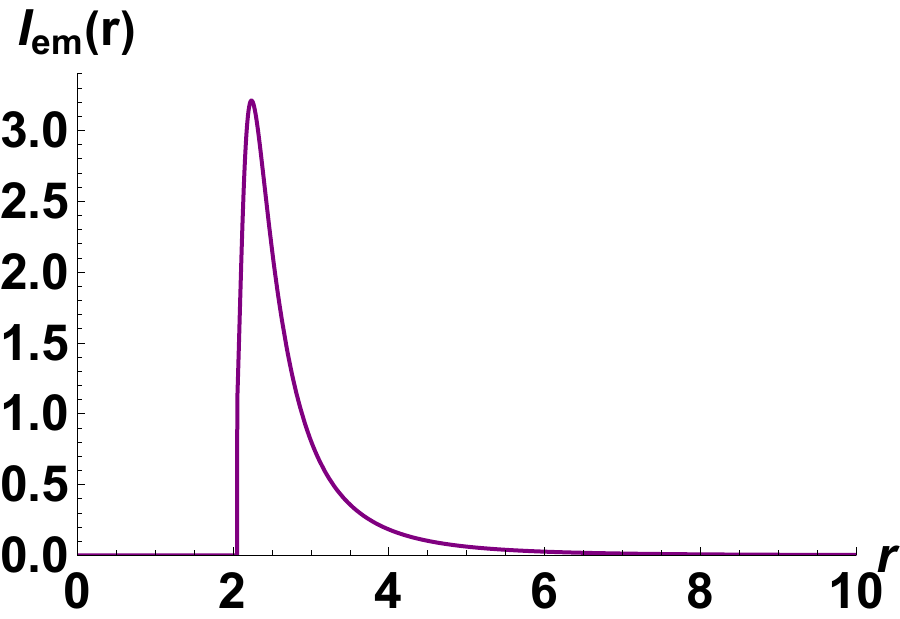}}
\caption{\textbf{Case II}: the emitted intensities $I_{em}(r)$ as a function of the radius $r$ with $Q=0, 0.3, 0.5$ for $M=1$.}
\label{emitted-inten2}
\end{figure}

\begin{figure}[!h]
\centering
\subfigure[Schwarzschild:Q=0]{
\includegraphics[scale=0.24]{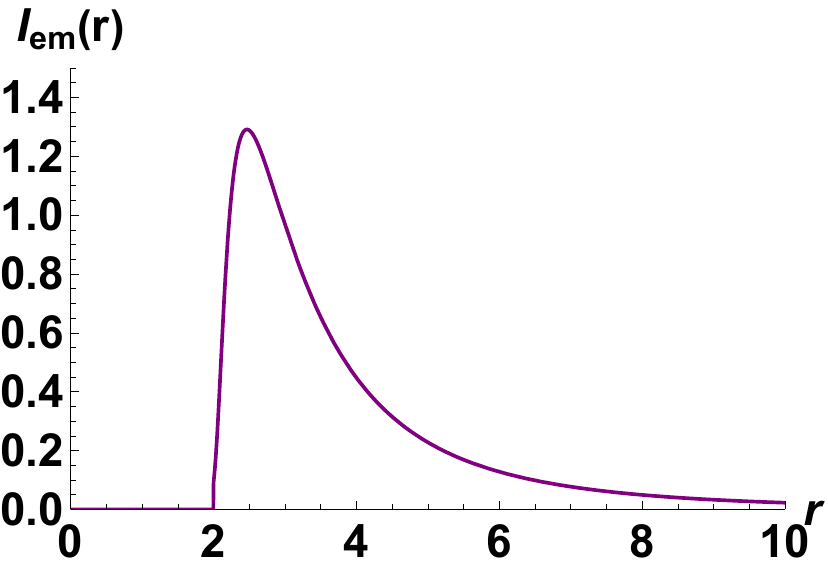}}
\subfigure[RN:Q=0.3]{
\includegraphics[scale=0.24]{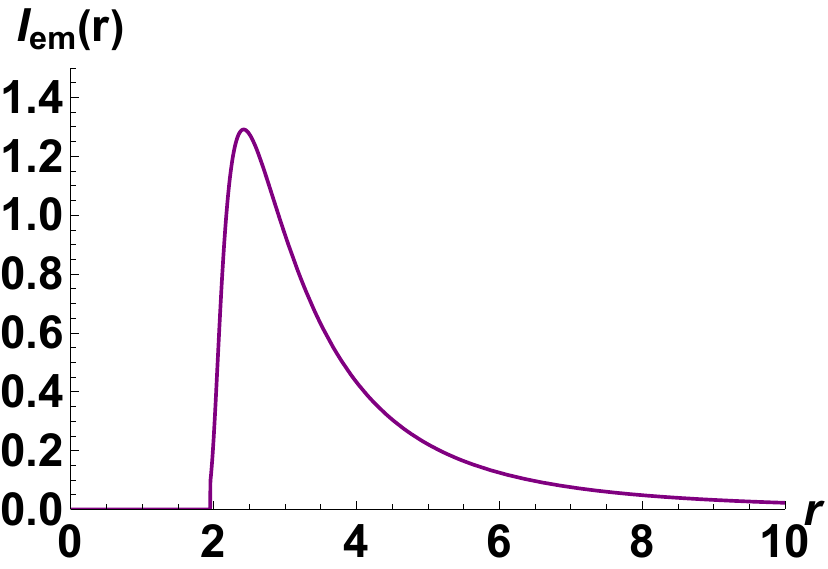}}
\subfigure[RN:Q=0.5]{
\includegraphics[scale=0.24]{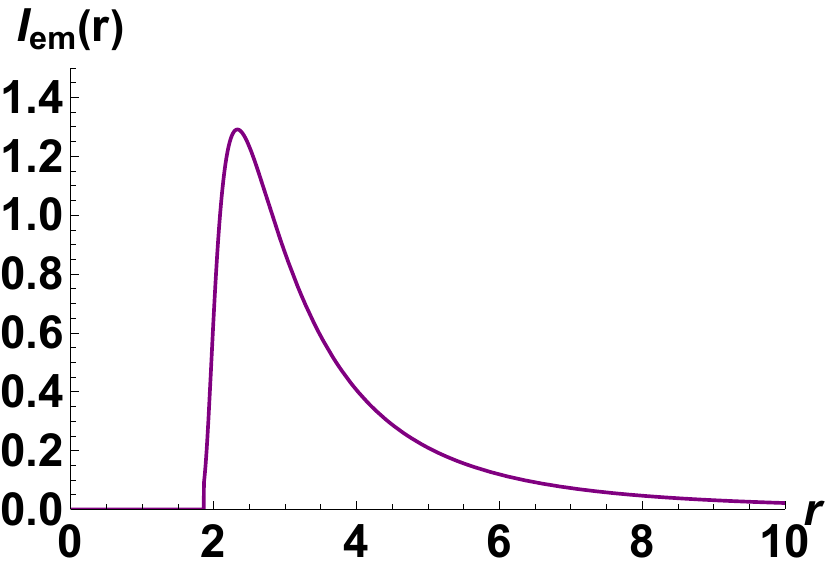}}
\subfigure[Horndeski:Q=0.3]{
\includegraphics[scale=0.22]{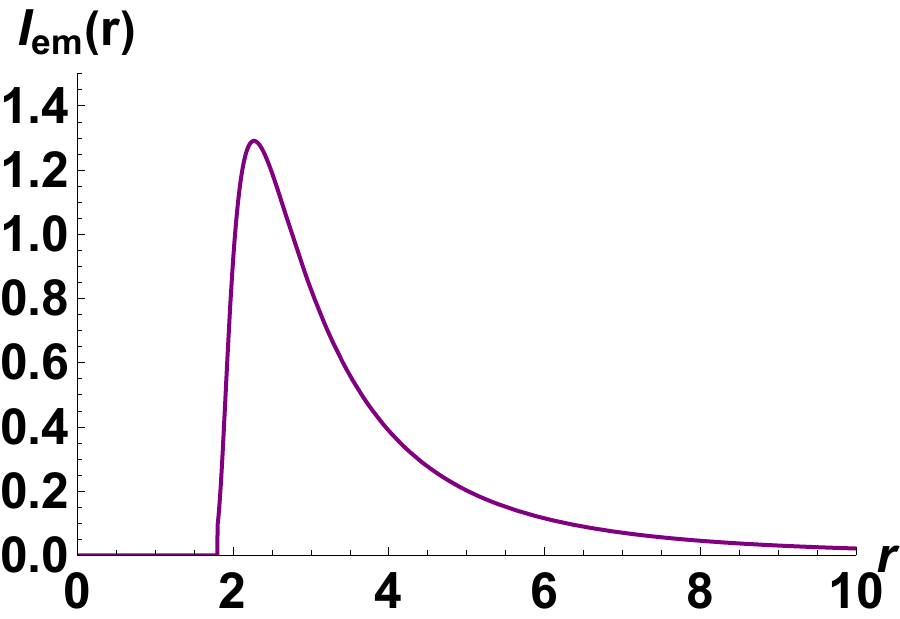}}
\subfigure[Horndeski:Q=0.5]{
\includegraphics[scale=0.22]{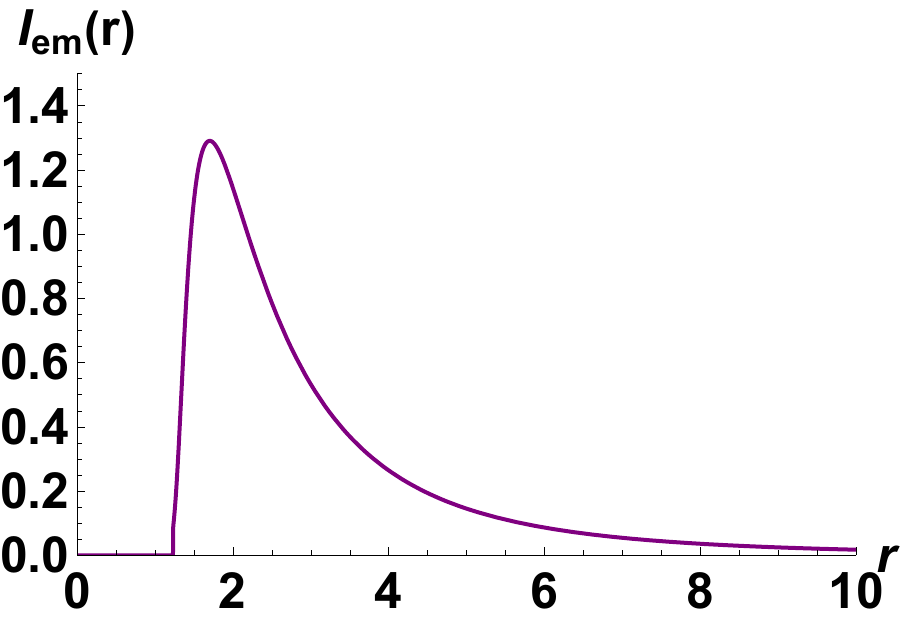}}
\caption{\textbf{Case III}: the emitted intensities $I_{em}(r)$ as a function of the radius $r$ with $Q=0,0.3, 0.5$ for $M=1$.}
\label{emitted-inten3}
\end{figure}

\begin{figure}[!h]
\centering
\subfigure[\, Schwarzschild:$Q=0$]
{\includegraphics[width=5.cm]{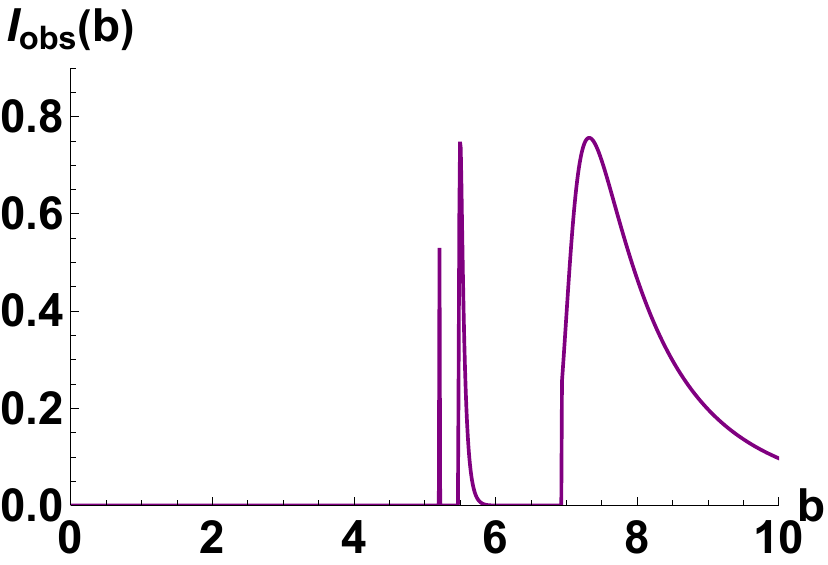} \label{}\hspace{1mm} \includegraphics[width=4.5cm]{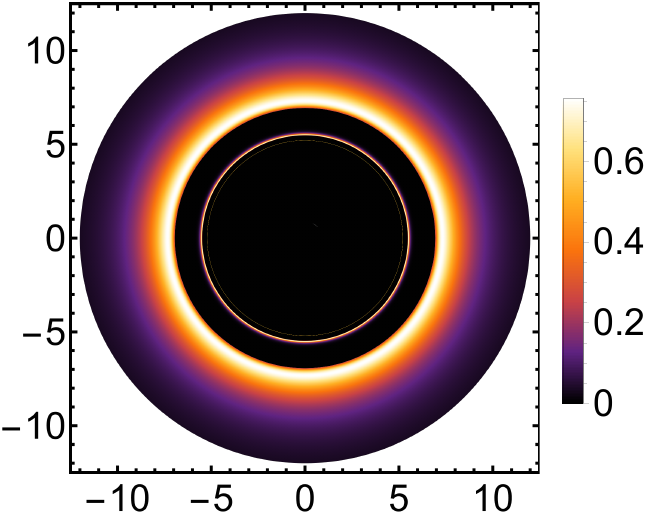}\hspace{1mm} \includegraphics[width=3.5cm]{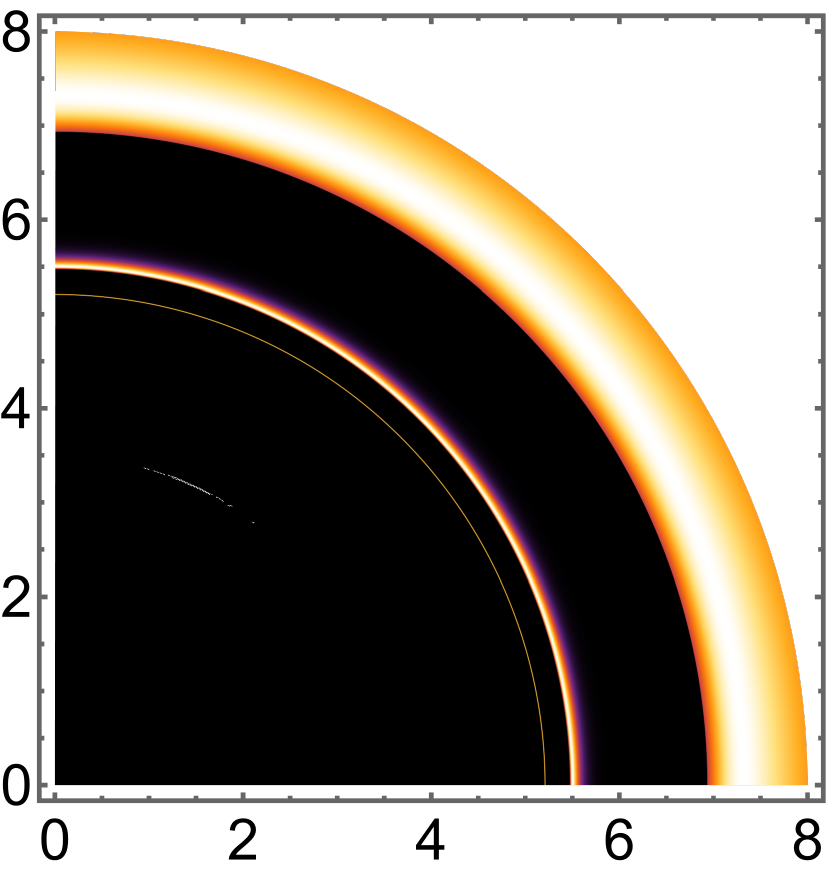}}\\
\subfigure[\, RN:$Q=0.3$]
{\includegraphics[width=5.cm]{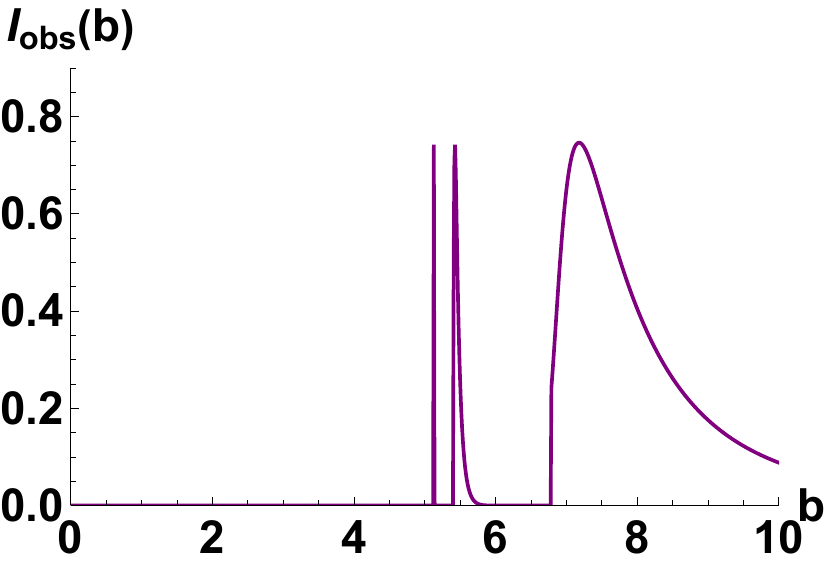} \label{}\hspace{2mm} \includegraphics[width=4.5cm]{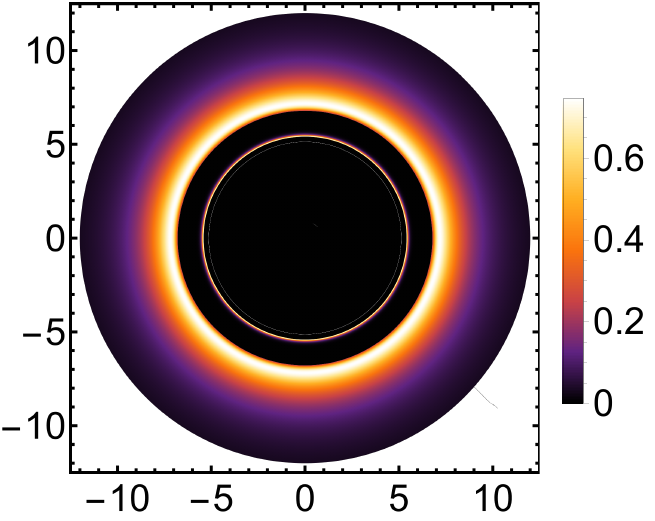}\hspace{2mm} \includegraphics[width=3.5cm]{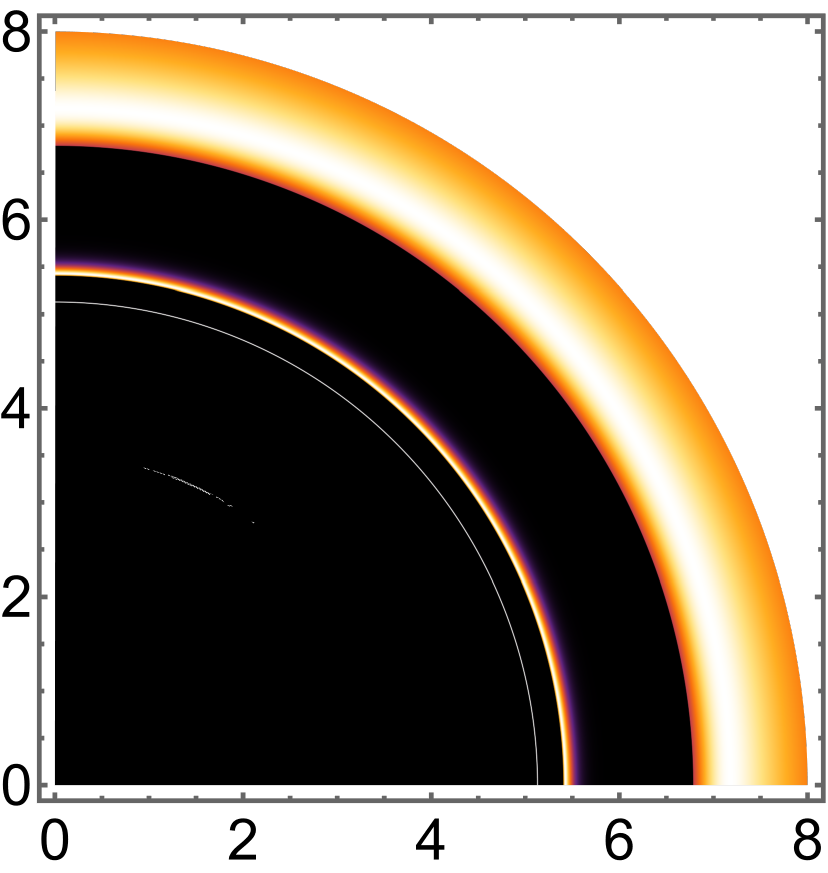}}\\
\subfigure[\, RN:$Q=0.5$]
{\includegraphics[width=5.cm]{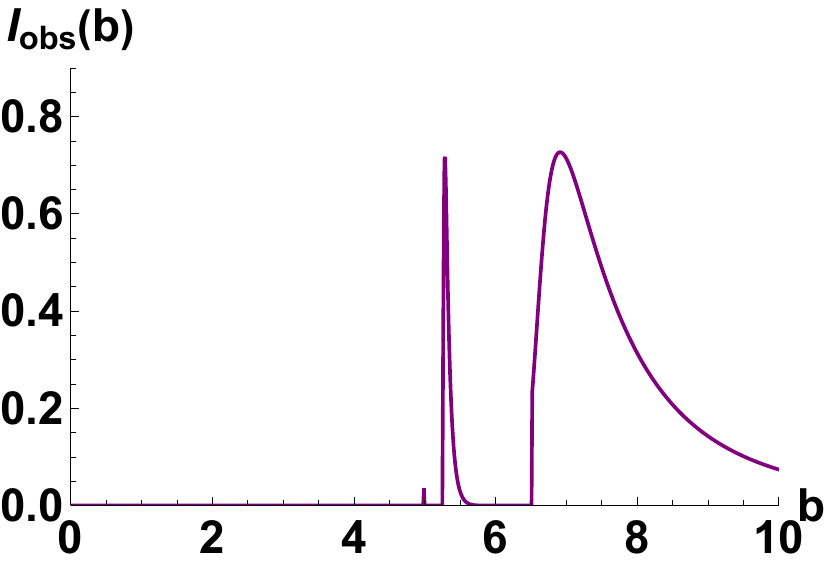} \label{}\hspace{2mm} \includegraphics[width=4.5cm]{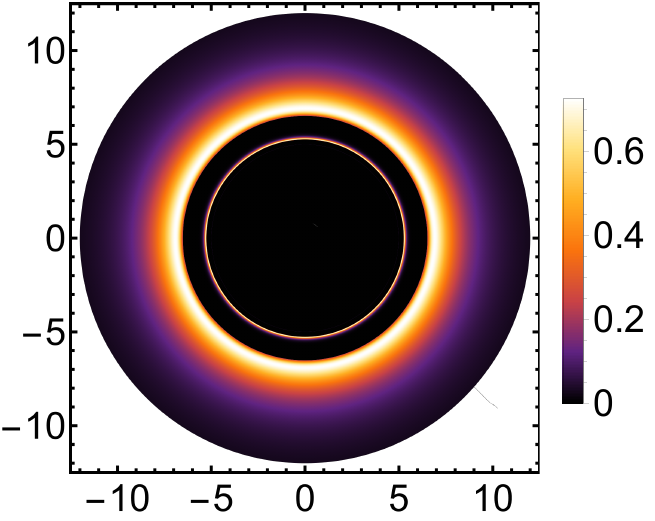}\hspace{2mm} \includegraphics[width=3.5cm]{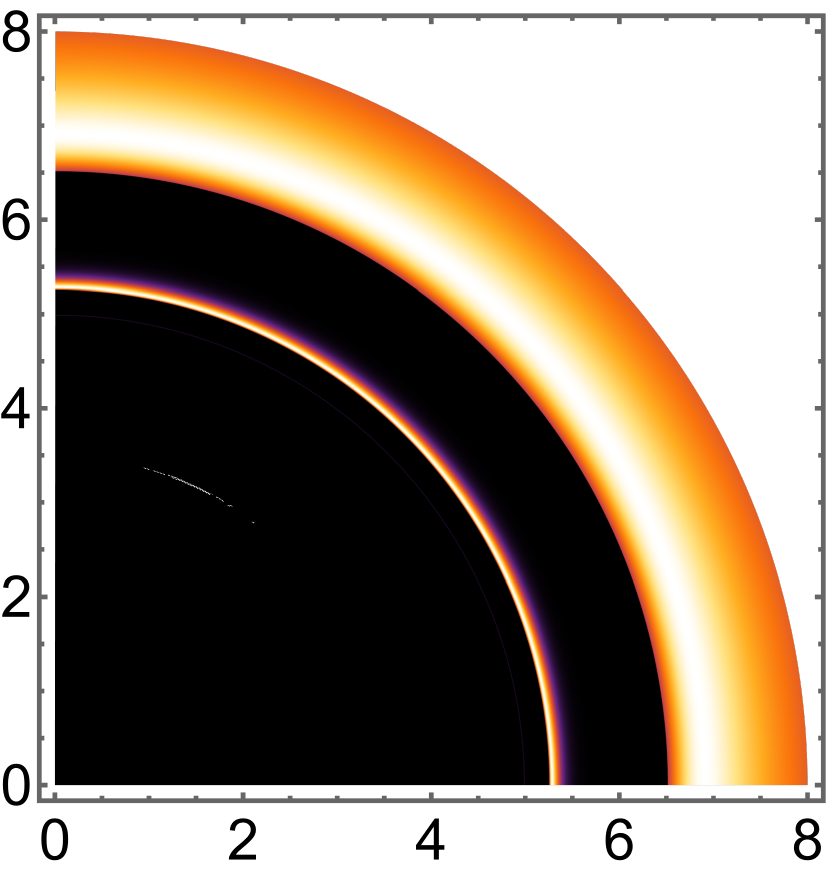}}\\
\subfigure[\, Horndeski:$Q=0.3$]
{\includegraphics[width=5.cm]{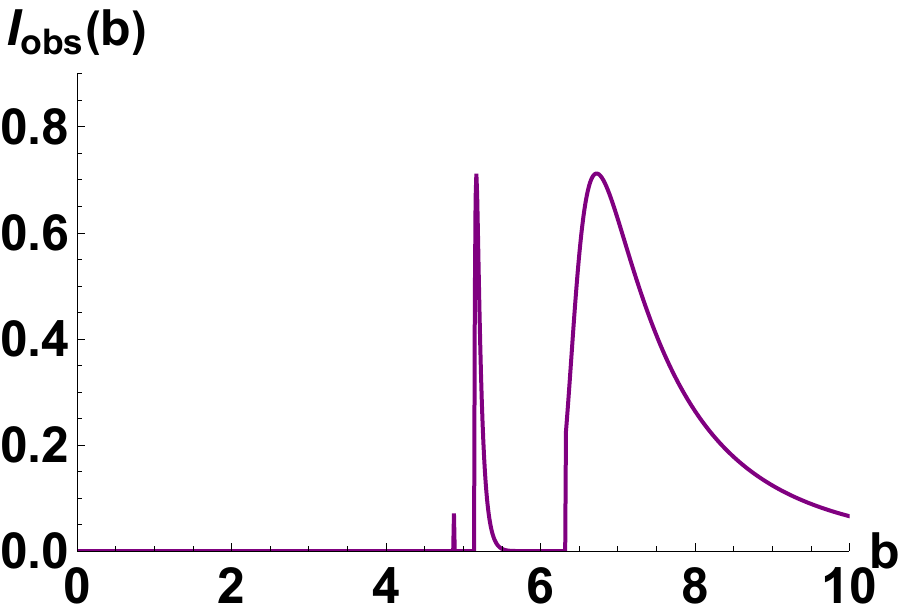} \label{}\hspace{2mm} \includegraphics[width=4.5cm]{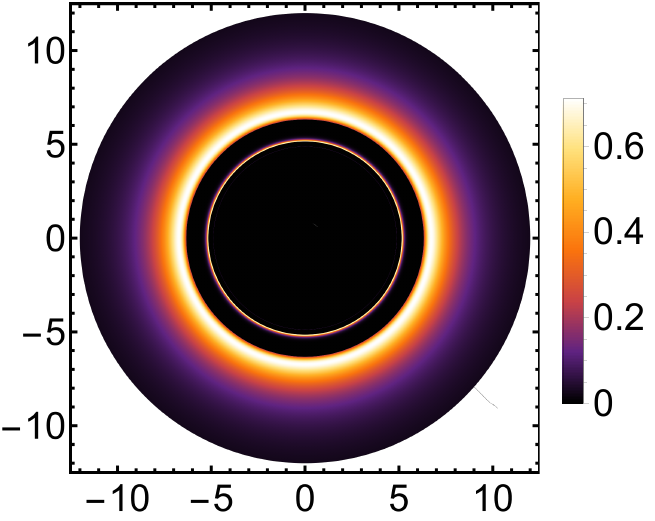}\hspace{2mm} \includegraphics[width=3.5cm]{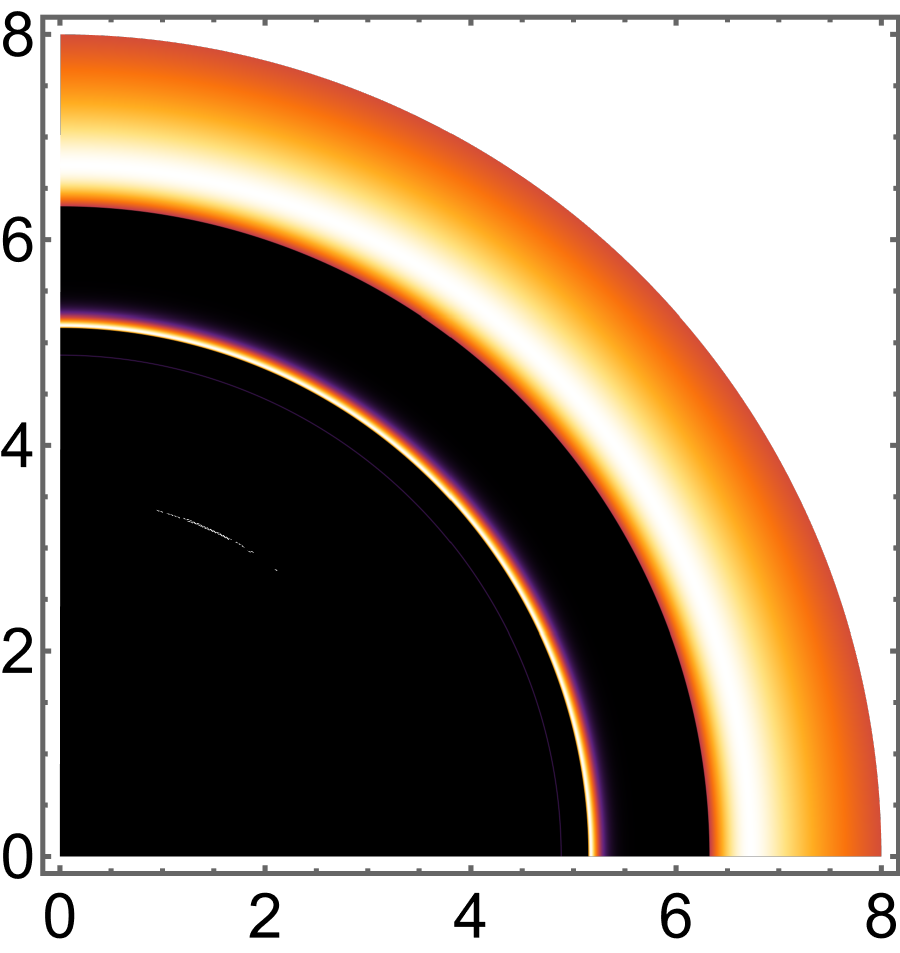}}\\
\subfigure[\, Horndeski:$Q=0.5$]
{\includegraphics[width=5.cm]{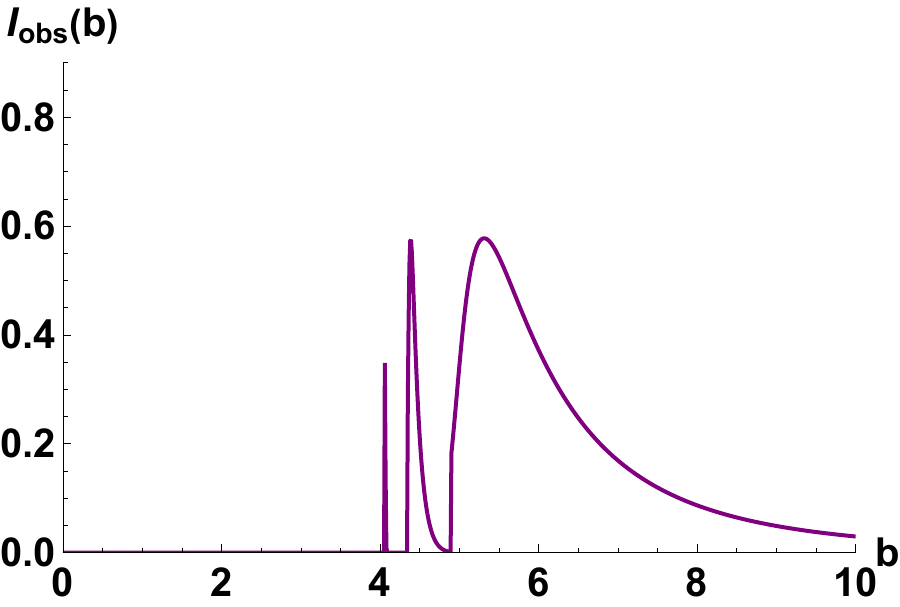} \label{}\hspace{2mm} \includegraphics[width=4.5cm]{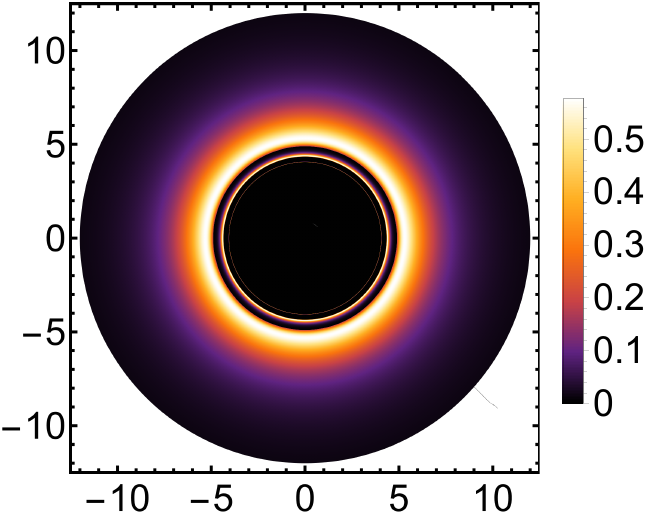}\hspace{2mm} \includegraphics[width=3.5cm]{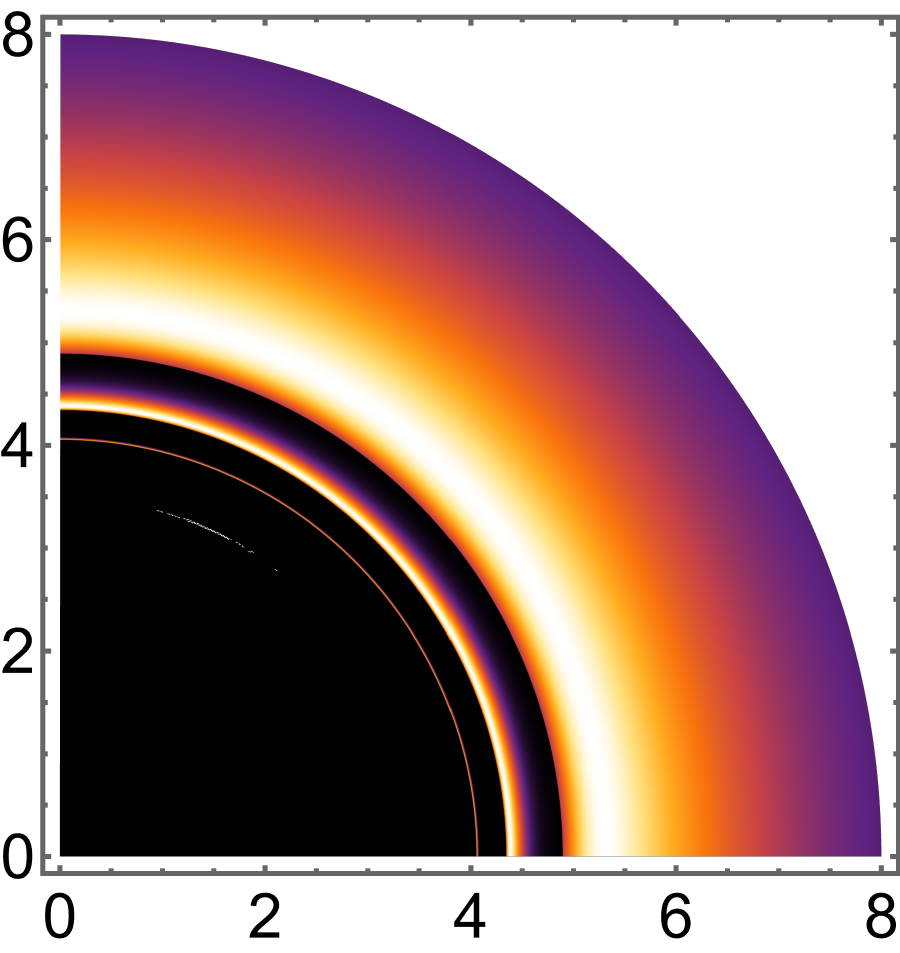}}\\
\caption{Optical  appearance of the black holes for \textbf{the case I} with different $Q$. \textbf{First column}: the total observed intensities $I_{obs}(b)$ as a function of the impact parameter. \textbf{Second column}:  the two-dimensional shadows cast of the black holes in the celestial coordinates. \textbf{Third column}: the zoomed in sectors. The black hole mass is set to 1.}
\label{modelone-isco}
\end{figure}

\begin{figure}[!h]
\centering
\subfigure[\, Schwarzschild:$Q=0$]
{\includegraphics[width=5.cm]{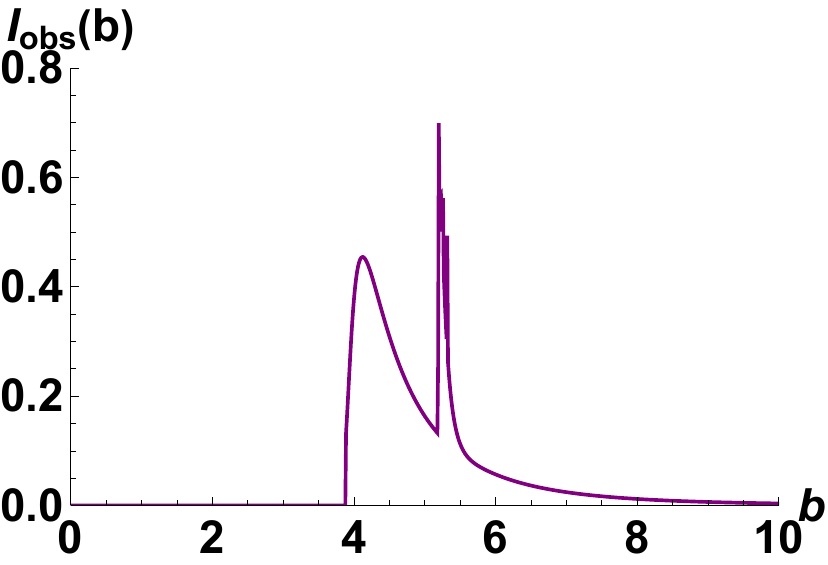} \label{}\hspace{2mm} \includegraphics[width=4.5cm]{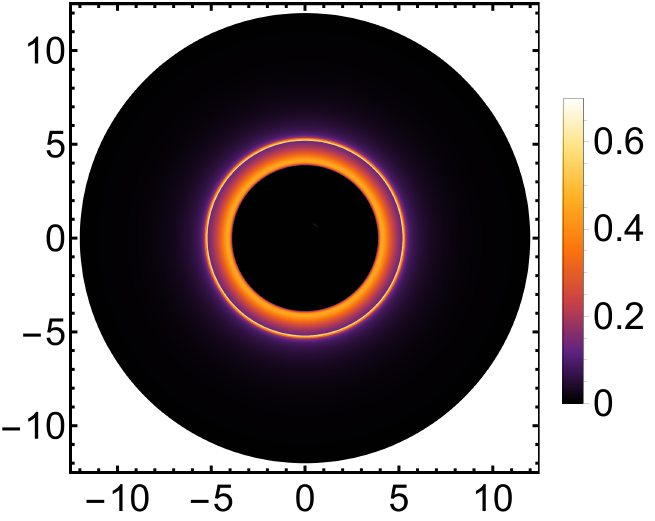}\hspace{2mm} \includegraphics[width=3.5cm]{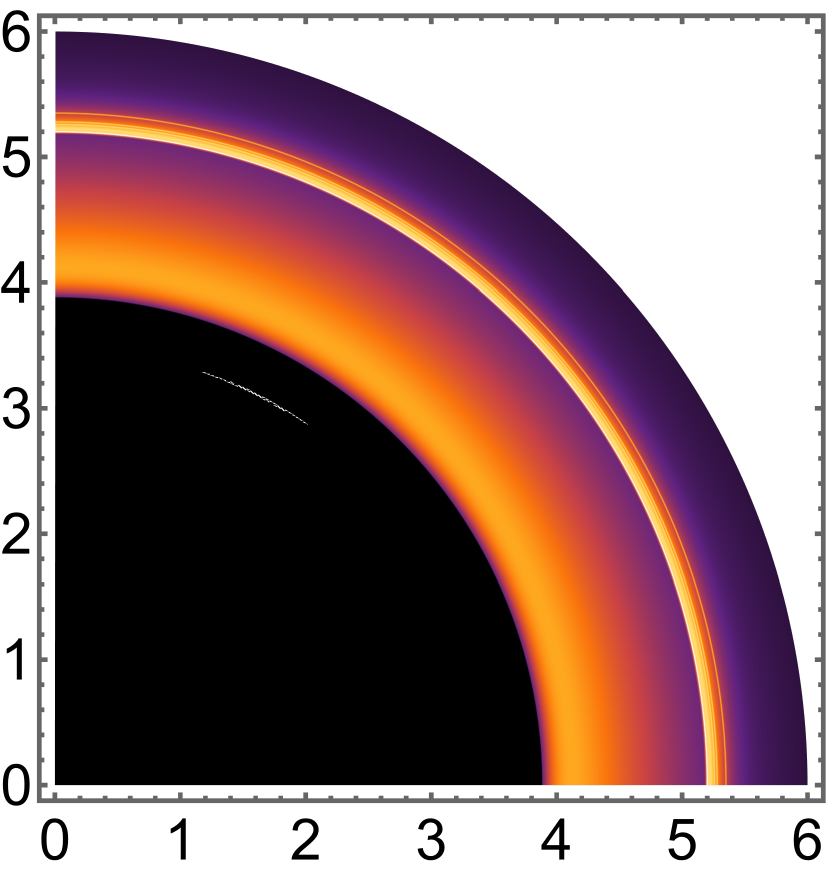}}\\
\subfigure[\, RN:$Q=0.3$]
{\includegraphics[width=5.cm]{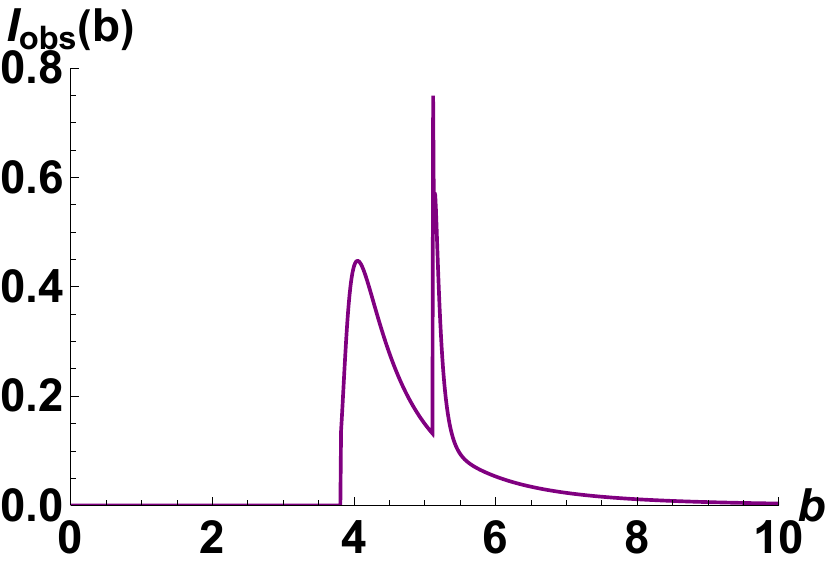} \label{}\hspace{2mm} \includegraphics[width=4.5cm]{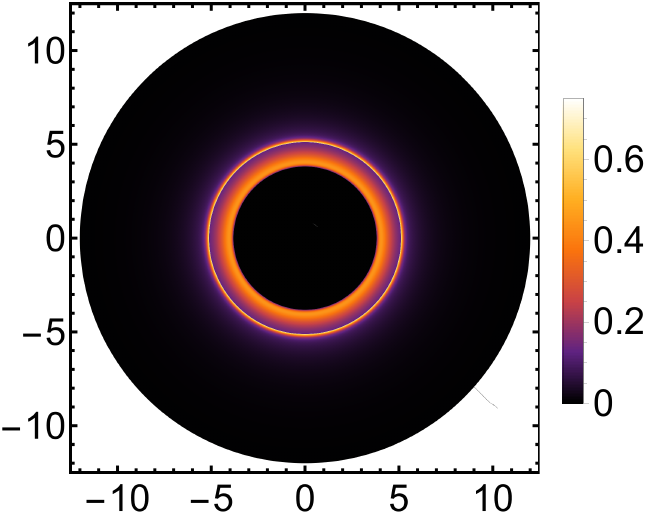}\hspace{2mm} \includegraphics[width=3.5cm]{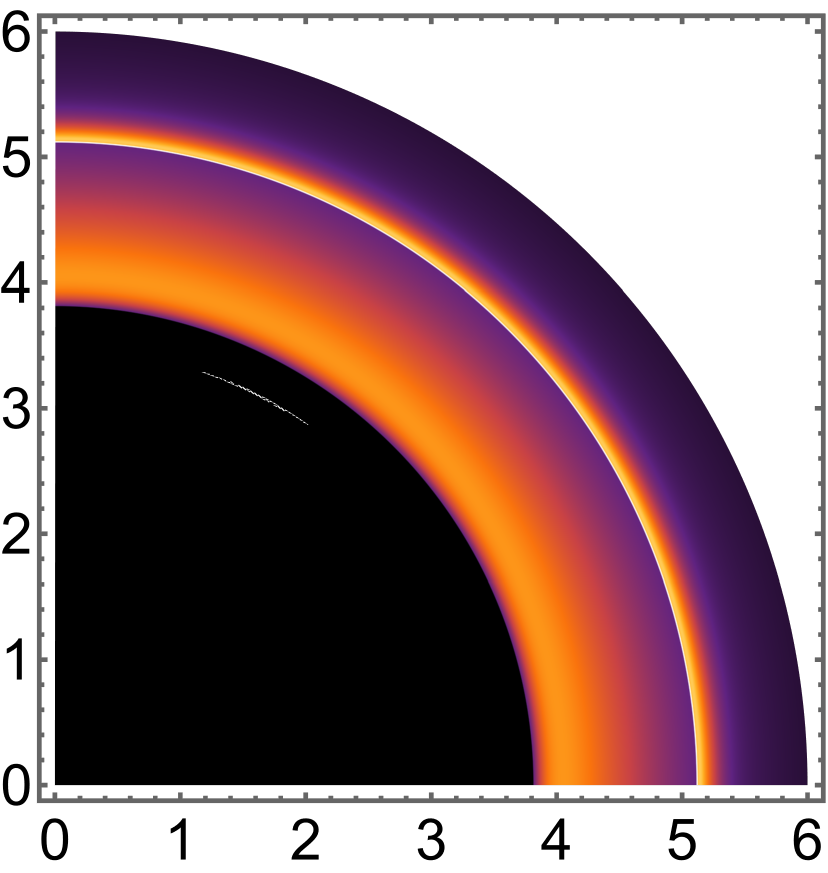}}\\
\subfigure[\, RN:$Q=0.5$]
{\includegraphics[width=5.cm]{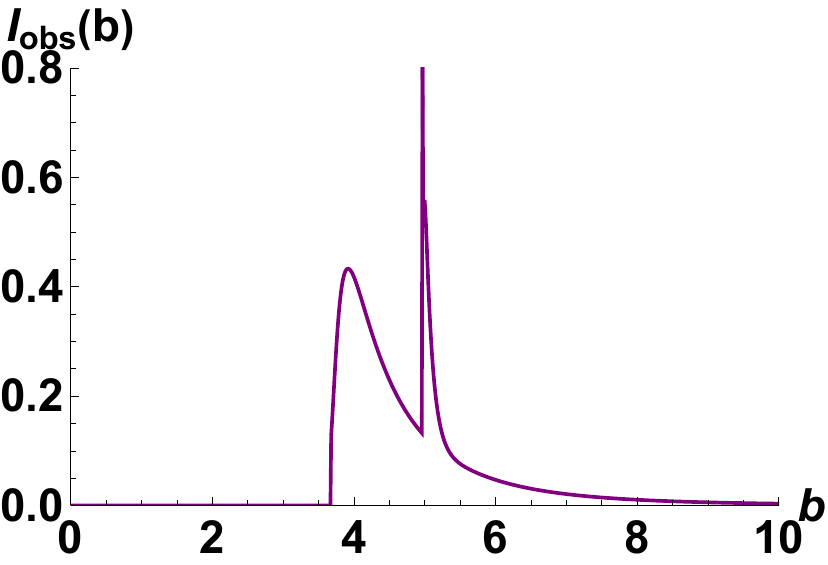} \label{}\hspace{2mm} \includegraphics[width=4.5cm]{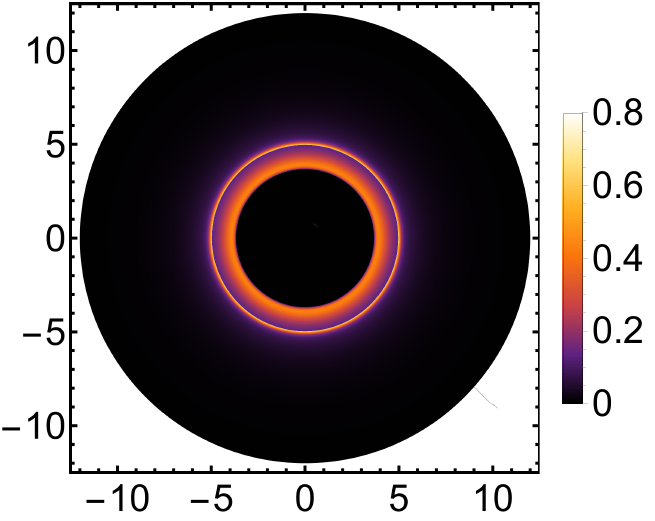}\hspace{2mm} \includegraphics[width=3.5cm]{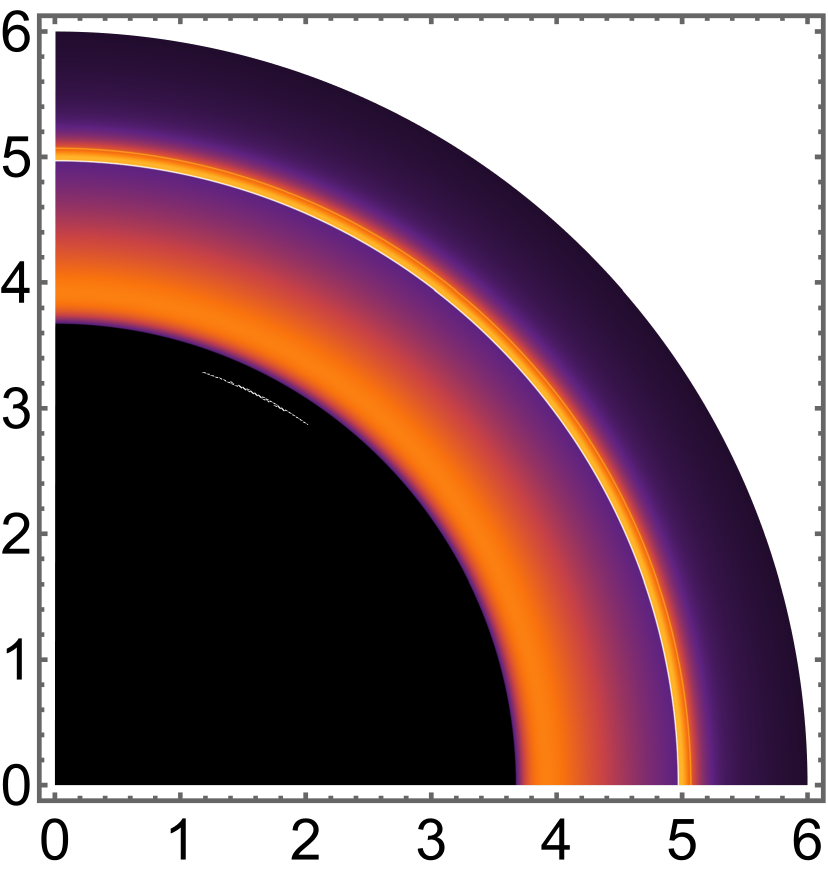}}\\
\subfigure[\, Horndeski:$Q=0.3$]
{\includegraphics[width=5.cm]{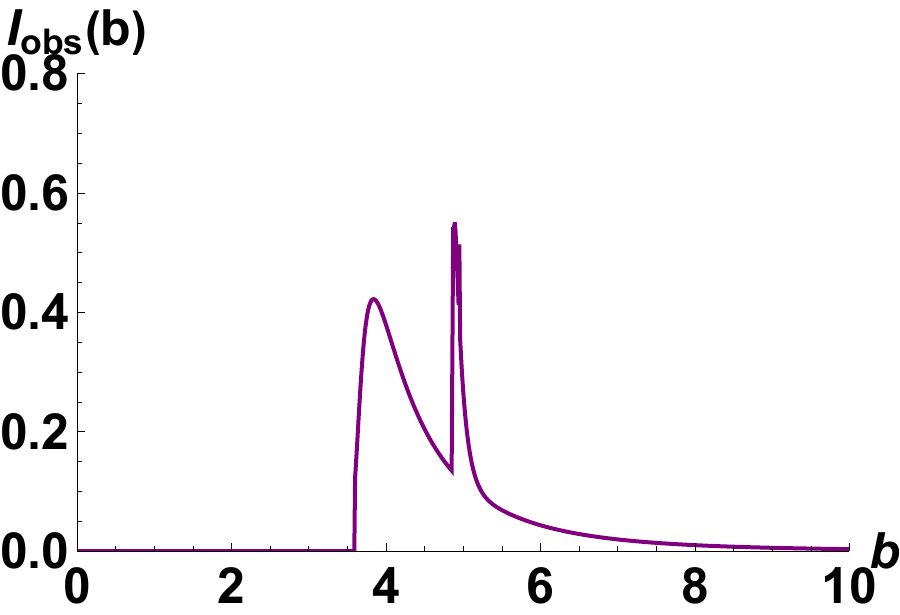} \label{}\hspace{2mm} \includegraphics[width=4.5cm]{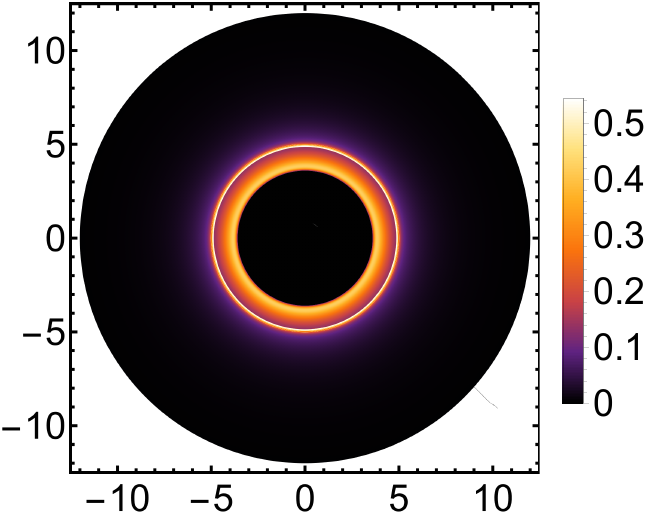}\hspace{2mm} \includegraphics[width=3.5cm]{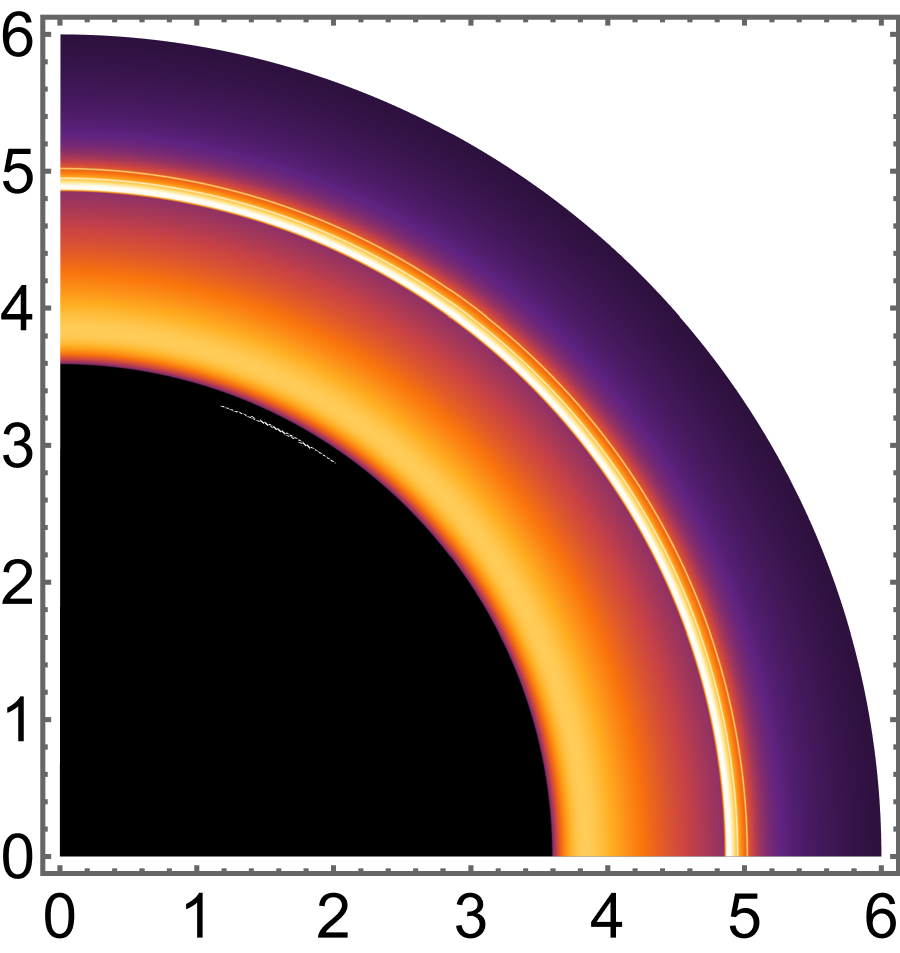}}\\
\subfigure[\, Horndeski:$Q=0.5$]
{\includegraphics[width=5.cm]{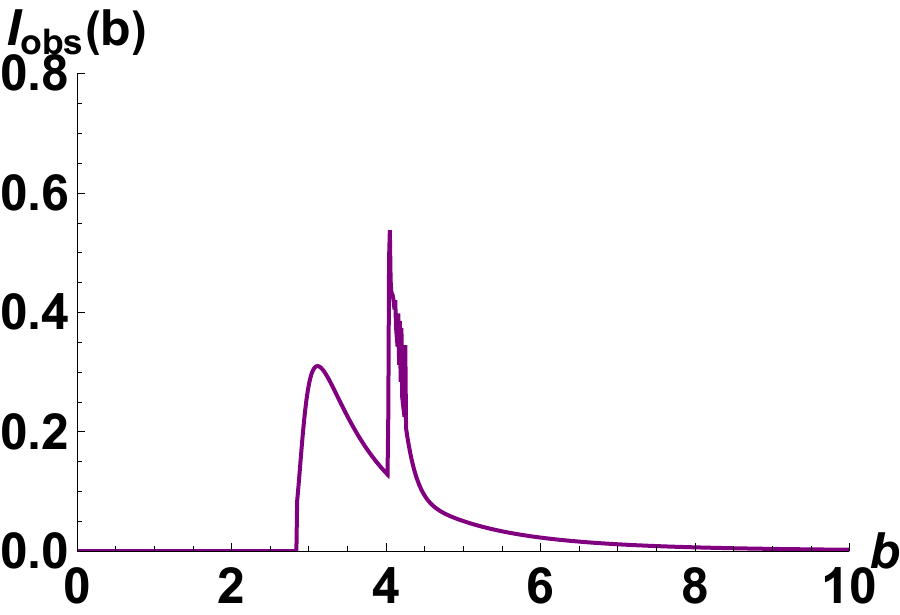} \label{}\hspace{2mm} \includegraphics[width=4.5cm]{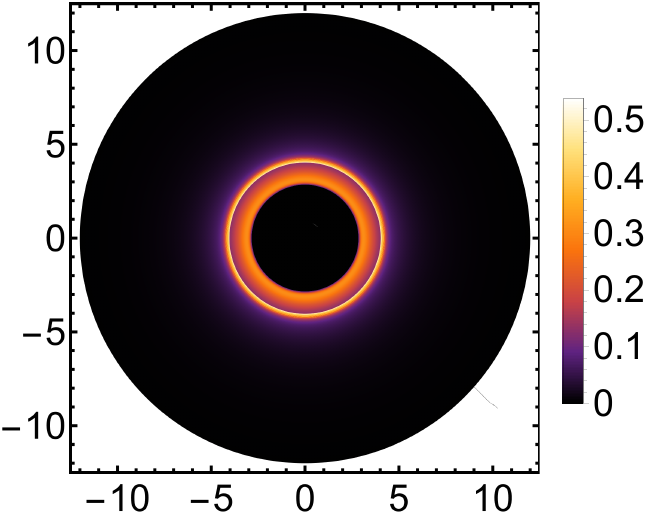}\hspace{2mm} \includegraphics[width=3.5cm]{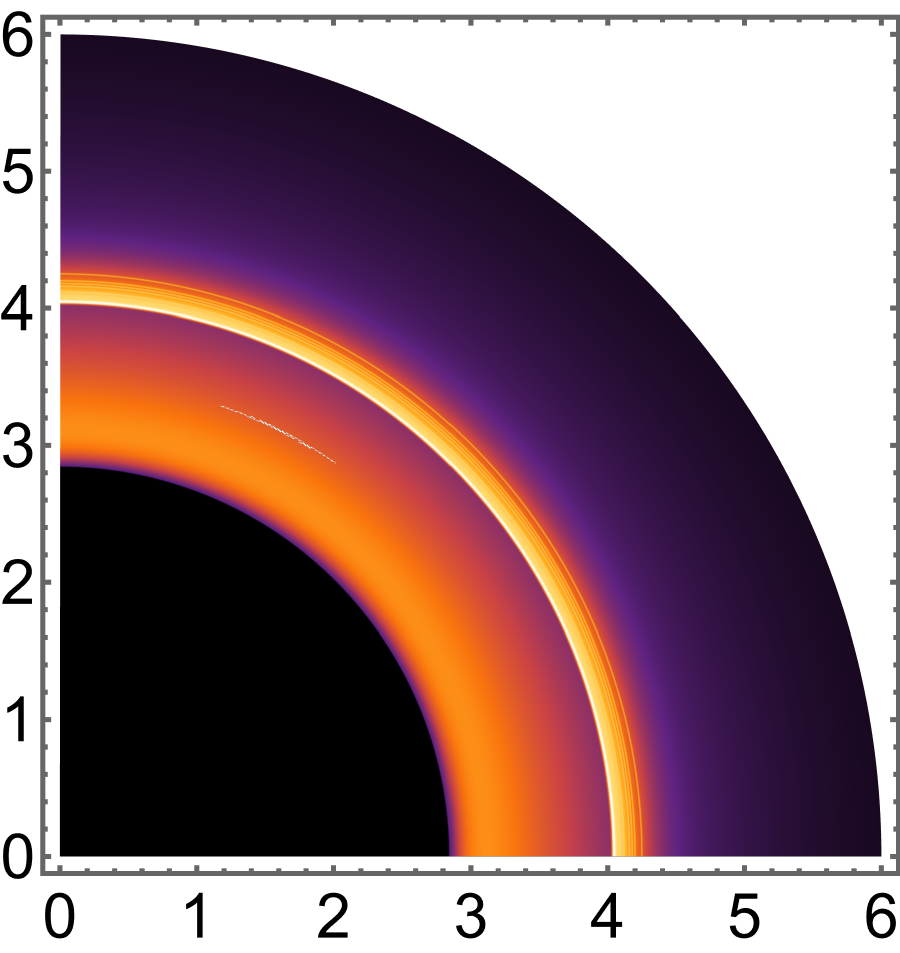}}\\
\caption{Optical  appearance of the black holes for \textbf{the case II} with different $Q$. \textbf{First column}: the total observed intensities $I_{obs}(b)$ as a function of the impact parameter. \textbf{Second column}:  the two-dimensional shadows cast of the black holes in the celestial coordinates. \textbf{Third column}: the zoomed in sectors. The black hole mass is set to 1.}
\label{modeltwo-ph}
\end{figure}

\begin{figure}[!h]
\centering
\subfigure[\, Schwarzschild:$Q=0$]
{\includegraphics[width=5.cm]{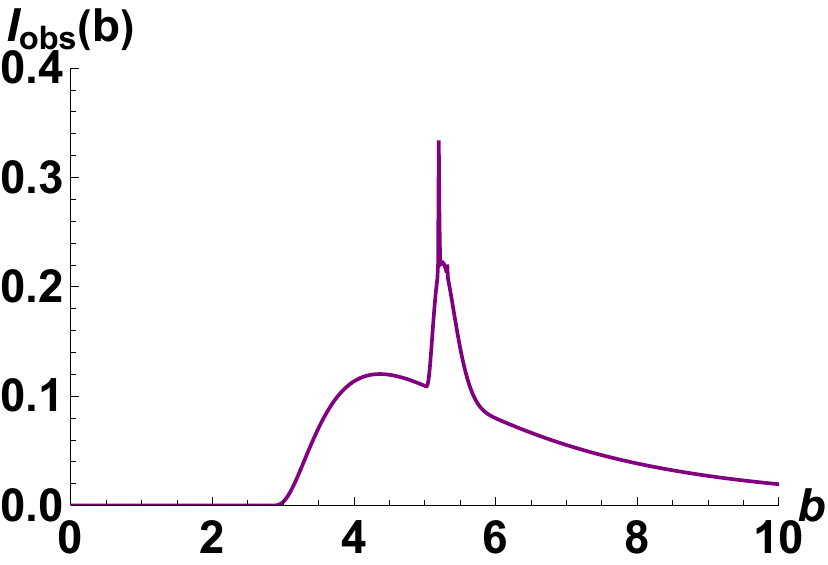} \label{}\hspace{2mm} \includegraphics[width=4.5cm]{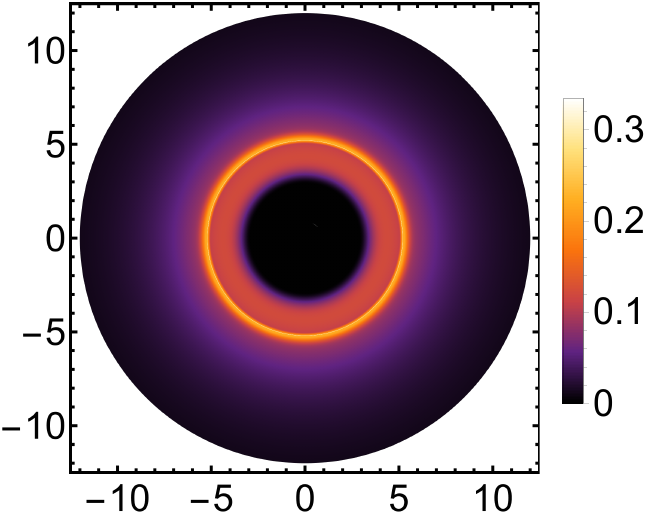}\hspace{2mm} \includegraphics[width=3.5cm]{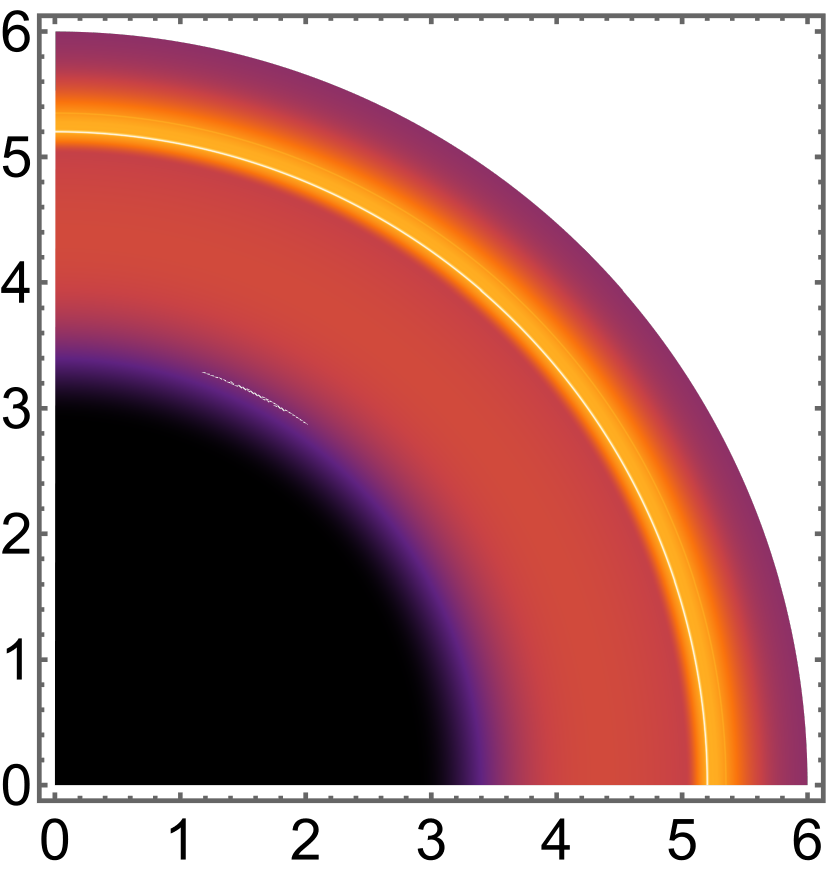}} \\
\subfigure[\, RN:$Q=0.3$]
{\includegraphics[width=5.cm]{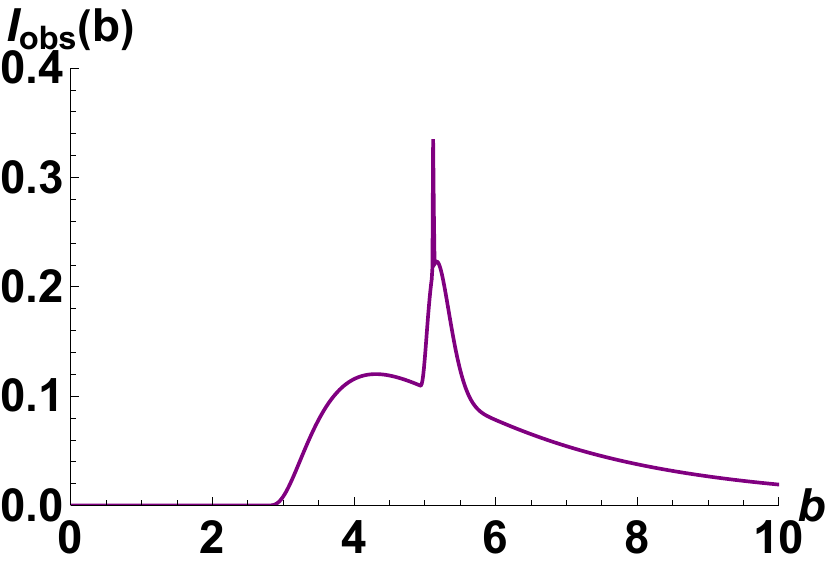} \label{}\hspace{2mm} \includegraphics[width=4.5cm]{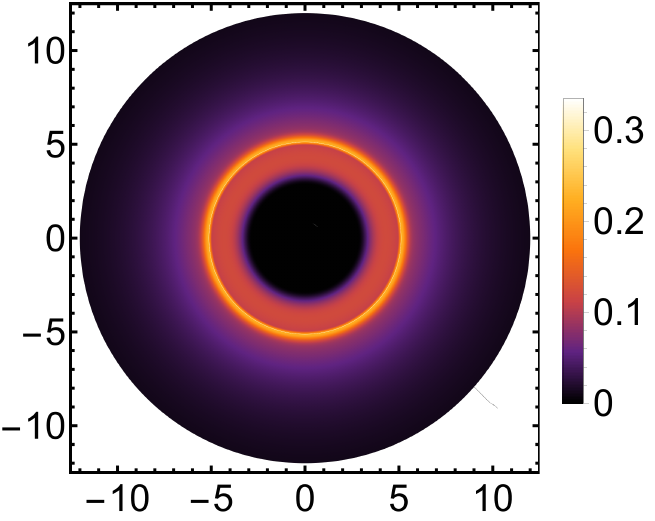}\hspace{2mm} \includegraphics[width=3.5cm]{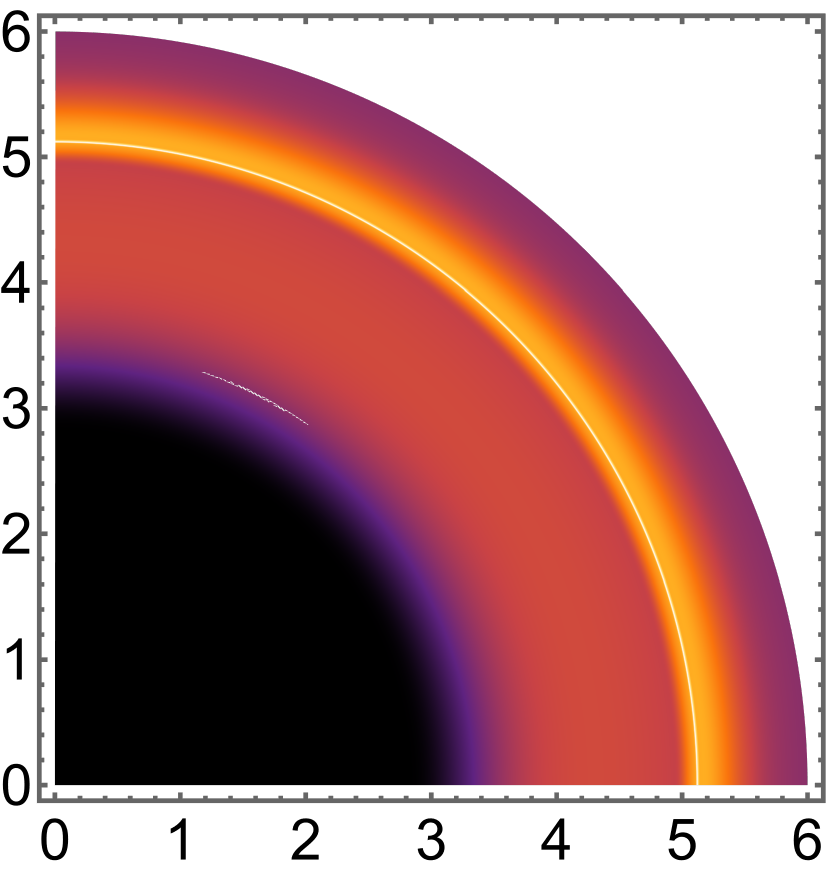}}\\
\subfigure[\, RN:$Q=0.5$]
{\includegraphics[width=5.cm]{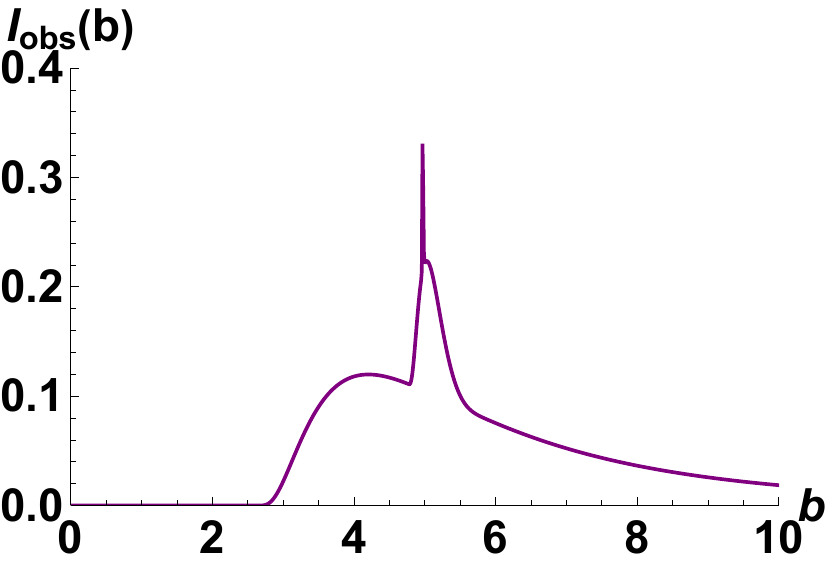} \label{}\hspace{2mm} \includegraphics[width=4.5cm]{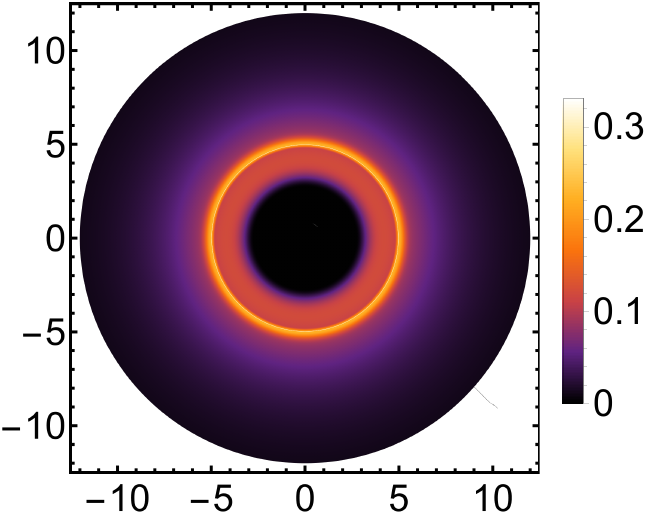}\hspace{2mm} \includegraphics[width=3.5cm]{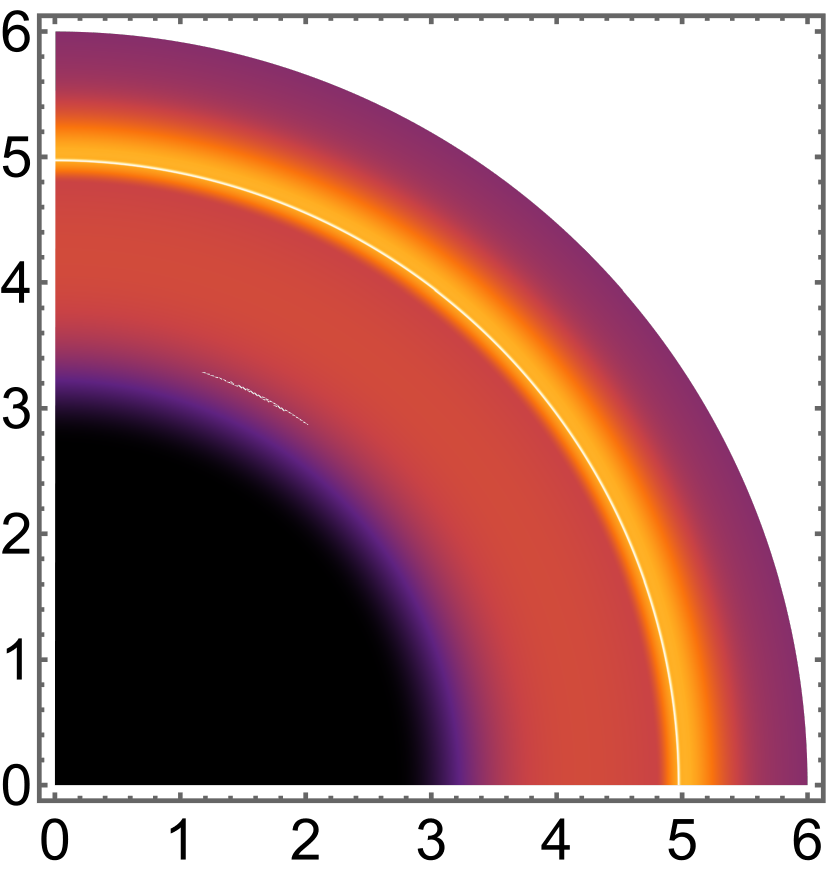}}\\
\subfigure[\, Horndeski:$Q=0.3$]
{\includegraphics[width=5.cm]{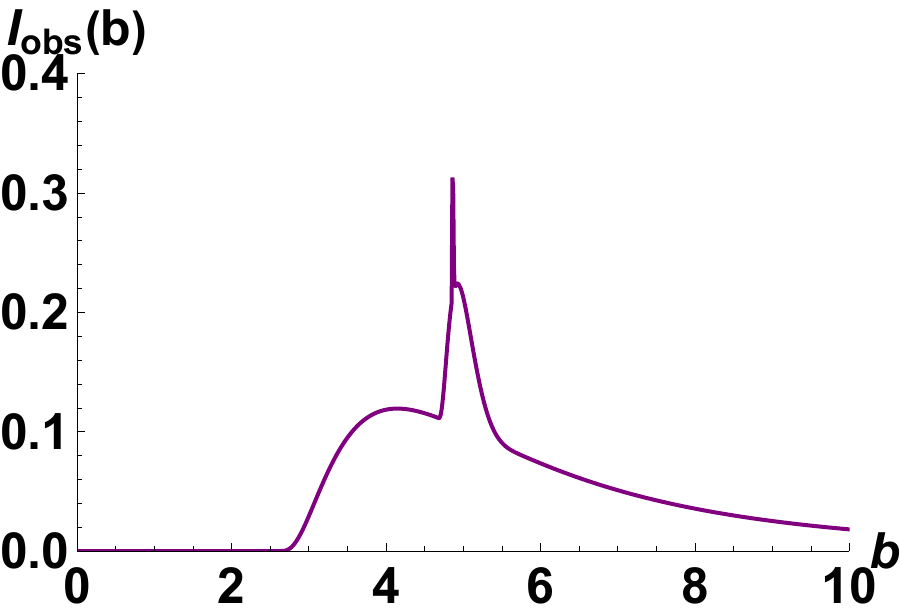} \label{}\hspace{2mm} \includegraphics[width=4.5cm]{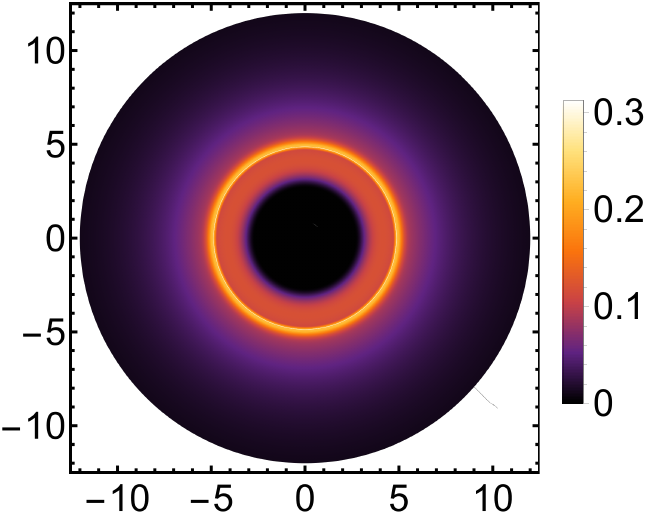}\hspace{2mm} \includegraphics[width=3.5cm]{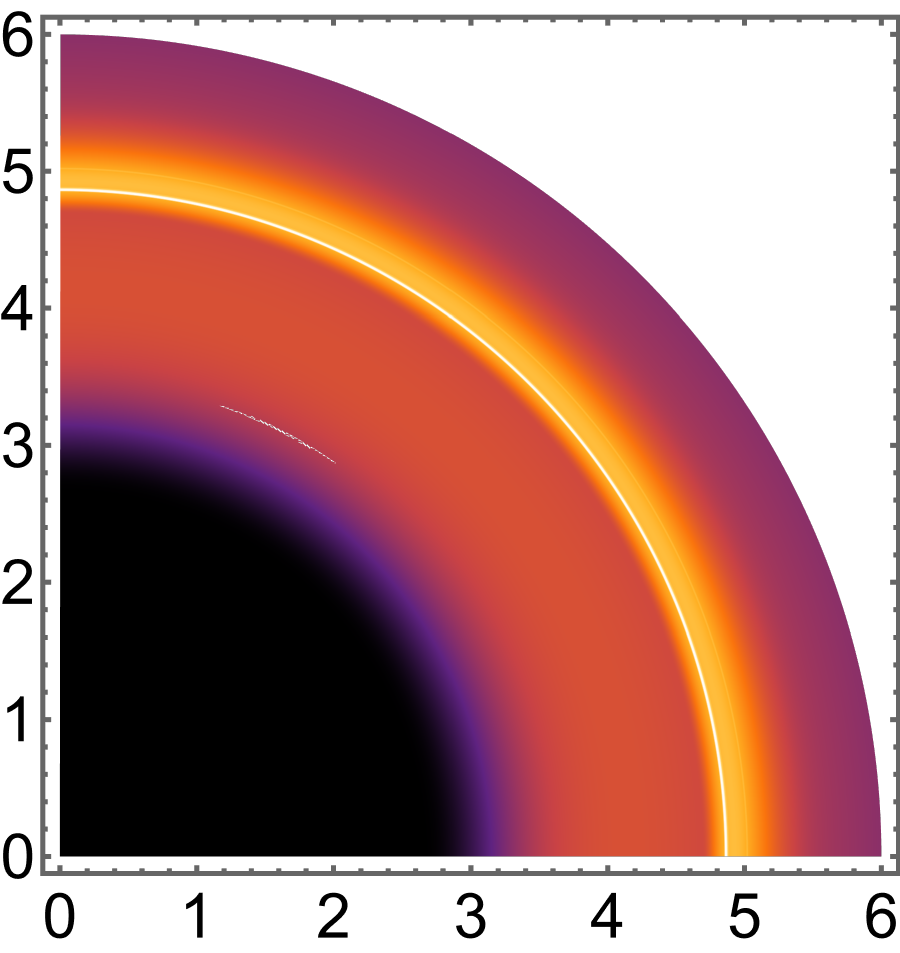}}\\
\subfigure[\, Horndeski:$Q=0.5$]
{\includegraphics[width=5.cm]{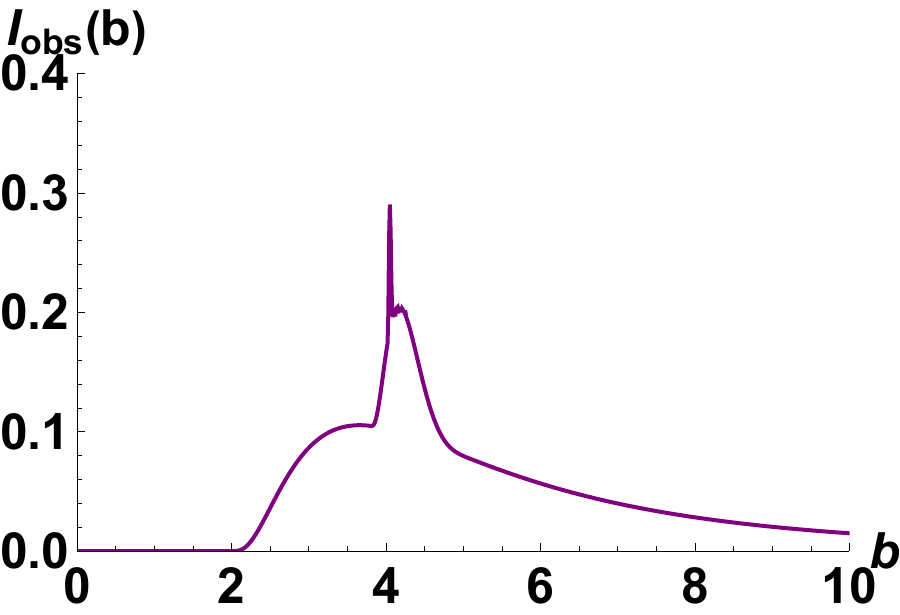} \label{}\hspace{2mm} \includegraphics[width=4.5cm]{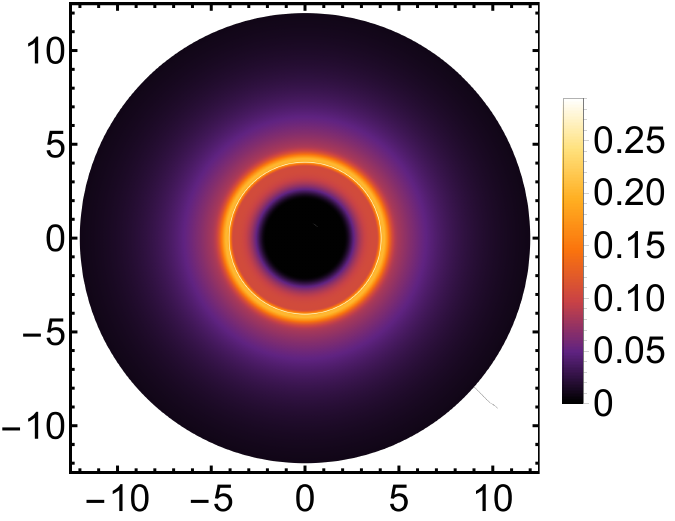}\hspace{2mm} \includegraphics[width=3.5cm]{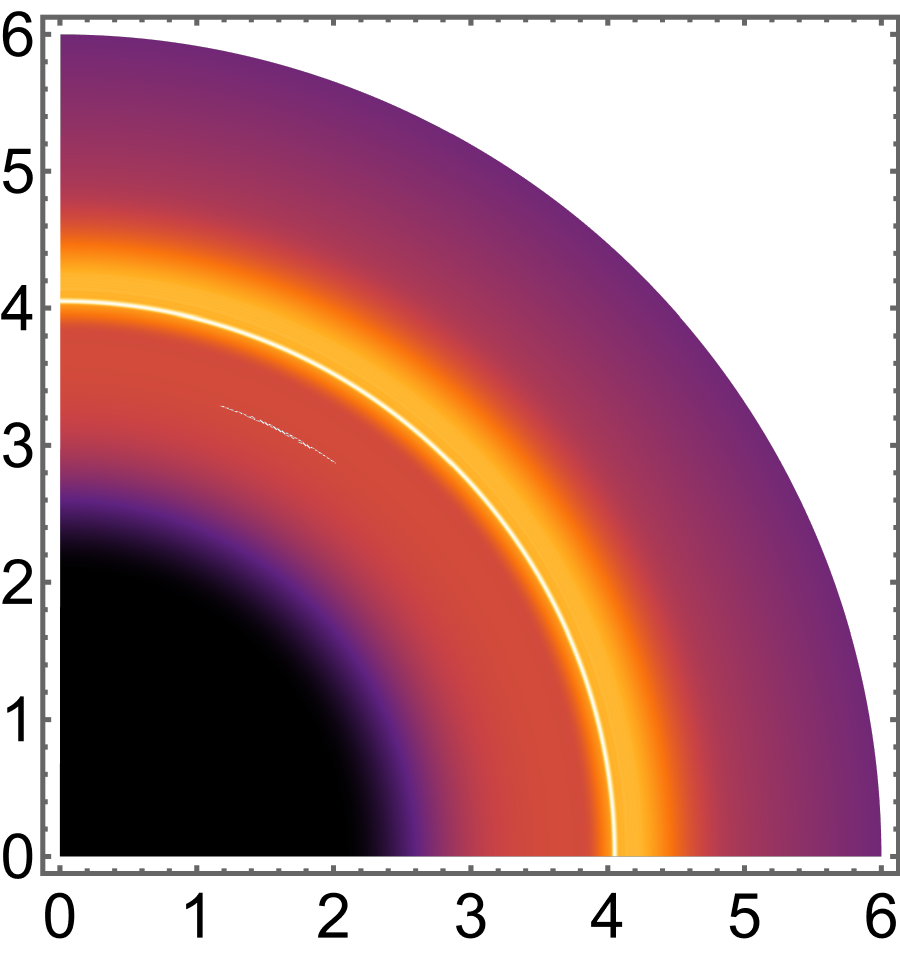}}\\
\caption{Optical  appearance of the black holes for \textbf{the case III} with different $Q$. \textbf{First column}: the total observed intensities $I_{obs}(b)$ as a function of the impact parameter. \textbf{Second column}:  the two-dimensional shadows cast of the black holes in the celestial coordinates. \textbf{Third column}: the zoomed in sectors. The black hole mass is set to 1.}
\label{modelthree-out}
\end{figure}

Utilizing the relation between the transfer functions and the total observed intensity (\ref{nbtotalinten}), we move on to investigate the optical appearance of the charged Horndeski (or RN) black hole. The total emission intensity $I_{em}(r)$ is determined by the emission position $r$ on the accretion disk. Thus our aim is to analyze the influence of the accretion disk radiation position on the observation appearance of the black hole. To end this purpose, we parameterize the radiations intensity profile of the accretion disk as the recently introduced Gralla-Lupsasca-Marrone (GLM) disk model \cite{Gralla:2020srx,Rosa:2023hfm,Rosa:2023qcv}, which has been illustrated to be in a close agreement with the observational predictions of general relativistic magneto-hydrodynamics simulations of astrophysical accretion disks.
The radiations intensity profile of the GLM model is written as \cite{Vincent:2022fwj}
\begin{align}
I_{em}(r,\gamma,\alpha,\beta)=\dfrac{exp\Big\{-\dfrac{1}{2}\Big[\gamma+arcsinh\left(\dfrac{r-\alpha}{\beta}\right)\Big]^2\Big\}}{\sqrt{(r-\alpha)^2+\beta^2}},\label{GLM}
\end{align}
where the shape of the radiations intensity profile $I_{em}$ depends on these free parameters $\gamma$, $\alpha$ and $\beta$: $\gamma$ is related to the rate of increase of the intensity profile from infinity down to the peak; $\alpha$ adjusts the radial position translation of the whole intensity profile; $\beta$ controls the dilation of the intensity profile as a whole. Hence, we can adjust the three free parameters  to select adequate radiations intensity profiles for the models under study. In this paper, we mainly investigate that three kinds of innermost radiation position of the accretion disk: the innermost stable circular orbit $r_{isco}$, the photon sphere radius $r_{ph}$ and the outside event horizon $r_+$. The result for these three situations are plotted in Figs.\ref{emitted-intenCase1}-\ref{modelthree-out}, and the more detailed discussions are as follows.

\textbf{Case I}: $\gamma=-2$, $\alpha=r_{isco}$ and $\beta=M/4$. Firstly, one assume that the innermost stable circular orbit $r_{isco}$ as the radiation stop position, which is the the bounder between test particles circling the black hole and test particles falling into the black hole. It can be obtained by \cite{Guo:2021bhr}
\begin{align}
r_{isco}=\dfrac{3A(r_{isco})A'(r_{isco})}{2A'(r_{isco})^2-A(r_{isco})A''(r_{isco})},\label{riscoEq}
\end{align}
which is derived in detail in the Appendix~\ref{A}.

Using (\ref{nbtotalinten}) and (\ref{GLM}), we respectively depict $I_{em}(r)$ as the function of $r$ in Fig.\ref{emitted-intenCase1} and $I_{obs}(b)$ as the function of $b$ in Fig.\ref{modelone-isco} for different values of $Q$, and the Fig.\ref{modelone-isco} shows the two-dimensional shadows cast of the black holes in the celestial coordinates as well. From the Fig.\ref{modelone-isco}, one can see that the direct emission has a big range of $b$, the lensed ring is limited to a small range of $b$, and the photon ring is an extremely narrow rang of $b$. Combining with the Fig.\ref{emitted-intenCase1} and Fig.\ref{modelone-isco}, the emitted positions on thin disk, the total observed intensity and the shadow size of the black hole decrease with the increase of the $Q$ in the charged Horndeski black hole, while these results decrease more slowly in the RN black hole when the $Q$ increases.

\textbf{Case II}: $\gamma=-2$, $\alpha=r_{ph}$ and $\beta=M/8$. Then, one assume that the region of emission is well outside the photon sphere. The Fig.\ref{emitted-inten2} and Fig.\ref{modeltwo-ph} describe that the $I_{em}(r)$ as a function of $r$ and $I_{obs}(b)$ as a function of $b$ with distinctive $Q$, respectively. One can see that the most important difference from the first model is that the positions of direct emissions are always located inside the lensed and photon ring intensities in Fig.\ref{modeltwo-ph}. Similar to previous analysis, 
the Fig.\ref{modeltwo-ph} shows the regions of the direct emission are much more than the regions of the lensed ring and the photon ring. From the Fig.\ref{emitted-inten2} and Fig.\ref{modeltwo-ph}, we can observe that the emitted positions on thin disk, the total observed intensity and 
the shadow size of the charged Horndeski black hole are smaller as a larger $Q$. In the same way, they are more weakly affected in the RN black hole as a larger $Q$ value.

\textbf{Case III}: $\gamma=-3$, $\alpha=r_{+}$ and $\beta=M/8$. Finally, one assume that the position of emission is extended all the way down to the outside horizon. For the different $Q$, the $I_{em}(r)$ as a function of $r$ and $I_{obs}(b)$ as a function of $b$ are already depicted in Fig.\ref{emitted-inten3} and Fig.\ref{modelthree-out}, respectively. The lensing ring and photon ring are superimposed on the direct image as well. From the Fig.\ref{modelthree-out}, the lensed ring is more prominent, but the direct emission remains dominant. The photon ring continues to be entirely negligible. Similarly, the the emitted positions on thin disk, the total observed intensity and the shadow size of the charged Horndeski black hole are smaller than those of the Schwarzschild black hole. In addition, the results are more significant in the charged Horndeski black hole in comparison with the RN black hole in Fig.\ref{emitted-inten3} and Fig.\ref{modelthree-out}.

\section{Conclusion and discussion}
\label{section5}
In this paper, we have investigated the shadow images and rings of the charged Horndeski black hole illuminated by spherical and thin disk-shaped accretion flows. We found that the larger electric charge results in a smaller peak effective potential at a larger charged Horndeski black hole radius, and the corresponding event horizon radius, shadow radius and critical impact parameter are gradually decrease with the increase of electric charge. In addition, we observed that the three functions are more sensitive to $Q$ in the charged Horndeski black hole than those of the RN black hole with the increase of $Q$ value.

Then, we studied the observation characteristics of the charged Horndeski black hole shadows and rings on two toy models of spherical accretion. For the charged Horndeski black hole illuminated by the static spherical accretion flow, we found that the total observed intensity increases at a smaller $b_{ph}$ with the increase of the electric charge. It has been found that the two-dimensional shadows in celestial coordinates are not a totally dark shadow with zero intensity due to part of the radiation of the accretion flow behind the black hole can escape to infinity. The shadows size and rings brightness around the charged Horndeski black hole are smaller and brighter than those of the RN black hole as a larger $Q$ value, respectively. Similarly, for the infalling spherical accretion flow, we found that there are have similar to results as well. It has shown that the brightness of rings in infalling accretion model are much less than those of static accretion model due to the Doppler effect, and the infalling accretion flow doesn't affect the shadow size of the back hole, implying that the black hole shadow size depends on the spacetime geometry and the ring luminosity depends on the accretion flow models.

Finally, we analyzed the shadow images and rings of the charged Horndeski black hole illuminated by thin disk-shaped accretion flows with different emission profiles. It has shown that the bright region near the charged Horndeski black hole can be divided into the direct emissions, lensed rings and photon rings according to the intersection times between the light line and the thin disk, and the contribution of the direct emission on the total observed intensity is dominating, the lensed ring is small, and the photon ring is hardly neglected. We found that the shadows size and the rings brightness around the charged Horndeski black hole are decrease with the increase of the electric charge, while the results are more obvious to $Q$ in the charged Hondeski black hole than those of the RN black hole. It has been demonstrated that the radiation positions of the accretion flows affect the shadows size and the rings brightness around the charged Horndeski black hole. In addition,  one can believe that the radius of the observed shadow is related to the boundary of the direct emissions. Therefore, we further investigate the relationship between the starting point of the direct emissions and the critical impact parameter, which may help us to utilize the EHT observations to measure the critical impact parameter, and then to test GR.

\section{Acknowledgements}
We are grateful to Profs. Hongsheng Zhang and Drs. Yihao Yin, Shi-Bei Kong, Arshad Ali and Yu-Ting Zhou for
interesting and stimulating discussions. This work is supported by
National Natural Science Foundation of China (NSFC) under Grant
nos. 12175105, 11575083, 11565017, 12147175, 12375043, Top-notch Academic
Programs Project of Jiangsu Higher Education Institutions (TAPP).

\appendix
\section{Details for the derivation of (\ref{potential}) and (\ref{riscoEq})}\label{A}
In this appendix, we give the detailed derivation of (\ref{potential}) and (\ref{riscoEq}). We set the $ds^2=\epsilon$, and then substituting (\ref{twoconserved}) into (\ref{4Dspacetime}), and one get
\begin{align}
B(r)\dot{r}^2+r^2\dot{\theta}^2=\epsilon+\dfrac{E^2}{A(r)}-\dfrac{L^2}{r^2}\sin^2\theta.\label{BrEq}
\end{align}
where $E$ and $L$ are the energy and the angular momentum of a particle; a dot denotes the derivative with respect to the affine parameter $\lambda$ of the geodesics; the $\epsilon=-1,0,1$ are corresponding to timelike, nulllike, spacelike geodesic. For simplicity, we choose the equatorial plane
$\theta=\pi/2$ to investigate the path of the particle. Therefore, the (\ref{BrEq}) can be simplified into the form
\begin{align}
\dot{r}^2=\dfrac{\epsilon}{B(r)}+\dfrac{E^2}{A(r)B(r)}-\dfrac{L^2}{B(r)r^2}\equiv f(r)^2.\label{massiveveffEq}
\end{align}

According to the Newtonian mechanics, one can obtain for the particle
\begin{align}
\dfrac{d\dot{r}}{d\lambda}=-\dfrac{\partial V_{eff}(r)}{\partial r},\label{Newmech}
\end{align}
and then the effective potential $V_{eff}(r)$ of the particle is derived as follows
\begin{align}
V_{eff}(r)=-\dfrac{1}{2}f(r)^2+C.\label{particleVeff}
\end{align}
We choose the $V_{eff}(r)|_{r=\infty}=0$ at infinity, and further using
(\ref{massiveveffEq}), we determine the constant
\begin{align}
C=\dfrac{1}{2}(E^2+\epsilon).
\end{align}
Finally, the effective potential is given by
\begin{align}
V_{eff}(r)=-\dfrac{1}{2}\left[\dfrac{\epsilon}{B(r)}+\dfrac{E^2}{A(r)B(r)}-\dfrac{L^2}{B(r)r^2}\right]+\dfrac{1}{2}(E^2+\epsilon).\label{FinparticleVeff}
\end{align}
The (\ref{FinparticleVeff}) is equivalent to (\ref{potential}) as $\epsilon=0$ for the photon. 

However, the $\epsilon=-1$ for the massive particle, hence the (\ref{FinparticleVeff}) is rewritten as
\begin{align}
V_{eff}(r)=\dfrac{1}{2}\left[\dfrac{1}{B(r)}-\dfrac{E^2}{A(r)B(r)}+\dfrac{L^2}{B(r)r^2}\right]+\dfrac{1}{2}(E^2-1).\label{massiveveff}
\end{align}
The conditions of the stable circular orbits for a massive particle in the equatorial plane of black hole are given as
\begin{align}
V_{eff}(r)=\dfrac{1}{2}(E^2-1), ~~V'_{eff}(r)=0.\label{massivecond}
\end{align}
Utilizing (\ref{massivecond}), $E$ and $L$ of the massive particle moving in a circular orbit in black hole can be obtained
\begin{align}
E=\dfrac{\sqrt{2}A(r)}{\sqrt{2A(r)-rA'(r)}},~~L=&\dfrac{r\sqrt{rA'(r)}}{\sqrt{2A(r)-rA'(r)}}.\label{massiveL}
\end{align}
The innermost stable circular orbits around black hole can be determined from the condition
\begin{align}
\dfrac{d^2V_{eff}(r)}{dr^2}\Big|_{r=r_{isco}}=0\label{massiverisco1}
\end{align}
Substituting (\ref{massiveveff}) into (\ref{massiverisco1}) and then using (\ref{massiveL}), we obtain the following equation
\begin{align}
\dfrac{3A(r)A'(r)-2rA'(r)^2+rA(r)A''(r)}{A(r)B(r)r[2A(r)-rA'(r)]}\Big|_{r=r_{isco}}=0.
\end{align}
Finally, we give the radius of the innermost stable circular orbit
\begin{align}
r_{isco}=\dfrac{3A(r_{isco})A'(r_{isco})}{2A'(r_{isco})^2-A(r_{isco})A''(r_{isco})}.\label{appenriscoEq}
\end{align}

\end{document}